\documentclass[a4paper,11pt]{article}
\usepackage[utf8]{inputenc}
\usepackage{fullpage}
\usepackage{amsmath,amssymb,amsthm}
\usepackage{hyperref}
\usepackage{mciteplus}
\usepackage{tikz}
\usetikzlibrary{arrows.meta,calc,decorations.markings,decorations.pathmorphing,patterns,math}
\usepackage{authblk}

\date{}

\title{\bf Quantum spectral problems and isomonodromic deformations}

\newcommand{\tr}{\operatorname{tr}}
\newcommand{\be}{\begin{equation}}
\newcommand{\ee}{\end{equation}}
\newcommand{\ba}{\begin{aligned}}
\newcommand{\ea}{\end{aligned}}
\newcommand{\re}{{\rm{e}}}
\newcommand{\ri}{{\rm{i}}}
\usepackage{color}

\author{Mikhail Bershtein, Pavlo Gavrylenko, Alba Grassi }

\begin{document}

  \newcommand\lb{\left(}
  \newcommand\rb{\right)} 
    \newcommand\rd{{\rm{d}}} 
\def\IZ{{\mathbb Z}}
\def\IR{{\mathbb R}}
\def\IC{{\mathbb C}}
\def\ICP{{\mathbb{CP}}}
\def\IN{{\mathbb N}}
\def\IP{{\mathbb P}}
\def\IT{{\mathbb T}}
\def\IS{{\mathbb S}}
\def\IF{{\mathbb F}}
\def\cN{{\mathcal N}}

\newtheorem{Example}{Example}

\newcommand\TT{{\mathbb T}}

\maketitle

\abstract{
We develop a self-consistent approach to study the spectral properties of a class of quantum mechanical operators by using the knowledge about monodromies of  $2\times 2$ linear systems (Riemann-Hilbert correspondence). Our technique applies to a variety of problems, though in this paper we only analyse in detail two examples. First we review the case of the (modified) Mathieu operator, which corresponds to a certain linear system on the sphere and makes contact with the Painlev\'e $\mathrm{III}_3$ equation. Then we extend the analysis to the 2-particle elliptic Calogero-Moser operator, which corresponds to a linear system on the torus. By using the Kiev formula for the isomonodromic tau functions, we obtain the spectrum of such operators in terms of self-dual Nekrasov functions  ($\epsilon_1+\epsilon_2=0$). Through blowup relations, we also find Nekrasov-Shatashvili type of quantizations ($\epsilon_2=0$). 

In the case of the torus with one regular singularity we obtain certain results which are interesting by themselves.  Namely, we derive blowup equations (filling some gaps in the literature) and we relate them to the bilinear form of the isomonodromic deformation equations. In addition, we extract  the $\epsilon_2\to 0$ limit of the blowup relations from the regularized action functional and CFT arguments.
}

{\vspace{-12.5cm}
\begin{flushright} \small CERN-TH-2021-070
\end{flushright}
\vspace{12.5cm}}

\tableofcontents

\numberwithin{equation}{section}

\section{ Introduction}
\paragraph{1.1} \label{p11}
The results of this paper are based on the interplay  between different branches of  mathematical physics. The key objects are 1d quantum mechanical operators, Painlev\'e equations, monodromies  of  \(2\times 2\) linear systems \cite{Gar,Fuchs,Sch,FN, Jimbo:1981zz, Malgrange1982}, 4d Nekrasov partition functions \cite{n,no2}, blowup relations \cite{ny1,Nakajima:2003uh} and conformal blocks \cite{BELAVIN1984333,DiFrancesco:639405,Zamolodchikov:1989mz}. At the center of this circle of connections lie \emph{linear systems}. 
We usually denote such system as 
\begin{equation}\label{eq: Linear system Intro}
	\frac{\mathrm{d}}{\mathrm{d}z}Y(z)=A(z) Y(z),
\end{equation}
where \(A(z)\) is a  \(2\times 2\) matrix, \(Y(z)=(Y_1(z),Y_2(z))^t\). 
 
One can also rewrite the linear system \eqref{eq: Linear system Intro}  as a second order differential equation in \(Y_1(z)\).  
To remove the first derivative term from this equation we also need to rescale \(Y_1(z)\) by switching to \(\widetilde{Y}_1(z)=\left( A_{12}(z) \right)^{-1/2}Y_1(z)\).
This  function satisfies a Schr\"{o}dinger--type equation of the form
\be( -\partial^2_z+W(z))\widetilde{Y}_1(z)=0.\ee
However, \(W(z)\) is not yet a good quantum mechanical potential because, first of all, it has extra singularities at the zeros of \(A_{12}(z)\) (so-called apparent singularities).
To work with an actual  quantum mechanical problem we demand that apparent singularities are hidden inside the actual singularities of \(A(z)\).
In this case the potential simplifies. We call the simplified version  \(U(z)\). Then we get the actual Schr\"odinger equation we are interested in:
\begin{equation}
\label{qm}
( -\partial^2_z+U(z))\widetilde{Y}_1(z)=0.
\end{equation}
Under  some special conditions on the monodromies of the linear system \eqref{eq: Linear system Intro} the function $\widetilde{Y}_1(z)$ becomes square integrable on some one-dimensional domain of the complex plane. Hence we get the eigenfunctions for a certain \emph{1d quantum mechanical operator}
 (and the corresponding formulas for its discrete spectrum) in terms of the solution to the \emph{isomonodromic deformation equations}. Indeed, the monodromy data of $A(z)$ are encoded in these equations.

In the simplest case these isomonodromic deformation equations are \emph{Painlev\'e equations}, see \cite{its-book} for a review and a list of references.  Among the corresponding quantum mechanical operators we recover the cubic, quartic and hyperbolic cosine potentials as well as (confluent) Heun's equation whose appearance in the context of Painlev\'e equations was also discussed in \cite{nok2, nok,Bender:2015bja, Lukyanov:2011wd,  dm2,dm1,Litvinov:2013sxa,Zabrodin:2011fk, Barragan-Amado:2020pad,Anselmo:2018zre,Barragan-Amado:2018pxh,CarneirodaCunha:2019tia,daCunha:2015ana,novaes2014,Lencses:2017dgf, Novaes:2018fry,Kashani-Poor:2013oza}.

Another case, which is also considered in this paper, is the linear system on the torus with one regular singular point of $A(z)$. The corresponding operator is  the 2-particle elliptic Calogero-Moser operator~\eqref{scw}.

\paragraph{1.2}  In the seminal paper \cite{gil1} Gamayun, Iorgov and Lisovyy suggested a formula for the tau function of Painlev\'e $\rm VI$  as a  sum of $c=1$\emph{ Virasoro conformal blocks} \footnote{Connection between isomonodromic deformations and two-dimensional quantum field theory was noticed before in \cite{SMJ1,SMJ2,SMJ3,SMJ4,SMJ5,WMT,Knizhnik_1989,Moore:1990aa}.}. In the last ten years this relation was proven and generalized to many other isomonodromic deformation problems, see for example \cite{gil,Bershtein:2014yia,Iorgov:2014vla,Gavrylenko:2016zlf,Gavrylenko:2017lqz,blmst,Nagoya:2015cja,nagoya2018remarks}. In particular, the generalization to the isomonodromic problem on the torus was recently worked out in \cite{Bonelli:2019boe, Bonelli:2019yjd,DelMonte:2020wty}. 

Due to the AGT correspondence \cite{agt} conformal blocks  essentially coincide with \emph{Nekrasov partition functions}. Hence, the aforementioned result can be stated as a formula expressing the solution to the isomonodromic deformation problem in terms of the self-dual (i.e.~$\epsilon_1+\epsilon_2=0$) Nekrasov partition functions\footnote{The relation between Painlev\'e equations and supersymmetric  gauge theories (their Seiberg-Witten curves) was noticed before in \cite{Mizoguchi:2002kg,Kajiwara:2004ri}.}. To be more precise, the isomonodromic tau function is equal to the Nekrasov-Okounkov dual partition function \cite{no2}. 
This correspondence is sometimes referred to as Isomonodromy/CFT/gauge theory correspondence. The formula for the tau function is usually called "Kiev formula", named after \cite{gil1}.

Using this correspondence, and the discussion of   paragraph 1.1, we get the exact formulas for the quantization conditions of the operators \eqref{qm} in terms of self-dual Nekrasov  functions.
More precisely the spectrum of  \eqref{qm} will be obtained by imposing (among other things) the vanishing of a suitable combination of isomonodromic tau functions \footnote{   
It is interesting to note that the existence of such expressions agrees with predictions from (the limit of) Topological String/Spectral Theory duality \cite{Grassi:2014zfa}.}. 

This also connects with the observation \cite{nok2, nok, Lukyanov:2011wd,  dm2,dm1,Litvinov:2013sxa}  that movable poles  in Painlev\'e $\rm III_3$, $\rm  I I$ and $\rm I$ are closely connected to the spectrum of a class of quantum mechanical operators.  Within our framework this observation can be straightforwardly generalised to other isomonodromic deformation problems, the corresponding  quantum  operator is simply obtained from \eqref{qm}.

\paragraph{1.3}  
There is another remarkable way to write down the discrete spectrum of these operators due to Nekrasov and Shatashvili (NS). In this approach the main ingredient is the NS limit (i.e. $\epsilon_2\rightarrow 0$)   \cite{Braverman:2004cr,ns,Nekrasov:2014yra} of Nekrasov  functions.  Compatibility between these two approaches follows from a special limit of \emph{Nakajima-Yoshioka blowup relations} \cite{ny1,Nakajima:2003uh}.  From this perspective one can view our results as an independent derivation of the NS formulas \cite{ns,Nekrasov:2014yra}. 

Let us note however that  the NS approach to spectral theory has some restrictions, for example when it comes to study the edges of the bands in periodic potentials, see for instance \cite{Basar:2015xna},  or the spectrum of relativistic integrable systems, see for instance  \cite{Grassi:2014zfa}.  On the contrary thinking in terms of vanishing of isomonodromic tau functions provides a unifying framework which naturally extends also to these situations. We discuss this briefly at the end of Sec~\ref{sec:others}.

The connection between the  self-dual and the NS limits of   Nekrasov  functions (or $c=1$ and $c=\infty$ conformal blocks) has been discussed in various contexts over the past few years.  
 For example, by using quantization conditions as motivation, a five dimensional version of  such relations was first proposed in  \cite{Sun:2016obh}.
The idea to use the $\epsilon_2\to 0$ limit of blowup relations for such problem can be found  in \cite{ggu}\footnote{This  was later used in  \cite{Huang:2017mis} and in several follow-up papers leading to the formulation of blowup equations for new class of geometries, see for instance \cite{Gu:2019pqj}.}. 
More recently, in \cite{Grassi:2019coc,Gavrylenko:2020gjb,OlegMovie}, a four dimensional version of these relations has been applied in the context of Painlev\'e equations and spectral theory.  We will discuss this further in the main text. 
Finally, in \cite{jtalk,1805497,Jeong:2020uxz} blowup equations for Nekrasov function with defects have been used to provide a direct link between the work of \cite{gil1} and the work of \cite{Litvinov:2013sxa} which relates the $c\to\infty$ limit of the $N_f=4$  BPZ  equation  to the Hamilton-Jacobi equation of Painlev\'e VI, see also \cite{Kashani-Poor:2013oza}. This has provided an alternative derivation for \cite{gil1} as well as a gauge theoretical meaning of the monodromy parameter $\eta$ appearing in the Kiev formula.

\paragraph{1.4}  This paper is structured as follows. 

In Sec.~\ref{main} we  accurately formulate the relationship between $2\times2$ linear systems and quantum mechanical operators. In order to get such operators (and the corresponding spectrum) we have to fulfil three constraints: the singularities matching condition, the reality condition and the square integrability of the solution.

In Sec.~\ref{sec:p3} we apply this procedure  to the example of Painlev\'e ${\rm III_3}$ whose associated \(2\times 2\) linear system is \eqref{Amat}.
The corresponding operator is the (modified) Mathieu 
\be \label{mat}-\partial^2_x+\sqrt{t}\left(\re^x+\re^{-x}\right)\,.\ee
We find that the operator spectrum is given by
\be\label{introe}{ E_n(t)=-t\partial_t\log\mathcal{T}_0 (\sigma_n,\eta_n,t)} \, , \ee
where $\mathcal{T}_0(\sigma,\eta,t)$ is the Painlev\'e ${\rm III_3}$ tau function, and $(\sigma_n,\eta_n)$ are solutions of\footnote{We can actually fix without loss of generality $\eta_n=0$ $\forall n$.} 
\be\label{introsing}{\mathcal{T}_0(\sigma+{1\over 2},\eta,t)=0 } \, , \ee
\be\label{introvai} \sin {\eta\over 2}=0\,.\ee
The variables $(\sigma, \eta)$ are  the monodromy data of the associated $2\times 2$ linear system given in \eqref{Amat} and they specify the initial conditions for the Painlev\'e \({\rm III_3}\) equation. In this language equation \eqref{introsing} corresponds to the singularity matching condition while \eqref{introvai} is the normalizability condition. See Sec.~\ref{sec:p3} for the details.
 Thanks to the Kiev formula \eqref{taup3}, $ \mathcal{T}_0$ is computed explicitly by using $c=1$ conformal blocks.  Hence \eqref{introe}-\eqref{introvai} completely determine the spectrum of \eqref{mat} in terms of the self-dual ($\epsilon_1+\epsilon_2=0$) Nekrasov function. The results of this section also overlap with \cite{Gavrylenko:2020gjb,Grassi:2019coc}.

 In  Sec.~\ref{sec:toro} we extend the analysis to the case of isomonodromic deformation on the one punctured torus. The associated $2\times 2$ linear system is given in \eqref{eq:torusSystem} and the isomonodromy equation corresponds to an elliptic form of  Painlev\'e VI  (the non-autonomous classical elliptic Calogero-Moser system \eqref{eq:Calogero}). 
 In this example our procedure leads naturally 
  to two quantum operators 
 \be\label{qecm} {\rm O}_{\mp}=-\partial_z^2+m(m\mp 1)\wp(z|\tau), \ee
which correspond to the 2-particle quantum elliptic Calogero-Moser operator. As in the previous case, the operator spectrum is obtained by asking some particular constraints on monodromy data of linear system  \eqref{eq:torusSystem}. 
 However, unlike in the previous case, here we have two charts parametrising the monodromy data of the linear system. We denote the corresponding coordinates  as \((\sigma, \eta) \) and \((\sigma, \tilde{\eta}).\)
In addition, to satisfy the condition of reality here we have several different options which require independent considerations, see Tab.~\ref{tab:homologyClasses}.  
Schematically, the spectrum of \eqref{qecm} is given by \footnote{Here $\sim$ means up to overall factors and shifts by $\eta_1(\tau)$ which depend on the reality condition we chose. The precise expression for each case is shown in the last column of Tab.~\ref{tab:homologyClasses}}
\be \label{eint}\ba E_n^{\rm \mp} \sim& H_\star^{\mp}(\sigma, \eta, \tau)\Big|_{(\sigma,\eta)=(\sigma_n^{\mp}, \eta_n^{\mp})}\\
&=\Big(\left.2\pi \ri\partial_{\tau}\log Z_0^D(\sigma,m,\eta,\tau)+2\pi \ri\partial_{\tau}\log \frac{\eta(\tau)}{\theta_3(0|2\tau)}{\mp }2 m \frac{\theta_3''(0|2\tau)}{\theta_3(0|2\tau)} \Big)\right|_{(\sigma,\eta)=(\sigma_n^{\mp}, \eta_n^{\mp})}\ea\ee
 where $Z_0$ is essentially the isomonodromic tau function on the torus as in \eqref{eq:tauFunction}. We also denoted by $(\sigma_n^{\mp}, \eta_n^{\mp})$  the set of values which satisfies the
  singularity matching condition \eqref{eq:apparentSingularityCancellation}
 \be\label{nointro} \theta_2(0|2\tau)Z_0^D\left(\sigma,m,\eta,\tau\right)-\theta_3(0|2\tau)Z_{1/2}^D\left(\sigma,m,\eta,\tau\right)=0\, ,
 \ee 
 as well as the normalizability condition listed in Tab.~\ref{tab:normalizabilityConditions}.  
 Note that the solutions to such equations contain simultaneously   the spectrum of $\mathrm{O}_+$ and $\mathrm{O}_-$.  Hence one still has to disentangle such solutions and map them either to $\mathrm{O}_+$ or to $\mathrm{O}_-$.  
 The Kiev formula \eqref{eq:dualNekrasov} give us $Z_{1/2}$ and $Z_{0}$ in terms of $c=1$ conformal blocks.  Hence the spectrum of \eqref{qecm} is completely determined in terms of self-dual ($\epsilon_1+\epsilon_2=0$) Nekrasov function.

In Sec.~\ref{sec:ns} we discuss the compatibility between our results and the Nekrasov-Shatashvili  (NS) exact quantization. For the Painlev\'e ${\rm III_3}$ example we essentially follow \cite{Gavrylenko:2020gjb,Grassi:2019coc} with slight improvement. For the example of the torus these results are new. Even blowup relations for the four-dimensional $\mathcal{N}=2^*$ theory were not written explicitly in the literature (see \cite{Gu:2019pqj} for the 5d version of some of these equations). For example, by using blowup equations we show that the solutions $\eta_\star^{\pm}$ to the singularity matching condition \eqref{nointro}  can be expressed as
\be {\eta_\star^{\pm}}=-\ri \partial_\sigma F^{\rm NS}(\sigma, m\mp{1\over 2}, {\tau}),\ee
where $F^{\rm NS}$ is the $c\to \infty$ conformal blocks on the torus. 

In Sec.~\ref{Sec:-2blowup} 
we  derive some new results for the isomonodromic  problem associated to the linear system on the torus with one regular singular point.  More precisely,  we show that the isomonodromic equation for the corresponding tau function takes the form of a very simple bilinear relation which is written in  equation  \eqref{eq: bilin dual Nekrasov tildei} and reads ( \(\mathfrak{q}=\re^{2\pi \ri\tau}\))
\begin{multline}
	 (\tilde{{Z}}_{0}^D)^2 \partial_{\log \mathfrak{q}}^2 \log \tilde{{Z}}_{0}^D+ 	(\tilde{{Z}}_{1/2}^D)^2 \partial_{\log \mathfrak{q}}^2 \log \tilde{{Z}}_{1/2}^D
	\\= 
	2\left(\frac{\partial_{\log \mathfrak{q}} \theta_3(0|\tau))}{\theta_3(0|\tau))} \Big(\partial_{\log \mathfrak{q}}-\frac{\partial_{\log \mathfrak{q}} \theta_3(0|\tau))}{\theta_3(0|\tau))}\Big)-m^2 \partial_{\log \mathfrak{q}}^2\log(\theta_3(0|\tau))\right) \big(\tilde{{Z}}_{0}^D \tilde{{Z}}_{0}^D
	+\tilde{{Z}}_{1/2}^D \tilde{{Z}}_{1/2}^D\big),
\end{multline}
where \(\tilde{{Z}}_{\epsilon}^D(\sigma,m, \eta,\tau)= \eta(\tau) {Z}_{\epsilon}^D(\sigma,m, \eta,\tau).\)
Such relation generalise to the torus setup the well known Hirota-like equations characterising  the Painlev\'e $\rm III_3$ tau functions. We use them, as well as the $\mathcal{N}=2^*$ blowup relations, to provide an alternative proof for the result of \cite{Bonelli:2019boe}.

In Sec.~\ref{functional} we deduce the NS limit of the blowup relations from the regularized action functional and CFT arguments. This is done by following the method developed in \cite{Gavrylenko:2020gjb}, which was also inspired by the works  of \cite{Litvinov:2013sxa,Lukyanov:2011wd}.

Finally, in Sec.~\ref{sec:others} we conclude by discussing  some other examples and generalisations.

There are five appendices which contain definitions (App.~\ref{conv} and App.~\ref{torusconvention}), additional tests (App.~\ref{sec:nstest}), technical details (App.~\ref{app:monodromies}), and some proofs (App.~\ref{proofs}).

 \section{General idea}\label{main}

The main idea can be outlined as follows.
We start from a \(2\times2\) linear system
\begin{equation}
\label{eq:linearSystem}
\frac{\mathrm{d}}{\mathrm{d}z}
\begin{pmatrix}
Y_1(z)\\Y_2(z)
\end{pmatrix}
=
\begin{pmatrix}
A_{11}(z)&A_{12}(z)\\
A_{21}(z)&-A_{11}(z)
\end{pmatrix}
\begin{pmatrix}
Y_1(z)\\Y_2(z)
\end{pmatrix}\,.
\end{equation}
For a given matrix
\be A(z)=\begin{pmatrix}
A_{11}(z)&A_{12}(z)\\
A_{21}(z)&-A_{11}(z)
\end{pmatrix}\ee
the global monodromy of the solution  \be {Y}(z)=\left(
\begin{array}{c}
 Y_1(z) \\
 Y_2(z) \\
\end{array}
\right)\ee is fixed. However, the opposite is generically not true. Given a solution ${Y}(z)$ with a corresponding monodromy, we can find a parametric family of matrices $A(z,t)$ realising such solution.  We refer to $A(z,t)$ as the set of isomonodromic deformations of $A(z)$. 
 One can then deduce that ${Y}(z,t)$ satisfies the following  system (see for instance \cite[Ch. 4]{its-book})
\be \label{iso}\ba  &\frac{\mathrm{d}}{\mathrm{d}z} {Y}(z,t) = A(z,t){Y}(z,t),\\
&\frac{\mathrm{d}}{\mathrm{d}t} {Y}(z,t) = B(z,t){Y}(z,t),
\ea\ee
where  $B(z,t)$ can be obtained with a suitable procedure once $A(z,t)$ is known.  The system \eqref{iso} comes together with a compatibility condition
\be \frac{\mathrm{d}}{\mathrm{d}z} \frac{\mathrm{d}}{\mathrm{d}t} {Y}(z,t)=\frac{\mathrm{d}}{\mathrm{d}t} \frac{\mathrm{d}}{\mathrm{d}z} {Y}(z,t),\ee
which can be written as
\be \label{comp} \frac{\mathrm{d}}{\mathrm{d}t} A(z,t)= \frac{\mathrm{d}}{\mathrm{d}z}B(z,t)+\left[B(z,t), A(z,t)\right].\ee
If $z$ is a coordinate on the $4$-punctured ${\ICP^1}$  and $A(z,t)\in sl(2, \IC)$ is suitably chosen, then \eqref{comp} takes the form of a  Painlev\'e equation, see for instance \cite[Ch. 4]{its-book}. The matrices  $A(z,t)$ and $B(z,t)$ are also known as Lax pairs. 
We claim that for a given isomonodromic problem, characterised by  $A(z,t)$, we can  associate a corresponding quantum mechanical operator \be \label{U0}- \partial_z^2+U(z, t)\ee
whose exact spectrum is computed using the tau function of the original isomonodromic problem. 

We proceed as follows. We wish to rewrite the linear system in the form of a 2nd order linear equation for \(Y_1(z,t)\). From the first equation in \eqref{eq:linearSystem} we have:
\begin{equation}
Y_2(z,t)=\frac{1}{A_{12}(z,t)}\left( Y_1'(z,t)-A_{11}(z,t)Y_1(z,t) \right)\, ,
\end{equation}
where by $'$ we denote the derivative w.r.t.~$z$.
By plugging \(Y_2(z,t)\) back into (\ref{eq:linearSystem}) we get \begin{equation}\ba
&Y''_1(z,t)+Y'_1(z,t)\left( -\frac{A_{12}'(z,t)}{A_{12}(z,t)}-\tr A(z,t) \right)+\\
&Y_1(z,t)\left( \det A(z,t)+\frac{A_{12}'(z,t)}{A_{12}(z,t)}A_{11}(z,t)-A_{11}'(z,t) \right)=0\,. \ea
\end{equation}
To remove the first derivative part we define:
\begin{equation}
\label{eq:Ytilde}
Y_1(z,t)=\sqrt{A_{12}(z,t)}\widetilde{Y}_1(z,t)\, .
\end{equation}
The resulting equation is
\begin{equation}
\label{eq:scalarEquation}
\left( -\partial^2_z+W(z,t) \right)\widetilde{Y}_1(z,t)=0\,,
\end{equation}
where
\begin{equation}
\label{eq:potential}
W(z,t)=   \left(-\det A+{A}_{11}'-\frac{A_{11}{A}_{12}'}{A_{12}}-\frac{2A_{12}{A}_{12}''-3({A}_{12}')^2}{4 A_{12}^2}\right).
\end{equation}
Furthermore, we want  \(\widetilde{Y}_1(z,t)\) to give the eigenfunction for some quantum mechanical problem. To achieve this we need to fulfil several requirements:
\begin{enumerate}
\item \label{item:1} { Let us denote the zero of \(A_{12}(z,t)\) by \(z_0\). Due to the change of variable \eqref{eq:Ytilde}, the equation (\ref{eq:scalarEquation}) has apparent singularity (singularity with trivial monodromy \((-1)\)) at \(z=z_0\).
We require that such apparent singularities match with the existing singularities of $A(z,t)$.  We refer to this constraint as {\it{ singularities matching condition}}. This requirement gives some restrictions on the matrix elements of \(A(z,t)\) and has two consequences. 
\begin{enumerate}
\item On one  hand such restriction  leads to a further simplification of the potential 
\be \label{ufin}W(z,t) \quad \xrightarrow[\text{condition}]{\text{  singularities matching}} \quad  U(z,t).\ee
\item \label{vanishing}On the other hand \(A(z,t)\)  are dynamical variables in the isomonodromic problem. Hence the aforementioned  condition can be written as a vanishing condition involving some particular combination of  isomonodromic tau functions. 
\end{enumerate}}

\item \label{item:2}  We want the operator \( \re^{\ri\alpha}\left(\partial^2_z-U(z,t)\right) \) to be self-adjoint on some one-dimensional domain \(\mathcal{C}\) in the variable \(z\) and for 
 some values of  \(\alpha\).
This requirement gives some reality conditions for the parameters of the potential and for the domain of $z$.

\item\label{item:3} We also demand that \(\widetilde{Y}_1(z,t)\) is normalizable.
For periodic potentials we don't need this condition.
For confining potentials this condition, together with point (b) above, gives an equation for the spectrum.
More precisely, it gives some constraints on the monodromy data: the transport matrix between two singular points should map regular solutions to regular solutions.
This is a very standard idea from quantum mechanics, but in contrast to the usual quantum mechanical problems, here the monodromies of \(\widetilde{Y}_1(z,t)\) are known by construction. 
\end{enumerate}

In this way we get a self-consistent approach which allows us to study the spectrum of some quantum mechanical operators by using the knowledge about isomonodromic deformations. 
In Sec.~\ref{sec:p3} and Sec.~\ref{sec:toro} we illustrate this procedure in details for the example of Painlev\'e ${\rm III_3}$ and for the  isomonodromic deformation on the torus.

\section{Modified Mathieu equation and Painlev\'e $\rm III_3$}\label{sec:p3}

In this section we will apply the strategy presented in Sec.~\ref{main} to the isomonodromic problem leading to the Painlev\'e $\rm III_3$ equation. As explained below, the relevant quantum operator in this context is the modified Mathieu, or 2-particle quantum Toda Hamiltonian.
Connection between the spectrum of modified Mathieu and the poles of Painlev\'e $\rm III_3$ have been observed for instance in \cite{nok, PCL, Lukyanov:2011wd} at the level of asymptotic expansions as well as numerically. 
This interplay was recently revisited in \cite{Grassi:2019coc,Gavrylenko:2020gjb} from the optic of the $\Omega$ background and blowup equations.

We follow \cite{Gavrylenko:2017lqz} (some formulas for Painlev\'e are taken from \cite{Bershtein:2016uov}). The linear system associated to  Painlev\'e   $\rm III_3$ has the form 
\begin{align}\label{linear}
	\frac{\mathrm{d}}{\mathrm{d}z}Y(z,t)=A(z,t) Y(z,t), 
\end{align}
where $z\in \ICP^1$ and
\begin{align}\label{Amat}
	A(z,t)=z^{-2}	\begin{pmatrix}
		0 & 0 \\w &0
	\end{pmatrix}
	+z^{-1}\begin{pmatrix}
		-p/w & t/w \\-1 & p/w
	\end{pmatrix}
	-\begin{pmatrix}
		0 & 1 \\0 &0
	\end{pmatrix},
\end{align}
where $w=w(t)$ and $p=p(t)$.
The compatibility condition of this isomonodromy problem is given by (see e.g \cite[eq. (2.10)]{Gavrylenko:2017lqz})
\be \label{hamform}
 \begin{cases} t{{\rm d} w \over \rd t}  =2p+w,\\
 t{{\rm d} p \over \rd t}=\frac{2p^2}{w}+p+w^2-t,
 \end{cases}
 \ee
  which can be written as the known Painlev\'e $\rm III_3$ equation:
 \be\label{PIIID8}
 {{\rm d^2} w \over \rd^2 t} =\frac{1}{w}\left( {{\rm d} w\over \rd t} \right)^2-\frac{1}{t}{{\rm d} w\over \rd t} +\frac{2w^2}{t^2}-\frac2t.
 \ee
By using 
\be w(t)=-(r/8)^2\re^{\ri u(r)}, \quad t=(r/8)^4\ee
 we can write \eqref{PIIID8} as the radial sin-Gordon equation
 \be\label{sinh} {{\rm d^2} u \over \rd^2 r}+r^{-1}{{\rm d} u \over \rd r}+\sin u =0. \ee
 We also introduce the  Hamiltonians $H_i$ as
 \begin{equation}\label{zetataudef}
	 H_0=H_1+\frac{p}{w}+\frac{1}{4} ,\quad   H_1=\frac{p^2}{w^2}-w-\frac{t}{w}.\,
\end{equation}
Note that here $p$ and $w$ are not canonical coordinates, we have $\{w,p\}=w^2$. Transformation to canonical coordinates from \cite{Bershtein:2016uov} is $p \mapsto pw^2-w/2, w\mapsto w$. There is a B\"acklund transformation of the Painlev\'e \(\rm III_3\) equation which permutes $H_{0}, H_{1}(t)$, it maps $w\mapsto t/w$.

The tau functions are defined as
\be \label{hm} H_i(t)=t\frac{\mathrm{d} \log\mathcal{T}_i(t)}{\mathrm{d}t}. \ee
The Painlev\'e transcendent $w$ can be expressed as 
\begin{equation} \label{eq:w through Tau}
	w=\frac{-1}{\partial^2_{\log t}\mathcal{T}_0 }=t^{1/2}\frac{\mathcal{T}_0^2}{\mathcal{T}_1^2}. 
\end{equation}
The tau functions $\mathcal{T}_{0}(t), \mathcal{T}_{1}(t)$ are holomorphic on the universal covering of $\mathbb{C}\setminus\{0\}$. Let $\{t_n(\sigma,\eta)\}_{n\geq 0}$ denote the zeros  of $\mathcal{T}_{1}(t)$, they correspond to movable poles of the Painlev\'e transcendent $w$.

Remarkably,  the tau function of Painlev\'e $\rm III_3$ has been computed in  \cite{gil, gil1} for generic initial conditions. They found
\be \label{taup3}\ba& \mathcal{T}_0(\sigma,\eta,t)=\sum_{n\in \mathbb{Z}}\re^{\ri n \eta} \frac{t^{{(\sigma+n)^2}}}{G\left(2\sigma+2n+1\right) G\left(1-{2\sigma}-2n \right)} Z(\sigma+n , t),\\
&\mathcal{T}_1(\sigma,\eta,t)=\sum_{n\in \frac12+
 \mathbb{Z}}\re^{\ri n \eta} \frac{t^{{(\sigma+n)^2}}}{G\left(2\sigma+2n+1\right) G\left(1-{2\sigma}-2n \right)} Z(\sigma+n , t),\ea\ee
where $G(z)$ denotes the Barnes $G$ function and $Z(\sigma, t)$ is the irregular $c=1$ Virasoro conformal block \footnote{Via AGT \cite{agt} this is equivalent to Nekrasov partition function of pure $\mathcal{N}=2$, $SU(2)$ SYM in the four-dimensional self-dual phase ($\epsilon_1+\epsilon_2=0$) of the $\Omega$ background \cite{n}.} whose precise definition can be found for instance in  \cite[eqs. (3.4)-(3.6)]{ilt}. The first few terms read
\be \label{zmath}
	Z(\sigma, t)=1 +\frac{ t}{2{\sigma^2}}+{8 \sigma^2+1\over 4 \sigma^2(4\sigma^2-1)^2}t^2+\mathcal{O}(t^3).
\ee
Higher order terms can be computed systematically by using combinatorics and Young diagrams, we refer to \cite{ilt} for the details.  
The parameters  $(\sigma, \eta)$ are  related to the monodromies  of the linear system \eqref{linear} around $z=0,\infty$ and parametrise the space of initial conditions (see for instance  \cite[Sec. 2]{Gavrylenko:2017lqz}  or \cite[Sec. 2]{ilt}).
It was proven in \cite{ilt} that, as long as $2 \sigma\not \in \mathbb{Z}$, the series \eqref{taup3} converges uniformly and absolutely on every bounded subset
of the universal cover of $\mathbb{C}\setminus \{0\}$.

The expression \eqref{taup3}  is also known as Kiev formula for Painlev\'e ${\rm III_3}$.

\subsection{Singularities matching condition}

 It is convenient to introduce $x=\log z$, as well as $\tilde{Y_1}(z)=\re^{x/2}\Psi(x)$. Then  \eqref{eq:scalarEquation}  reads
\be\label{eq:potential2} \left( \partial_x^2 - V(x,t)\right)\Psi(x) =0 \, ,\ee
where
\be V(x,t)=\re^{2x}W(\re^x,t) +{1\over 4}.\ee
By using the explicit expression \eqref{Amat} we get
\begin{equation}\label{eq:2nd from 1nd Nf=0}
 V(x,t)= \frac{p^2+p w-w \left(t+w^2\right)}{w^2}+\frac{t (p+w)}{w (w \re^x-t)}+\frac{3 t^2}{4 (t-w \re^x)^2}+\frac{t}{\re^x}+\re^x+\frac{1}{4}.
\end{equation}\newline 
The linear system  \eqref{linear} has singularities at $x=\pm \infty$. 
However, since \be \label{cancMath}A_{12}(x)=-(\re^x-t/w), \ee we have an auxiliary pole in the equation \eqref{eq:potential2} at the point $x=\log(t/w)$. 
We do not have this pole if $w=\infty$ or $w=0$. Hence, we need to be at such points (singularities matching condition). Let us analyse these two cases in more detail. \newline  

\textbf{Case $w=\infty$} at some time $t_{\star}$.  Using \eqref{hamform} we have
\be \label{wsol}
w \sim \frac{t_\star^{2}}{(t-t_\star)^{2}}+\mathcal{O}(1),\quad p\sim \frac{-t_\star^{3}}{(t-t_\star)^{3}}-\frac{t_\star^{2}}{2(t-t_\star)^{2}}+\mathcal{O}(1).
\ee 
It follows from these expressions, or from \eqref{eq:w through Tau}, that
\be  \label{eq:cancMath}
\mathcal{T}_1(t_\star)=0,
\ee
as well as $\mathcal{T}_0(t_\star)\neq 0$, and $H_0$ is finite.

\textbf{Case $w=0$} at some time $t_\star$. 
We have 
\[w \sim t_\star^{-1}(t-t_\star)^{2}+\mathcal{O}((t-t_\star)^{3}),\quad p \sim (t-t_\star) +\mathcal{O}((t-t_\star)^{2}).   \]  It follows from these expressions, or from \eqref{eq:w through Tau}, that
\be \mathcal{T}_0(t_\star)=0, \ee
as well as $\mathcal{T}_1\neq 0$, $H_1$ is finite. 

The two cases  $w=0$ and  $w=\infty$ are actually related by B\"acklund transformation and are equivalent. In the rest of the work we will focus without loss of  generality on  the constraints coming from imposing   $w=\infty$.

\subsection{Quantum mechanical operator} 
It is easy to see that if we expand the potential \eqref{eq:2nd from 1nd Nf=0} around $w=\infty$ , we obtain
\begin{equation}\label{PU3}
V(x,t)\quad \to \quad U(x,t)=\re^{x}+t\re^{-x}+H_0.
\end{equation}
Here and below we use \(t\) instead of \(t_{\star}\) for simplicity.
The corresponding spectral problem is\footnote{We shifted here \(x\mapsto x+\frac12\log t\).}
\be \label{mathieu}\left( \partial_x^2 - \left(\sqrt{t} \re^x+\sqrt{t} \re^{-x}-E\right)\right)\Psi(x)=0, \ee
which is the well known (modified) Mathieu operator.
Moreover, from \eqref{hm} and \eqref{PU3} we have \be\label{ez1} E=-H_0=-t\frac{\mathrm{d} \log\mathcal{T}_0(\sigma, \eta,t)}{\mathrm{d}t}. \ee
 If $\sqrt{t} >0$,  $x\in \IR$  this operator is self-adjoint with a positive discrete spectrum on $L^2(\IR)$.

\subsection{Quantization conditions and spectrum}

According to our general approach illustrated in Sec.~\ref{main}, the exact quantization condition for the operator \eqref{mathieu} is obtained by asking simultaneously  the singularities matching condition as well as the normalizability   of the associated linear problem.
The singularities matching condition is given in equation \eqref{eq:cancMath}: \be \label{ttv}\mathcal{T}_1(\sigma, \eta,t)=0.\ee 
The condition  that ${Y_1}$ is normalisable can be expressed in terms of the connection matrix $ \mathcal{E}$ for the Painlev\'e ${\rm III_3}$ equation. We follow \cite[Sec.2]{Gavrylenko:2017lqz} and use a gauge transformation together with the two-fold covering \(z=\zeta^2\) to write \eqref{linear} in  the form 
\be \label{linear2}\partial_{\zeta}\widehat Y(\zeta,t)=\widehat A(\zeta,t) \widehat Y(\zeta, t)\ee
with
\be \widehat Y (\zeta,t)= \left(
\begin{array}{cc}
 \frac{\ri}{\sqrt{2} \sqrt{\zeta }} & \frac{\ri \sqrt{\zeta }}{\sqrt{2}} \\
 \frac{\ri}{\sqrt{2} \sqrt{\zeta }} & -\frac{\ri \sqrt{\zeta }}{\sqrt{2}} \\
\end{array}
\right)  Y (\zeta^2,t)\ee
and \be \widehat A(\zeta,t)= {1\over \zeta^2} \left(\left(w+{t\over w}\right)\sigma_3+\left(w-{t\over w}\right)\ri\sigma_2\right)-{1\over \zeta}\left({2p\over w}+{1\over 2}
\right)\sigma_1-2\sigma_3, \ee
where $\sigma_i$ are the Pauli matrices. %
The reason for such rewriting is that the matrices multiplying $\zeta^{-2}$ and $\zeta^{0}$ in \eqref{linear2} are diagonalizable,
hence we can easily write  formal solutions around $\zeta=0$ and $\zeta=\infty$.   Around $\zeta=0$ we have
  \be \widehat Y^{(0)}_{\rm form}(\zeta,t)=  \left(-{w\over\sqrt{t}}\right)^{-\sigma_1/2}\left( {1\!\!1}+ \mathcal{O}(\zeta)\right)\re^{2\sqrt{t} \sigma_3/\zeta}.\ee
Likewise  around $\zeta=\infty$ we have 
\be \widehat Y^{(\infty)}_{\rm form}(\zeta,t)=\left( {1\!\!1}+ \mathcal{O}(\zeta^{-1})\right)\re^{-2\sigma_3 \zeta}.\ee
The connection matrix $ \mathcal{E}$ relates solutions around $0$ to  solutions around $\infty$ as
$\widehat Y ^{(0)}\mathcal{E} =\widehat Y ^{(\infty)}$. We have (see \cite[eq. (2.7)]{Gavrylenko:2017lqz}):
\be \mathcal{E}={1\over \sin (2 \pi \sigma)}\left(\begin{matrix} \sin (\eta/2)& -\ri \sin (2 \pi \sigma+\eta/2)\\
\ri \sin (2 \pi \sigma-\eta/2)&\sin (\eta/2)
\end{matrix}\right).\ee
Normalizability of $ Y$
requires that we  map decaying solutions around $\zeta= 0$ to decaying solutions around $\zeta= \infty$. Hence the diagonal elements of $\mathcal{E}$ have to vanish:    \be \label{nor}\sin \left({\eta\over 2}\right)=0.\ee
By combining \eqref{ttv}, \eqref{nor}, and the B\"acklund transformation \(\mathcal{T}_1(\sigma,\eta,t)\sim\mathcal{T}_0(\sigma+\frac12,\eta,t)\) from \cite{Bershtein:2014yia}
we get the quantization condition for modified Mathieu:
\be{\label{taum} \mathcal{T}_0(\sigma+{1\over 2},0,t)=0.} \ee
From the point of view of spectral theory we think of \eqref{taum} as a quantization condition for $\sigma$. If we denote the solutions to such quantization condition by \be \label{sigman}\{\sigma_n\}_{n\geq1},\ee then the spectrum $\{E_n(t)\}_{n\geq 1}$ of modified Mathieu is obtained from
\eqref{ez1} and reads
\be\label{eeni}{ E_n(t)=-t\partial_t\log\mathcal{T}_0 (\sigma_n,0,t)}.\ee 
We also cross-checked against explicit (numerical) computations that \eqref{eeni} and \eqref{taum} indeed reproduce the correct spectrum of modified Mathieu. Hence, from that point of view, the exact quantization condition of modified Mathieu follows from the  Kiev formula for the tau function of Painlev\'e ${\rm III_3}$ \eqref{taup3} and can be expressed entirely by using $c=1$ Virasoro conformal blocks.

Some comments.

\begin{itemize}
\item In the work \cite{nok2}, which was later made more precise in \cite{Lukyanov:2011wd, PCL}, the Author considers the semi-classical Bohr-Sommerfeld quantization for the  Mathieu operator as a quantization for the variable $t$ in \eqref{mathieu}.  Then he connects such solutions $\{t_n(E)\}$ (in the limit $n\to \infty)$ to the poles in the time variable $t$ of the function $u$ satisfying \eqref{sinh}. Roughly speaking one has $u\sim\log (\sqrt{t}-\sqrt{t_n})$. These poles are the zeros of the tau function.

Here instead we are  using the inverse analysis. 
 We do not start from the quantization for the  Mathieu operator: we derive it from tau function of Painlev\'e $\rm III_3$ as computed in \cite{gil}. 

\item Note that \eqref{taum} is precisely the condition found in \cite[Sec.~6]{Grassi:2019coc} even though the derivation of \cite{Grassi:2019coc} is different from the approach presented in this section. Moreover in \cite{Grassi:2019coc} one still needs to relay on Matone relation (hence $c=\infty$ conformal blocks). Instead in our perspective we have \eqref{eeni}. We will see in Sec.~\ref{sec:buMa} that \eqref{eeni} and Matone relation are connected via blowup equations.

\item Some of the results presented in this section  overlap with \cite{Gavrylenko:2020gjb}. 
\end{itemize}

\section{Weierstrass  potential and  isomonodromic deformations on the torus}\label{sec:toro}

In this section we will apply the strategy presented in Sec.~\ref{main} to the isomonodromic problem on the one-punctured torus which was studied recently in \cite{Bonelli:2019boe}, and then in \cite{DelMonte:2020wty}.  As explained below, the relevant quantum operator in this context is the 2-particle  quantum elliptic Calogero-Moser Hamiltonian.

We follow \cite{Bonelli:2019boe}.
We start from the following  linear system: 
\begin{equation}\label{eq:torusSystem}
	{\rd \over \rd z} Y(z,\tau)=
	A(z,\tau)
	Y(z,\tau)\,,
\end{equation}
with
\begin{equation}\label{At} 
	A(z,\tau)= 
	\begin{pmatrix}
		p&m x(2Q,z)\\
		m x(-2Q,z)&-p
	\end{pmatrix},
\end{equation}
where 
\[
	x(u,z)=\frac{\theta_1(z-u|\tau)\theta_1'(0|\tau)}{\theta_1(z|\tau)\theta_1(u|\tau)}
\]
is the Lam\'e function, $\theta_1$ is the Jacobi theta function, and $\theta_1'(0|\tau)=\partial_z\theta_1(z|\tau)\Big|_{z=0}$. See Appendix~\ref{conv} for the conventions.
The coordinate $z$ is on the torus $\IT ^2$ with modular parameter $\tau$. Note that \eqref{At} has a simple pole at  $z=0$. 
It was shown in  \cite{levin1999hierarchies,takasaki1999elliptic,Levin2000,manin}, see also \cite{Bonelli:2019boe} and reference therein, that the compatibility condition of the linear system \eqref{eq:torusSystem}
leads to 
\be \label{pdef}p=2 \pi \ri \partial_{\tau}Q,\ee
\begin{equation}
\label{eq:Calogero}
(2\pi \ri)^2{ \rd^2 \over \rd \tau^2} Q=m^2 \wp(2 Q|\tau)',
\end{equation}
where $'$ refers to the derivative w.r.t.~the first argument, and $\wp$ is the Weierstrass function  defined in  \eqref{wpdef}.
We see that the potential of this system is the Weierstrass \(\wp\)-function, where the time \(\tau\) is identified with the modular parameter: this is  the classical non-autonomous 2-particle elliptic Calogero-Moser system.

Notice that \eqref{eq:Calogero} has to be supplied by two integration constants $(\sigma, \eta)$. These are such that at $m=0$ we have
\be \label{eq:Q for m=0}
	Q\Big |_{m=0}= \tau \sigma  +{\eta\over 4 \pi}. 
\ee
The variables $(\sigma, \eta)$  are also related to the monodromies of $Y(z,t)$ around the A and B cycles of the torus, whereas $m$ determines the monodromy around the singularity $z=0$.  This is explained in \cite[Sec.~4.2]{Bonelli:2019boe}. We report some of these results in Appendix \ref{app:monodromies}.

One also defines the Hamiltonian associated to \eqref{eq:Calogero}, \eqref{pdef}  as \cite[eq. (3.9)]{Bonelli:2019boe}
\begin{equation}
\label{eq:1}
H=p^2-m^2\left(\wp(2Q|\tau)+2 \eta_1(\tau)\right),
\end{equation}
where $\eta_1(\tau)$ is defined in \eqref{eta1def}.
The tau function $\mathcal{T}$ corresponding to the linear system \eqref{eq:torusSystem} is then defined following \cite{Jimbo:1981zz,
Jimbo:1982zz, Malgrange1982,2010CMaPh} 
as
\be\label{hdef} 
	H=2\pi \ri \partial_\tau \log \mathcal{T}. 
\ee
It is very convenient to introduce the functions $Z^D_{0}, Z^D_{1/2}$ as in  \cite{Bonelli:2019boe} by the formula
\begin{equation}
	\label{eq:tauFunction}
	\mathcal{T}(\sigma,m,\eta,\tau)=\frac{\eta(\tau)Z^D_{1/2}(\sigma,m,\eta,\tau)}{\theta_2(2Q|2\tau)}=\frac{\eta(\tau)Z^D_{0}(\sigma,m,\eta,\tau)}{\theta_3(2Q|2\tau)},
\end{equation}
where $\eta(\tau)$ is the Dedekind's $\eta$ function defined in \eqref{deddek}, and  $Q= Q(\sigma, m, \eta, \tau)$ is a solution of \eqref{eq:Calogero}.
These formulas express indirectly both the tau function \(\mathcal{T}\) and the transcendent \(Q\) in terms of some functions \(Z_0^D\) and \(Z_{1/2}^D\).
Note that the function $Q$ is determined by $Z^D_{0}$ and $Z^D_{1/2}$ (up to a sign and shifts by $\mathbb{Z}+\mathbb{Z}\tau$) via the equation 
\begin{equation}
	\label{eq:solutionTorus}
	\frac{\theta_2\left(2Q(\sigma,m,\eta,\tau)|2\tau\right)}{\theta_3\left(2Q(\sigma,m,\eta,\tau\right)|2\tau)}=\frac{Z_{1/2}^D\left(\sigma,m,\eta,\tau\right)}{Z_0^D\left(\sigma,m,\eta,\tau\right)}\, .
\end{equation}
Indeed, suppose that $\tilde{Q}$ solves the same equation, then using the relation 
\be\label{Quni}
	\frac{\theta_2(2\tilde{Q}|\tau)}{\theta_3(2\tilde{Q}|\tau)}-\frac{\theta_2(2Q|\tau)}{\theta_3(2Q|\tau)}=\frac{\theta_1(Q-\tilde{Q}|\tau)\theta_1(Q+\tilde{Q}|\tau)}{\theta_3(2\tilde{Q}|2\tau)\theta_3(2Q|2\tau)}
\ee
we get \(\tilde{Q}=\pm Q+n\tau+\ell\), with $n,\ell \in \IZ$.

In order to write the isomonodromic deformation equations in terms of $Z^D_{0}, Z^D_{1/2}$ it is convenient to introduce 
	\begin{equation}
	\label{eq: dual Nekrasov tilde}
	\tilde{{Z}}_{\epsilon}^D(\sigma,m, \eta,\tau)= \eta(\tau) {Z}_{\epsilon}^D(\sigma,m, \eta,\tau).
\end{equation}
We will show in Sec.~\ref{Sec:-2blowup} that if
 $Q$ defined by \eqref{eq:tauFunction} satisfies the isomonodromic deformation equations \eqref{eq:Calogero},
then $\tilde{Z}^D_{0}, \tilde{Z}^D_{1/2}$ satisfy 
\begin{multline}\label{eq: bilin dual Nekrasov tildei}
	 (\tilde{{Z}}_{0}^D)^2 \partial_{\log \mathfrak{q}}^2 \log \tilde{{Z}}_{0}^D+ 	(\tilde{{Z}}_{1/2}^D)^2 \partial_{\log \mathfrak{q}}^2 \log \tilde{{Z}}_{1/2}^D
	\\= 
	2\left(\frac{\partial_{\log \mathfrak{q}} \theta_3(0|\tau)}{\theta_3(0|\tau)} \Big(\partial_{\log \mathfrak{q}}-\frac{\partial_{\log \mathfrak{q}} \theta_3(0|\tau)}{\theta_3(0|\tau)}\Big)-m^2\partial^2_{\log \mathfrak{q}}\log\theta_3(0|\tau)  \right) \big(\tilde{{Z}}_{0}^D \tilde{{Z}}_{0}^D
	+\tilde{{Z}}_{1/2}^D \tilde{{Z}}_{1/2}^D\big),
\end{multline}
where \(\mathfrak{q}=\re^{2\pi \ri\tau}\).

The main proposal of \cite{Bonelli:2019boe} is the explicit expression of $Z^D_{0}, Z^D_{1/2}$ as dual Nekrasov partition functions. They found that
\begin{equation}
	\label{eq:dualNekrasov}
	Z_{\epsilon}^D(\sigma,m, \eta,\tau)=\sum\limits_{n\in \mathbb{Z}{+}\epsilon}\re^{\ri n\eta}\frac{\prod_{\epsilon'=\pm}G(1-m+2\epsilon'(\sigma{+}n))}{\prod_{\epsilon'=\pm}G(1+2\epsilon'(\sigma{+}n))} \mathfrak{q}^{(\sigma{+}n)^2-1/24} Z(\sigma+n,m,\mathfrak{q})\,,\;
\end{equation}
where $Z(\sigma,m,\mathfrak{q})$  denotes  the $c=1$ conformal block on the torus, i.e.~Nekrasov partition function for the $SU(2)$, $\mathcal{N}=2^*$ theory in the four-dimensional self-dual phase of the $\Omega$ background,  see eq.~\eqref{zsd2s} for the definition.
This proposal was proved recently in a more rigorous and mathematical way in \cite{DelMonte:2020wty} using the techniques of Fredholm determinants.
 Notice that if $m=0$ we have a very simple expression 
 \be\label{simple}\mathcal{T}(\sigma,0,\eta,\tau)
 =\re^{2 \ri \pi  \sigma ^2 \tau  }.\ee
   This formula can be easily deduced from equations \eqref{eq:Q for m=0} and \eqref{hdef}, or from the formula \eqref{eq:dualNekrasov} using  
   $Z(\sigma,0,\mathfrak{q})=\mathfrak{q}^{1/24}{\eta(\tau)^{-1}}$.
       In Sec.~\ref{Sec:-2blowup} we will demonstrate  that \eqref{eq:dualNekrasov} indeed satisfy \eqref{eq: bilin dual Nekrasov tildei}, providing in this way another proof of the isomonodromy-CFT correspondence for the 1-punctured torus, alternative to \cite{Bonelli:2019boe,DelMonte:2020wty}.

\

\noindent\textbf{Remark.}
One might wonder how could it happen that \eqref{eq:dualNekrasov} is naively non-symmetric under \(m\mapsto-m\), while \eqref{eq:Calogero} is symmetric.
The answer is that one needs to accompany this transformation by the transformation of the Barnes functions
\be 
\frac{G(1-\nu+n)}{G(1-\nu)}=(-1)^{\frac{n(n-1)}2}\frac{G(1+\nu-n)}{G(1+\nu)}\left(\frac{\pi}{\sin\pi\nu}\right)^n\,,
\ee
which leads to the transformation \((m,\eta)\mapsto(-m,\tilde{\eta})\), where \(\tilde{\eta}\) is defined by\footnote{The transformation \(\eta\to\tilde{\eta}\) or \(m\to-m\) also appears for other isomonodromic systems in the context of transition between different topologies of the spectral networks \cite{Coman:2018uwk,Coman:2020qgf}, or as a very close companion of the complex conjugation, used to construct ``physical'' correlation function \cite{Gavrylenko:2018cai}.
}
\begin{equation}
\label{eq:etaTilde}
\re^{\ri \frac{\tilde{\eta}}2}=\re^{\ri \frac{\eta}2} \frac{\sin\pi(2\sigma+m)}{\sin\pi(2\sigma-m)}\,.
\end{equation}
The latter transformation is not well defined  if \(\sigma=\frac12(\pm m+k),$ $k\in \mathbb{Z}\).
In this case we have to chose either  $\eta$ or  $\tilde \eta$ to be finite.
Different choices will correspond to different charts on the monodromy manifold, see also Appendix \ref{app:TraceCoordinates}.

\subsection{Singularities matching condition}

In the current example we have
\begin{equation}
\label{eq:A12}
A_{12}(z)=m \frac{\theta_1(z-2Q|\tau)\theta_1'(0|\tau)}{\theta_1(z|\tau)\theta_1(2Q|\tau)}.
\end{equation}
Hence $A_{12}(z)$ admits several zeroes unless\footnote{To be precise, one should require \(Q=\frac{n}{2}+\frac{\tau k}{2}\). However, due to \cite[Appendix C]{Bonelli:2019boe}, such a shift can be achieved by a simple B\"acklund transformation \((\sigma,\eta)\mapsto (\sigma+k/2,\eta+2\pi n)\), so it is sufficient to consider only \(Q=0\).
}
\begin{equation}\label{eq:Qzero}
	Q=Q(\sigma, m, \eta, \tau)=0.
\end{equation}
 This is our condition of singularities matching. By using \eqref{eq:solutionTorus} we see that
\eqref{eq:Qzero} is equivalent~to
\begin{equation}
\label{eq:apparentSingularityCancellation}
\theta_2(0|2\tau)Z_0^D\left(\sigma,m,\eta,\tau\right)-\theta_3(0|2\tau)Z_{1/2}^D\left(\sigma,m,\eta,\tau\right)=0\,.
\end{equation}

\subsection{Quantum mechanical operator}

Let us denote by \(\tau_{\star}\) the solution to  \eqref{eq:Qzero}
\begin{equation}
	\label{eq:20}
	Q(\sigma, m,\eta,\tau_{\star})=0.
\end{equation}
By solving  
{ \eqref{eq:Calogero} around $Q=0$}  we
get
\begin{equation}\label{eq:Q approx} 
	Q\approx \frac{\exp \left(\mp\frac{\ri \pi }{4}\right) \left(\sqrt{m} \sqrt{\tau-\tau_\star}\right)}{\sqrt{2 \pi }}. 
\end{equation}
One is actually free to choose any of the two signs.
We keep them both in order to see possible symmetries.
We will also denote the quantities corresponding to the two different solutions by~\(^{\mp}\): the upper sign always corresponds to the upper sign in \eqref{eq:Q approx}, and vice versa. 

By using  \eqref{pdef} it is easy to see that  
\begin{equation}
	\label{eq:momentumSingular}
	p=2\pi i\partial_{\tau}Q={ \pm} {m\over 2 Q}+\mathcal{O}(Q^0).
\end{equation}
Hence the Hamiltonian \eqref{eq:1} is finite at the point \(\tau_\star\)\footnote{Another way to get this is to notice that  \(\frac{\mathrm{d}}{\mathrm{d}\tau}H=-m^2\partial_{\tau}\left(\wp(2Q|\tau)+2\eta_1(\tau)\right)\) is regular when \(Q=0\) (note that the partial \(\tau\)-derivative is computed at fixed \(Q\)). Then one gets the behavior \eqref{eq:momentumSingular} from the finiteness of \(H_\star\).
}. Likewise  we can think of  \eqref{eq:Qzero} as an equation for $\sigma$ or $\eta$. In this case we will denote the corresponding solution by $\sigma_\star$ or $\eta_\star$.  The corresponding Hamiltonian will always be finite and we will denote its values by  \(H_{\star}^{\mp}\).

Let us now look at the quantum mechanical operator. After some algebra, we find that the potential \eqref{eq:potential} associated to the linear system \eqref{eq:torusSystem} can be written as 
\begin{equation}
\label{eq:ellipticPotential-0}
\ba
W(z,\tau)=&\, H+m^2\left(\wp(z)+2 \eta_1(\tau)\right)-p \left( \zeta(z-2Q|\tau)-\zeta(z|\tau)+{ 4 Q \eta_1(\tau) } \right)\\
&+\frac12  \left( \wp(z-2Q|\tau)-\wp(z|\tau) \right)+{ \frac14} \left( 
\zeta(z-2Q|\tau)-\zeta(z|\tau) +{ 4 Q \eta_1(\tau) }\right)^2 ,
\ea
\end{equation}
where  the  elliptic functions $\wp, \zeta, \eta_1$ are defined in Appendix \ref{conv}.
In deriving \eqref{eq:ellipticPotential-0} we also used several identities for the Lam\'e function $x(u,z)$ which can be found  in \cite[Appendix A]{Bonelli:2019boe}.

The potential \eqref{eq:ellipticPotential-0} is quite complicated, especially because it depends on $Q=Q(\sigma,m,\eta,\tau)$. 
However, when we impose the singularities matching condition   \eqref{eq:Qzero} the second line of \eqref{eq:ellipticPotential-0}  vanishes. In addition, by using \eqref{eq:momentumSingular} we can rewrite the first line in \eqref{eq:ellipticPotential-0} as 
\be -p \big( \zeta(z-2Q|\tau)-\zeta(z|\tau)+{ 4 Q \eta_1(\tau) } \big) ={\mp} {m } \big(  \wp (z|\tau)+ 2  \eta_1(\tau)  \big) +\mathcal{O}(Q).\ee 
 It follows that the relevant potential at $Q=0$ is 
\be \label{eq:ellipticPotential}\ba  U(z,\tau)
&= (m^2{ \mp }m) \wp(z|\tau) + \left(H^{\mp}_\star +2(m^2{\mp} m) \eta_1(\tau)\right). \ea\ee
Hence the quantum operator arising from  isomonodromic deformations on the torus is the  2-particle {\it{quantum}} elliptic  Calogero-Moser system with potential
\( (m^2{ \mp }m) \wp(z|\tau).\)

To have a physically well-defined spectral problem, in this paper we will restrict without loss of generality to $|m|>1$.
Note that  $H^{\mp}_\star $ can be computed explicitly from \eqref{hdef}, \eqref{eq:tauFunction} with the help of \eqref{eq:momentumSingular}.  It reads:
\begin{equation}\label{eq:energyd}
H^{\mp}_\star =\Big(\left.2\pi \ri\partial_{\tau}\log Z_0^D(\sigma,m,\eta,\tau)+2\pi \ri\partial_{\tau}\log \frac{\eta(\tau)}{\theta_3(0|2\tau)}{\mp }2 m \frac{\theta_3''(0|2\tau)}{\theta_3(0|2\tau)} \Big)\right|_{Q=0}\,.
\end{equation}
Here we denote by $'$ the derivative w.r.t.~the first argument of the $\theta$ function.

\

\noindent\textbf{Remark.} By inverting \eqref{eq:Q approx} we get $\tau-\tau_{\star}\approx \pm\frac{2\pi \ri}{m}Q^2$. By substituting this into \eqref{eq:Calogero}, \eqref{eq:1} one can compute further terms
\begin{equation}
\label{eq:tauExpansion}
\tau-\tau_{\star}=\pm \frac{2\pi \ri}{m}\left( Q^2-\frac{ H^{\mp}_\star+2m^2\eta_1(\tau_{\star})}{m^2}Q^4 \right)+\mathcal{O}(Q)^6.
\end{equation}
To derive this formula it is sufficient to use the approximation \(\wp(x|\tau)\sim \frac1{x^2}\), since higher order terms in such expansion start to contribute from \(\mathcal{O}(Q)^6\). The upper and lower signs in \eqref{eq:tauExpansion} agree with the ones in \eqref{eq:Q approx}.  We will use \eqref{eq:tauExpansion} in Section \ref{functional}.

\

\noindent\textbf{Remark.} By using \eqref{eq:tauExpansion} and \eqref{eq:energyd} we can compute the first few terms in the \(\mathfrak{q}\)-expansion of \(H^{\mp}_{\star}\). We get 
\begin{equation}
\label{eq:Hminus}
H_{\star}^{\mp}=4\pi^2 \left(-\sigma^2 + \frac{2m^2(m\mp 1)^2}{1-4\sigma^2}\mathfrak{q}_{\star} + \mathcal{O}(\mathfrak{q}_{\star}^2) \right).
\end{equation}

\subsection{Reality condition}

Now we wish to fulfil another requirement: the reality of the potential in the Schr\"odinger equation
 \be\label{scw2} \left(-\partial_z^2+(m^2\mp m)\wp(z|\tau)+H^{\rm \mp}_\star+2(m^2\mp m)\eta_1(\tau)\right)\widetilde{Y}_1(z)=0\,.\ee
There are several ways to do this.
First we study the conjugation of the Weierstrass function: \(\overline{\wp(z|\tau)}=\wp(\overline{z}|-\overline{\tau})\).
This transformation reflects the fundamental domain of the modular group with respect to the vertical line.
There are two (actually intersecting) branches which are invariant  under such conjugation: \(\tau\in \ri \mathbb{R}_{>0}\) and \(\tau\in\frac12+\ri\mathbb{R}_{>0}\)~\footnote{The latter branch maps to \(|\tau|=1\) by the transformation \(\tau'=\frac{\tau}{1-\tau}\), in the Table \ref{tab:homologyClasses} we map it to the line $-\frac12+\ri\mathbb{R}_{>0}$.}.
The two special points with additional symmetry of the elliptic curve also lie on these branches: \(\tau=\ri\) and \(\tau=\re^{\ri\pi/3}\).

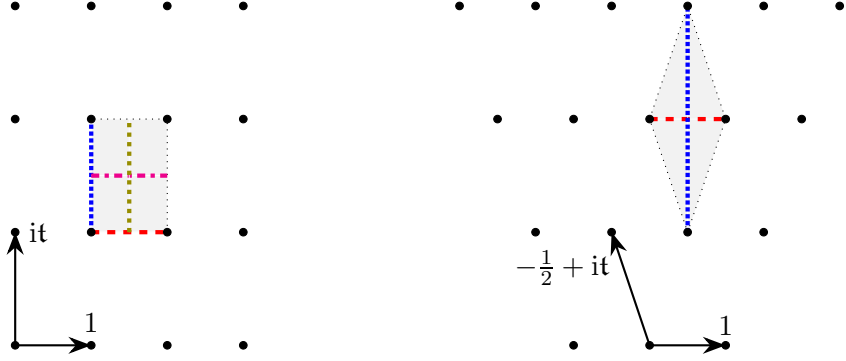
\begin{figure}[h!]
  \begin{center}
\begin{tabular}{cc}
\begin{tikzpicture}
\draw[-{Stealth[length=3mm]},thick](0,0)--(1,0);
\draw[-{Stealth[length=3mm]},thick](0,0)--(0,1.5);
\draw[white,fill=gray!10](1,1.5)--(2,1.5)--(2,3)--(1,3)--cycle;
\draw[dotted](2,1.5)--(2,3)--(1,3);

\draw[ultra thick,dashed,red,fill=white](1,1.5)--(2,1.5);
\draw[ultra thick,densely dotted,blue,fill=white](1,1.5)--(1,3);
\draw[ultra thick,dash dot,magenta,fill=white](1,2.25)--(2,2.25);
\draw[ultra thick,dotted,olive,fill=white](1.5,1.5)--(1.5,3);

\foreach \i in {0,...,3}
  \foreach \j in {0,...,3}
    \draw[fill=black](\i,\j*1.5) circle[radius=0.05,fill=black];

\node at (1,0.3){\(1\)};
\node at (0.3,1.5){\(\ri \mathfrak{t}\)};
\end{tikzpicture}
&\qquad \qquad \qquad
\begin{tikzpicture}
\draw[-{Stealth[length=3mm]},thick](0,0)--(-1/2,1.5);
\draw[-{Stealth[length=3mm]},thick](0,0)--(1,0);
\draw[dotted,fill=gray!10](0.5,1.5)--(1,3)--(0.5,4.5)--(0,3)--cycle;

\draw[ultra thick,dashed,red,fill=white](0,3)--(1,3);
\draw[ultra thick,densely dotted,blue,fill=white](0.5,1.5)--(0.5,4.5);

\foreach \j in {0,...,3}
  \foreach \i in {-\j,...,2}
    \draw[fill=black](\i+\j*0.5-1,\j*1.5) circle[radius=0.05,fill=black];

\node[anchor=north east] at (-0.4,1.4){\(-\frac12+\ri \mathfrak{t}\)};
\node[anchor=south] at (1,0){\(1\)};
\end{tikzpicture}
\end{tabular}
\end{center}

\caption[]{\label{fig:lattices} Lines corresponding to a real potential in \eqref{scw2}. Left figure: cases 1--4. Right figure: cases 5, 6.}
\end{figure}

If we want the potential to be real, then \(z\) should lie on some suitable domain \(\mathcal{C}\). In Table \ref{tab:homologyClasses} we give a list of all the possible options. In this Table
the complex  variables $\tau$ and $z$ are parametrised by the two real variables \be \label{realv} \mathfrak{t}\in \mathbb{R}_{>0}, \quad x\in (0,1)\, .\ee
The corresponding lines in the \(z\)-plane are shown in Figure
\ref{fig:lattices}. During the computations we also used the modular transformation for the Weierstrass function: \be \wp(z|\tau)=\tau^{-2}\wp(\frac{z}{\tau}|- \frac{1}{\tau})\,.\ee
After parametrising $z$ and $\tau$ via \eqref{realv}  we  write the Schr\"odinger equation \eqref{scw2} as
\begin{equation}
	\left(-\partial^2_x+u_\mp(x,\mathfrak{t})+ E(\mathfrak{t})\right)\psi(x,\mathfrak{t})=0~,
\end{equation}
where $E(\mathfrak{t})$ and  $u_\mp(x,\mathfrak{t})$  are reported in the last and the second to last  column of Table \ref{tab:homologyClasses}. 
Later we will also use the notation
\begin{equation}\label{scw} {\rm O}_\mp =-\partial^2_x+u_\mp(x,\mathfrak{t})\, .\end{equation}

\begin{table}[h!]
\begin{center}
\begin{tabular}{|*{8}{c|}}
\hline 
\#&\(\tau\)&\(z\)&\([\mathcal{C}]\)&Notation&Potential \(u_\mp(x,\mathfrak{t})\)&Energy \(E(\mathfrak{t})\)\\
\hline
\begin{minipage}{0.2cm}\vspace{0.1cm}1\vspace{0.2cm}\end{minipage}
&\(\ri \mathfrak{t}\)&\(x\)&\(A\)&\tikz{\draw[ultra thick, red, dashed](0,0)--(1,0);}&\((m^2\mp m)\wp(x|\ri \mathfrak{t})\)&\(-H^{\mp}_\star-2(m^2\mp m)\eta_1(\ri\mathfrak{t})\)\\
\hline
\begin{minipage}{0.2cm}\vspace{0.1cm}2\vspace{0.2cm}\end{minipage}
&\(\ri \mathfrak{t}\)&\(\ri \mathfrak{t} x\)&\(B\)&\tikz{\draw[ultra thick, blue, densely dotted](0,0)--(1,0);}&\((m^2\mp m)\wp(x|\frac \ri{\mathfrak{t}})\)&\(\mathfrak{t}^2 \left( H^{\mp}_\star+2(m^2\mp m)\eta_1(\ri\mathfrak{t}) \right)\)
\\
\hline
\begin{minipage}{0.2cm}\vspace{0.1cm}3\vspace{0.2cm}\end{minipage}
&\(\ri \mathfrak{t}\)&\(x+\frac{\ri\mathfrak{t}}{2}\)&\(A\)&\tikz{\draw[ultra thick, magenta, dash dot](0,0)--(1,0);}&\((m^2\mp m)\wp(x+\frac{\ri\mathfrak{t}}2|\ri \mathfrak{t})\)&\(-H^{\mp}_\star-2(m^2\mp m)\eta_1(\ri\mathfrak{t})\)\\
\hline
\begin{minipage}{0.2cm}\vspace{0.1cm}4\vspace{0.2cm}\end{minipage}
&\(\ri \mathfrak{t}\)&\(\ri\mathfrak{t}x+\frac12\)&\(B\)&\tikz{\draw[ultra thick, olive, dotted](0,0)--(1,0);}&\((m^2\mp m)\wp(x+\frac{\ri}{2\mathfrak{t}}|\frac{\ri}{\mathfrak{t}})\)&\(\mathfrak{t}^2 \left( H^{\mp}_\star+2(m^2\mp m)\eta_1(\ri\mathfrak{t}) \right)\)\\
\hline
\hline
\begin{minipage}{0.2cm}\vspace{0.1cm}5\vspace{0.2cm}\end{minipage}
&\(\ri \mathfrak{t}-\frac12\)&\(x\)&\(A\)&\tikz{\draw[ultra thick, red, dashed](0,0)--(1,0);}&\((m^2\mp m)\wp(x|\ri \mathfrak{t}-\frac12)\)&\(-H^{\mp}_\star-2(m^2\mp m)\eta_1(\ri\mathfrak{t}-\frac12)\)\\
\hline
\begin{minipage}{0.2cm}\vspace{0.1cm}6\vspace{0.2cm}\end{minipage}
&\(\ri \mathfrak{t}-\frac12\)&\(2\ri \mathfrak{t}x\)&\(C=2B+A\)&\tikz{\draw[ultra thick, blue, densely dotted](0,0)--(1,0);}&\((m^2\mp m)\wp(x|\frac{\ri}{4\mathfrak{t}}-\frac12)\)&\(4\mathfrak{t}^2 \left( H^{\mp}_\star+2(m^2\mp m)\eta_1(\ri\mathfrak{t}-\frac12) \right)\)\\
\hline
\end{tabular}
\end{center}
\caption{\label{tab:homologyClasses}The possible spectral problems associated to the linear system \eqref{eq:torusSystem}. The  Hamiltonian $H^{\mp}_\star$ is given in \eqref{eq:energyd}. The cases 3 and 4 correspond to the Bloch waves and will not be discussed here.}
\end{table}

\subsection{Normalizability conditions}\label{Normalizability conditions}

As we saw above, all the potentials with a discrete spectrum  that appear in this problem live on a line segment \(\mathcal{C}\) bounded by the point \(0\) and  its image, which we denote by \(a=0.\gamma_\mathcal{C}\). Normalizability of the linear problem means that we map normalizable solutions at one end to normalizable solutions at the other end. As in equation \eqref{nor}, we expect normalizability to give us some constraints on the monodromies $(\sigma,\eta)$ of the linear system.

Let us first study the solution of the linear system \eqref{eq:torusSystem} around \(z=0\). For the purposes of this paper we can focus on  the limit \(\tau\to \tau_{\star}\).
In order to do this we use \eqref{eq:momentumSingular} and further expand \eqref{At} at \(Q\to 0\):
\begin{equation}
\label{eq:2}
A(z,\tau)\approx
\begin{pmatrix}
\pm\frac{m}{2Q} & \frac{m}{2Q}-\frac{m}{z}\\
-\frac{m}{2Q}-\frac{m}{z} & \mp \frac{m}{2Q}.
\end{pmatrix}.
\end{equation}
By substituting \eqref{eq:2}  into  \eqref{eq:torusSystem}  we get
\begin{equation}
\label{eq:3}
Y(z,\tau)=
\begin{pmatrix}
1 & 1\\
1 & -1
\end{pmatrix}
\left( \mathbb{I}+
\frac{mz}{2Q(1\mp 2m)}
\begin{pmatrix}
0 & -1\pm 1\\
1\pm 1 & 0
\end{pmatrix}
+\mathcal{O}(z)^2 \right)
\begin{pmatrix}
z^{-m} & 0\\
0 & z^m
\end{pmatrix}\mathsf{C} ,
\end{equation}
for some $z$-independent matrix $\mathsf{C}$.
We see that this expression has a singularity at \(Q\to 0\), so it needs to be renormalized by choosing an appropriate diagonal matrix \(\mathsf{C}\).
This is done in different ways for the upper and for the lower sign:
\begin{itemize}
\item Upper sign, \(\mathsf{C}=\operatorname{diag}\left((1-2m)Q/m,1\right)\):
\be Y(z,\tau)\approx \left(
\begin{array}{cc}
 \frac{(1-2 m) Q z^{-m}}{m}+z^{1-m} +\mathcal{O}(z^{2-m})& z^m +\mathcal{O}(z^{2+m})\\
 \frac{(1-2 m) Q z^{-m}}{m}-z^{1-m}  +\mathcal{O}(z^{2-m}) & -z^m +\mathcal{O}(z^{2+m})\\
\end{array}
\right)+\mathcal{O}(Q^2) .\ee
Hence the leading asymptotics in the limit\footnote{We keep \(z\) fixed and send \(Q\to 0\), or in other words consider \(2Q\ll z\).
} \(Q\to 0\) are \be (z^{1-m},z^m). \ee

\item Lower sign, \(\mathsf{C}=\operatorname{diag}\left(1,(-1-2m)Q/m\right)\):
\be Y(z,\tau)\approx \left(
\begin{array}{cc}
 z^{-m} +\mathcal{O}(z^{2-m})& \frac{z^m (m z-(2 m+1) Q)}{m} +\mathcal{O}(z^{2+m})\\
 z^{-m} +\mathcal{O}(z^{2-m})& \frac{z^m (2 m Q+m z+Q)}{m} +\mathcal{O}(z^{2+m})\\
\end{array}
\right)+\mathcal{O}(Q^2) .\ee
Hence the leading asymptotics in the limit \(Q\to 0\) are \be (z^{-m},z^{m+1}). \ee
\end{itemize}
Notice that in the limit \(Q\to 0\) and  around \(z=0\), the function
\(\widetilde{Y}_1(z,\tau)\) differs from \(Y_1(z,\tau)\) only by a normalization factor, i.e.~by  \(\sqrt{2Q/m}\).  In particular they both have the same asymptotics.

{A similar analysis can be repeated for the point  \(a=0.\gamma_\mathcal{C}\).}
Hence for \(m>0\) a normalizable solution \(Y_{norm}(z)\) should have positive asymptotics near both boundaries. For the upper sign we find
\be Y(z)\sim z^m , \quad Y(z)\sim (z-a)^m ,\ee
while for the lower sign we get
 \be Y(z)\sim z^{1+m}, \quad Y(z)\sim (z-a)^{1+m} . \ee 
This means that normalizable solutions should have monodromies \(\re^{2\pi im}\) around both points (for both upper and lower signs):
\begin{equation}
\label{eq:61}
Y_{norm}(z.\gamma_0)=\re^{2\pi \ri m}Y_{norm}(z), \quad Y_{norm}(z.\gamma_a)=\re^{2\pi \ri m}Y_{norm}(z),
\end{equation}
where \(\gamma_0\) and \(\gamma_a\) are the contours encircling \(0\) and \(a\).
To fulfil the first requirement it is sufficient to project onto the column of \(Y(z)\) with appropriate asymptotics\footnote{We notice that one can capture the two asymptotics by their monodromies only for \(m\notin \frac12\mathbb{Z}\).
However, there are no singularities in \(m\) in the solution of the spectral problem at \(m\in \frac12 \mathbb{Z}\), so everything can be continued analytically to these points.
}:
\begin{equation}
\label{eq:7}
Y_{norm}(z)=Y(z)(M_0-\re^{-2\pi \ri m}\mathbb{I}).
\end{equation}

We now look at the second requirement in \eqref{eq:61}. Let $M_\mathcal{C}$ be the monodromy along the cycle $\mathcal{C}$ as defined in Table \ref{tab:homologyClasses}.
We want to  map normalisable solutions around 0 to normalisable solution around $a=0.\gamma_{\mathcal{C}}$ (and the other way around). Hence we ask\begin{equation}
\label{eq:8}
Y(z)\cdot { M_\mathcal{C}^{-1}}(M_0-\re^{2\pi \ri m}\mathbb{I})M_\mathcal{C}\cdot(M_0-\re^{-2\pi \ri m}\mathbb{I})=0.
\end{equation}
Using that \(M_{\mathcal{C}}^{-1}\) and \(Y(z)\) are non-degenerate we have
\begin{equation}
\label{eq:46}
(M_0-\re^{2\pi \ri m}\mathbb{I})M_{\mathcal{C}}(M_0-\re^{-2\pi \ri m}\mathbb{I})=0.
\end{equation}
This is the normalizability equation.

It is, of course, more convenient to rewrite this condition in the basis where \(M_0\) is diagonal:
\begin{equation}
\label{eq:50}
M_0^{(I,II)}=
\begin{pmatrix}
\re^{2\pi \ri m} & 0\\
0 & \re^{-2\pi \ri m}
\end{pmatrix}\, .
\end{equation}
The superscripts  \(^{(I)}\), \(^{(II)}\) denote different charts on the monodromy manifold with coordinates \((\sigma,\eta)\) and \((\sigma,\tilde{\eta})\), see Appendix~\ref{app:monodromies} for more details. The diagonalization of $M_0$ is performed by a matrix that depends on the chart, see Appendix~\ref{app:M0diag} for more details.
Hence \eqref{eq:46} means that for \(m>0\) it maps the column vector \((*,0)\) to itself.
For \(m<0\) instead it should map \((0,*)\) to itself. This means that \(M_{\mathcal{C}}\) is upper- or lower-triangular:
\begin{equation}
\label{eq:Mgamma}
M_{\mathcal{C}}^{(I,II)}=
\begin{pmatrix}
*&*\\
0&*
\end{pmatrix}, \text{ for } m>0,\quad \quad
M_{\mathcal{C}}^{(I,II)}=
\begin{pmatrix}
*&0\\
*&*
\end{pmatrix}, \text{ for } m<0\,.
\end{equation}
Hence if $m>0$ the normalizability condition   reads \begin{equation}
\label{eq:62}
\left(M_{\mathcal{C}}^{(I,II)}\right)_{21}=0.
\end{equation}
Likewise the condition for \(m<0\) is \be \left(M_{\mathcal{C}}^{(I,II)}\right)_{12}=0\, . \ee

To simplify our analysis it is convenient to notice that if \(2\sigma\in\pm m+\mathbb{Z}\), the matrix elements of \(M_{\cdots}^{(I,II)}\) are regular in one of the two charts with \(\eta\) or \(\tilde{\eta}\) finite\footnote{We are not talking here about the singularities at \(\sigma\in \frac12 \mathbb{Z}\).
These cannot be removed by switching to another chart and  they are completely forbidden in this framework.
The singularities at \(m\in \frac12+\mathbb{Z}\) are not visible in the spectral problem and will not appear in the normalizability conditions.
}, see also \eqref{eq:etaTilde}. In addition, the element \(12\) or \(21\) does not vanish simultaneously for any pair of matrices of our interest, \(M_A\), \(M_B\), and \(M_C\).
So we consider the ratios of the corresponding matrix elements:
\begin{equation}
\label{eq:63}
f^{\mathcal{C}/A}_{21}=\frac{\left( M_{\mathcal{C}}^{(I)} \right)_{21}}{\left( M_A^{(I)} \right)_{21}}=\frac{\left( M_{\mathcal{C}}^{(II)} \right)_{21}}{\left( M_A^{(II)} \right)_{21}},\quad
f^{\mathcal{C}/A}_{12}=\frac{\left( M_{\mathcal{C}}^{(I)} \right)_{12}}{\left( M_A^{(I)} \right)_{12}}=\frac{\left( M_{\mathcal{C}}^{(II)} \right)_{12}}{\left( M_A^{(II)} \right)_{12}}.
\end{equation}
These expressions are better because they are independent from the remaining diagonal conjugation.
Their explicit values are (see Appendix \ref{app:M0diag})
\begin{equation}
\label{eq:64}
f_{21}^{B/A}=\frac{\re^{\ri \frac{\eta}{2}}\frac{\sin\pi(2\sigma+m)}{\sin\pi(2\sigma-m)}-\re^{-\ri \frac{\eta}{2}}\frac{\sin\pi(2\sigma-m)}{\sin\pi(2\sigma+m)}}{-2\ri \re^{-\ri\pi m}\sin2\pi\sigma}=-\re^{\pi \ri m}\frac{\sin \frac{\tilde{\eta}}{2}}{\sin2\pi\sigma},
\end{equation}

\begin{equation}
\label{eq:65}
f_{12}^{B/A}=-\re^{-\pi \ri m}\frac{\sin \frac{\eta}{2}}{\sin2\pi\sigma}=\frac{\re^{\ri \frac{\tilde{\eta}}{2}}\frac{\sin\pi(2\sigma-m)}{\sin\pi(2\sigma+m)}-\re^{-\frac{\ri\tilde{\eta}}{2}}\frac{\sin\pi(2\sigma+m)}{\sin\pi(2\sigma-m)}}{-2\ri\re^{\ri\pi m}\sin2\pi\sigma},
\end{equation}

\begin{multline}
\label{eq:68}
f_{21}^{C/A}=\frac{\ri \re^{\ri\pi m-\ri\tilde{\eta}+2\pi \ri\sigma}}{\sin^22\pi\sigma}
\left( \re^{\ri\tilde{\eta}-2\pi \ri\sigma}\cos\pi(\sigma-\frac{m}{2})-\cos\pi(\sigma+\frac{m}{2}) \right)\times
\\\times
\left(\re^{\ri \tilde{\eta}-2\pi \ri\sigma}\sin\pi(\sigma-\frac{m}{2})+\sin\pi(\sigma+\frac{m}{2})  \right),
\end{multline}

\begin{multline}
\label{eq:67}
f_{12}^{C/A}=\frac{\ri \re^{-\ri \pi m-\ri \eta+2\pi \ri \sigma}}{\sin^22\pi\sigma}
\left( \re^{\ri \eta-2\pi \ri\sigma}\cos\pi(\sigma+\frac{m}{2})-\cos\pi(\sigma-\frac{m}{2}) \right)\times
\\\times
\left( \re^{\ri \eta-2\pi \ri\sigma}\sin\pi(\sigma+\frac{m}{2})+\sin\pi(\sigma-\frac{m}{2}) \right).
\end{multline}
We conclude that the normalizability condition for \(\mathcal{C}=A\) is 
\be 
\ba
\left(f^{B/A}_{21}\right)^{-1}=0, \quad \text{if } m>0,\\
\left(f^{B/A}_{12}\right)^{-1}=0, \quad \text{if } m<0\,.
 \ea
\ee
If \(\mathcal{C}=B\) or \(\mathcal{C}=C\) instead we have
 \be\ba
  f^{\mathcal{C}/A}_{21}=0, \quad \text{if } m>0~,\\
f^{\mathcal{C}/A}_{12}=0, \quad \text{if } m<0~.\\ \ea\, \ee
We write all these conditions explicitly in  Table \ref{tab:normalizabilityConditions}.

\begin{table}[h!]
\begin{center}
\begin{tabular}{|*{4}{c|}}
\hline
Cases \# &\(\mathcal{C}\)&\(m\)& \(\eta\), \(\sigma\) \\
\hline
1, 5 &\(A\)&\(m>0\)&\(\sigma\in\pm\frac m2+\frac12\mathbb{Z}\), \quad \(\eta\) --- finite \\
\hline
1, 5 &\(A\)&\(m<0\)&\(\sigma\in\pm\frac m2+\frac12\mathbb{Z}\), \quad \(\tilde{\eta}\) --- finite \\
\hline
2 &\(B\)&\(m>0\)&\(\tilde{\eta}\in { 2 \pi }\mathbb{Z}\) \\
\hline
2 &\(B\)&\(m<0\)&\(\eta\in  { 2 \pi } \mathbb{Z}\) \\
\hline
6 & \(C\)&\(m>0\)& \(\re^{\ri\tilde{\eta}}=\re^{2\pi \ri\sigma}\frac{\cos\pi(\sigma+\frac{m}{2})}{\cos\pi(\sigma-\frac{m}{2})}\), \quad \(\re^{\ri \tilde \eta}=-\re^{2\pi \ri\sigma}\frac{\sin\pi(\sigma+\frac{m}{2})}{\sin\pi(\sigma-\frac{m}{2})}\) \\
\hline
6 & \(C\)&\(m<0\)& \(\re^{\ri \eta}=\re^{2\pi \ri\sigma}\frac{\cos\pi(\sigma-\frac{m}{2})}{\cos\pi(\sigma+\frac{m}{2})}\), \quad \(\re^{\ri \eta}=-\re^{2\pi \ri\sigma}\frac{\sin\pi(\sigma-\frac{m}{2})}{\sin\pi(\sigma+\frac{m}{2})}\) \\
\hline
\end{tabular}
\end{center}
\caption{\label{tab:normalizabilityConditions}Normalizability conditions for different spectral problems. The Cases $\#$  are as in the first column of Tab.~\ref{tab:homologyClasses}.  }
\end{table}

\subsection{Quantization conditions and spectrum} \label{sec:quant}
Following the general approach presented in Sec.~\ref{main}, we want to test that the singularities matching condition \eqref{eq:apparentSingularityCancellation}, combined with the normalizability conditions of Table \ref{tab:normalizabilityConditions}, reproduces the correct spectrum of the operators in Table \ref{tab:homologyClasses}.
We work out in details the cases $\#$ 1 and $\#$ 2  of Table \ref{tab:homologyClasses}.  The other examples work analogously.

\subsubsection{Case $\#$ 2} 
Let us first focus on case $\#$ 2  of Table \ref{tab:homologyClasses}.  The operator we consider is \be \label{op2}{\rm O}_{\mp}=-\partial_x^2+ (m^2{\mp} m)\wp(x|\frac \ri{\mathfrak{t}}),  \quad x\in [0,1], \quad \mathfrak{t}\in \mathbb{R}_+ \ee
on $L^2[0,1]$.  

{ Let us first consider the case $ m<0$}. The relevant conditions can be written as ($\tau=\ri \mathfrak{t}$)
\begin{equation} \label{QMneg}{\theta_2(0|2\tau)Z_0^D\left(\sigma,m,{\eta}, \tau\right)-\theta_3(0|2\tau)Z_{1/2}^D\left(\sigma,m,{\eta},\tau\right)=0\,  \quad \text{with}\quad \eta \in 2 \pi \IZ \,.
}
\end{equation}
Notice 
 that there are two inequivalent values of  $ \eta $ in \eqref{QMneg}
\begin{equation}
\label{eq:etaQuantization}
    \eta=
    \begin{cases}
0 & \text{mod } 4 \pi \, , \\
        2\pi  & \text{mod } 4 \pi \, .
    \end{cases}
  \end{equation}
These correspond to even and odd eigenvalues of \eqref{op2}. Moreover, the solutions to \eqref{QMneg}  
reproduce  the spectrum of both ${\rm O}_+$ and ${\rm O}_-$. Hence 
  it is useful to introduce the notation
\be   \eta_n^+=
    \begin{cases}
    0 & \text{if  $n$ is even}  \\
  2\pi  & \text{if $n$ is odd }   
    \end{cases}, ~ \quad \eta_n^-=2\pi-\eta_n.
 \ee
We think of \eqref{QMneg} as a quantization condition for $\sigma$. 
More precisely, we can organise the zeroes of  \eqref{QMneg} in ascending order
\be\label{2o}(\sigma, \eta)\in  \{(\sigma_0, 0), (\sigma_1,2\pi), (\sigma_2,2\pi), (\sigma_3,0),\cdots\}.\ee
This sequence contains both the spectrum of ${\rm O}_+$ and  ${\rm O}_-$. Experimentally,  we find that a pattern to disentangle them is the following.
If we wish to study the operator ${\rm O}_-$ in \eqref{op2}, then we have to consider the subset of \eqref{2o} given by
\be(\sigma, \eta)\in  \{(\sigma_{2n+1}, \eta_{n}^-)\}_{n\geq 0}= \{(\sigma_{n}^{-}, \eta_{n}^-)\}_{n\geq 0}.\ee
Instead, if we wish to study the operator ${\rm O}_+$ in \eqref{op2}, then we have  to consider the subset of \eqref{2o} given by
\be(\sigma, \eta)\in   \{(\sigma_{2n}, \eta_n^+)\}_{n\geq 0}= \{(\sigma_{n}^{+}, \eta_n^+)\}_{n\geq 0}.\ee
To obtain the exact spectrum of \eqref{op2} one also needs the relation $E(\tau)$ reported in the last column of Table \ref{tab:homologyClasses}. More precisely, the energy levels $E_n^{\rm \mp}$ of  ${\rm O}_\mp$ are obtained  from $(\sigma_n^{\pm},\eta_n^{\pm})$ as \be\label{esigma}{  E_n^{\mp}(m,\mathfrak{t})=\mathfrak{t}^2 \left( (H^{\mp}_\star)^{(n)}+2(m^2{\mp} m)\eta_1(\ri \mathfrak{t}) \right),} \ee
with 
\begin{equation}
\label{eq:energy2} 
(H^{\mp}_\star)^{(n)}=\left(2\pi \ri\partial_{\tau}\log Z_0^D(\sigma_n^{\mp},m,\eta_n^{\mp},\tau)\right)\mid_{\tau=\ri t}+2\pi \ri\partial_{\tau}\left(\log \frac{\eta(\tau)}{\theta_3(0|2\tau)}\right)\Big|_{\tau=\ri \mathfrak{t}}{\mp }2 m \frac{\theta_3''(0|2\ri \mathfrak{t})}{\theta_3(0|2\ri \mathfrak{t})}\, ,
\end{equation}
where $'$ denotes the derivative w.r.t.~the first argument of the $\theta$ function. 
Some independent tests are provided in Table \ref{tab:twoqc}.\newline
  
  \begin{table} 
\centering
   \begin{tabular}{l | l  l}
  \\
Nb& $E_0^+$& $E_0^-$  \\
\hline
 1& \underline{48}.68163544578 & \underline{9}2.4799329375161  \\
3 & \underline{48.43513}749440 &  \underline{91.858766}0448900 \\
5 & \underline{48.435138199}47 &  \underline{91.8587662451}199  \\
 \hline
Num & 48.43513819950  & 91.8587662451245  \\
 \end{tabular}    
\caption{ The ground state energy of ${\rm O}_\mp$ in \eqref{op2} as computed from \eqref{QMneg} and \eqref{esigma} for $t=1$, $m=-\sqrt{6}$. We denote by  $\rm Nb$ the order $q^{\rm Nb}$ at which we truncate the instanton partition function $Z(\sigma,m, q)$ in \eqref{eq:dualNekrasov}. We  underline the stable digits. The numerical result reported in the last line  is performed as Appendix \ref{b1} , see also \cite{Takemura,Hatsuda_2018}.}
 \label{tab:twoqc}
  \end{table}

{If $ { m>0}$, instead of \eqref{QMneg} we have}
\begin{equation} \label{QMneg2}\ba
&\theta_2(0|2\tau)Z_0^D\left(\sigma,m,{\eta}, \tau\right)-\theta_3(0|2\tau)Z_{1/2}^D\left(\sigma,m,{\eta},\tau\right)=0,  \\
&\re^{\ri{\frac{\eta}2}}=\re^{\ri\frac{\tilde\eta}2} \frac{\sin\pi(2\sigma-m)}{\sin\pi(2\sigma+m)}, \quad \tilde\eta \in 2 \pi \IZ \,.
\ea
\end{equation}
One can perform the same analysis as before. In particular, the zeros of \eqref{QMneg2} are mapped to the spectrum of ${\rm O}_{\mp}$ by using \eqref{esigma}. The only subtlety is that, if $m>0$, we impose  $ \tilde\eta \in 2 \pi \IZ$ instead of $\eta \in 2 \pi \IZ$ as summarised in Table \ref{tab:normalizabilityConditions}.

 \subsubsection{Case $\#$ 1}
 We study
 \be \label{op3}{\rm O}_{\mp}=-\partial_x^2+(m^2\mp m)\wp(x|i \mathfrak{t}), \quad x\in [0,1], \quad \mathfrak{t}\in \mathbb{R}_+ . \ee
 We focus on $m>1$ without loss of generality. 
The normalizability condition in Table \ref{tab:normalizabilityConditions} gives 
\begin{enumerate}
\item $ \eta$ finite
\item $\sigma=\sigma_{1,2}$, where
\begin{equation}
\label{eq:sigmal}
2\sigma_1=m+k_1 \quad\text{or}\quad 2\sigma_2=-m+k_2
\end{equation}
with
$ k_{\ell} \in \mathbb{Z}$, $\ell=1,2$.
\end{enumerate}
Hence we look at the singularities matching condition   \eqref{eq:apparentSingularityCancellation} as an equation for $\eta$. 
More precisely, we should find a solution of \eqref{eq:apparentSingularityCancellation} in a form 
\be \eta=\eta_{\star}(\sigma,m,\tau).  \ee
To do this we substitute an Ansatz \be \re^{\ri \eta_{\star}/2}=\mathfrak{q}^{-\sigma} \re^{\ri \eta_0/2} \re^{\sum_{i=1}^{\infty} c_i\mathfrak{q}^i}. \ee
At the first non-trivial level such substitution gives the quadratic equation:
\begin{multline}
\label{eq:4}
2\frac{G(1-m-2\sigma)G(1-m+2\sigma)}{G(1-2\sigma)G(1+2\sigma)}-\re^{\ri \eta_0/2}\frac{G(1-m-2\sigma-1)G(1-m+2\sigma+1)}{G(1-2\sigma-1)G(1-2\sigma+1)}-\\-
\re^{-\ri \eta_0/2}\frac{G(1-m-2\sigma+1)G(1-m+2\sigma-1)}{G(1-2\sigma+1)G(1-2\sigma-1)}=0.
\end{multline}
Its two solutions are
\begin{equation}
\label{eq:5}
\re^{\ri \eta_0/2}=\frac{(2\sigma+m)\Gamma(-m-2\sigma)\Gamma(2\sigma)}{(2\sigma-m)\Gamma(-m+2\sigma)\Gamma(-2\sigma)},\quad
\re^{\ri \eta_0/2}=\frac{\Gamma(-m-2\sigma)\Gamma(2\sigma)}{\Gamma(-m+2\sigma)\Gamma(-2\sigma)}.
\end{equation}
By solving  \eqref{eq:apparentSingularityCancellation} iteratively as a power series in $\mathfrak{q}$, we get the full answers in both these  cases:
\begin{equation} \label{eq:etaMinus}
	\ba 
		\re^{\ri \eta_{\star}^-/2}=&-\mathfrak{q}^{-\sigma}\frac{\Gamma(1-m-2\sigma)\Gamma(2\sigma)}{\Gamma(1-m+2\sigma)\Gamma(-2\sigma)}\exp \left(\frac{8m^2(m-1)^2}{(1-4\sigma^2)^2}\mathfrak{q} + \mathcal{O}(\mathfrak{q}^2) \right)\\
		=&\exp\left(\partial_{\sigma}F^{\rm NS} ( \sigma,  m-{1\over 2}, \mathfrak{q})/2\right),
	\ea
\end{equation}
\begin{equation}
\label{eq:6}\ba
\re^{\ri \eta_{\star}^+/2}=&\mathfrak{q}^{-\sigma}\frac{\Gamma(-m-2\sigma)\Gamma(2\sigma)}{\Gamma(-m+2\sigma)\Gamma(-2\sigma)}\exp \left(\frac{8m^2(m+1)^2}{(1-4\sigma^2)^2}\mathfrak{q} + \mathcal{O}(\mathfrak{q}^2) \right)\\
=&\exp\left(\partial_{\sigma}F^{\rm NS} ( \sigma,  m+{1\over 2}, \mathfrak{q})/2\right),\ea\end{equation}
where we used the definition of $F^{\rm NS}$ in  \eqref{ns2s} \footnote{In principle, formulas \eqref{eq:6}, \eqref{eq:5} work for both cases $\#$ 1 and $\#$ 2 as we will also see in Sec.~\ref{sec:ns}. However, in $\#$ 2 it is more natural to think of the normalizability equation as an equation for $\sigma$ instead of an equation for $\eta$. But the two approaches are equivalent.}. The appearance of this quantity will  be clarified in Sec.~\ref{sec:ns}.
The two solutions $\re^{\ri \eta_{\star}^\pm/2}$ correspond to the two operators $ {\rm O}_{\mp}$ in \eqref{op3}. We can focus without loss of generality on ${\rm O}_-$.

If now we impose the normalizability condition for $\sigma$ \eqref{eq:sigmal} on this solution, we get
\begin{equation}
\label{eq:66}\ba
\re^{\ri\eta_\star^{-} /2}\mid_{\sigma=\sigma_{\ell}}=&-\mathfrak{q}^{(-1)^\ell m/2 - k_{\ell}/2}\frac{\Gamma(1-(-1)^\ell m-k_{\ell})\Gamma((-1)^{\ell+1} m+k_{\ell})}{\Gamma(1-(-1)^{\ell+1}  m+k_{\ell})\Gamma((-1)^\ell m-k_{\ell})}(1+\mathcal{O}(\mathfrak{q})). \ea
\end{equation}
Now we consider the condition that \( \eta\) is finite. This means that the gamma functions  should not have poles  nor zeroes.
For \(\sigma_1\) (i.e. $\ell=1$ in \eqref{eq:66}) this means that \be  k_1\geq 1\,. \ee 
For \(\sigma_2\) instead we have
\be  k_2\leq -1\,. \ee
In other words, this means
\begin{equation}
\label{eq:69}
2\sigma_{\ell}=(-1)^{\ell+1}(m+k),\quad k\geq1, \quad \ell=1,2\,.
\end{equation}
To compute the energy we use Table~\ref{tab:normalizabilityConditions} as well as \eqref{eq:energyd}
\begin{equation}
\label{eq:30}
E=-H_\star^-(\sigma, m, \eta_\star^-,\tau) -2 m(m-1)\eta_1(\tau) \Big |_{\sigma=\pm(m-k)/2}\,.
\end{equation}
The first terms of the expansion are
\begin{equation}
\label{eq:70}
E=-\frac{\pi^2}{3}m(m-1)+\pi^2(m+k)^2+8\pi^2m(m-1) \mathfrak{q}\left( { 1}+ \frac{m(m-1)}{(m+k)^2-1} \right)+\mathcal{O}(\mathfrak{q})^2,
\end{equation}
which coincides with perturbative calculation \eqref{eq:energyAcycle} (see also \cite[Sec.2]{Hatsuda_2018}).
 
 \section{Nekrasov-Shatashvili quantization from Kiev formula}\label{sec:ns}
 
The operators discussed above have an interpretation as (four-dimensional) quantum Seiberg-Witten curves.
In particular they  also  appear in the work of  Nekrasov and Shatashvili (NS) in the context of the Bethe/gauge correspondence for (non-relativistic) quantum integrable models \cite{ns, Nekrasov:2014yra}.
  In this section we show that the exact quantization condition proposed by  \cite{ns,Nekrasov:2014yra} can in fact be derived from the approach based on  the tau function of isomonodromic problems presented above. The key ingredients in this analysis are the
  Kiev formulas for tau functions \cite{gil, gil1, Bonelli:2019boe}, as well as   Nakajima-Yoshioka blowup equations \cite{ny1,Nakajima:2003uh}.
 The relation between Painlev\'e  and blowup equations has appeared before in the literature. 
 For example in \cite{Bershtein:2018zcz, Bershtein:2014yia} blowup equations on $\mathbb{C}^2/\mathbb{Z}^2$ were used to prove the Kiev formula \cite{gil} and its q-deformation \cite{Bershtein:2016aef}. 
 More recently an alternative proof for the Painlev\'e VI example was presented  in \cite{1805497,Jeong:2020uxz}  based on blowup equation {with defects}.  
 The interplay between  Painlev\'e and blowup equations appearing in this section is similar to the one of \cite{Grassi:2019coc, ggu, Gavrylenko:2020gjb} and does not require any defect. 

In order to write differential blowup relations we will use the Hirota differential operators $\mathrm{D}^{k}_{\epsilon_1,\epsilon_2}$ with respect to \(\log \mathfrak{q}\) which are defined by the formula 
\begin{equation}\label{eq:Hirota operators}
	F(\mathfrak{q} \re^{\epsilon_1 \hbar}) 	G(\mathfrak{q} \re^{\epsilon_2 \hbar})=\sum \frac{\hbar^k}{k!} \mathrm{D}^{k}_{\epsilon_1,\epsilon_2} (F,G). 
\end{equation}
For example $\mathrm{D}^{1}_{\epsilon_1,\epsilon_2}(F,G)=\epsilon_1  G~\partial_{\log \mathfrak{q}}F+\epsilon_2 F~\partial_{\log \mathfrak{q}}G$.

\subsection{Modified Mathieu} \label{sec:buMa}

In this section we prove that \eqref{eeni} and \eqref{taum} lead to the quantization condition obtained in \cite{ns, mirmor} and proven in \cite{kt2}.  

The starting point are Nakajima-Yoshioka blowup equations \cite{ny1} for 
Nekrasov partition function of pure $\mathcal{N}=2$, $SU(2)$ Seiberg-Witten theory in the four-dimensional $\Omega$ background. Such partition function is denoted by 
$\mathcal{Z}(a,\epsilon_1,\epsilon_2, t)$,
see for instance \cite[Sec 4.1]{Grassi:2019coc} for a complete definition and more references.
In this paper we are interested in two special limits of this function 

The first one is the
 self-dual limit where $\epsilon_2 \to - \epsilon_1$. In this case we have 
\be 
	\mathcal{Z}(a,\epsilon_1,\epsilon_2, t) \xrightarrow{\epsilon_2 =-\epsilon_1=-1, a=\sigma}  {\sim \mathfrak{q}^{\sigma^2}  \frac{1}{\prod_{\epsilon'=\pm}G(1+2\epsilon' \sigma)} } Z(\sigma,t)  \,,
\ee
where $ Z(\sigma,t)$ is the $c=1$ Virasoro conformal block appearing in \eqref{zmath},  and $\sim$ stands for the constant factor.

The second limit is the Nekrasov-Shatashvili limit $\epsilon_2\to 0$.
In this case we have \be \epsilon_2 \log \left(\mathcal{Z}(\sigma,1,\epsilon_2, t)\right) \xrightarrow{\epsilon_2 \to 0} -{F}_{\rm NS}( \sigma, t), \ee
where ${F}_{\rm NS}$ is the NS free energy for the pure $\mathcal{N}=2, SU(2)$ four-dimensional theory. More precisely we have
\be {F}_{\rm NS}(\sigma, t)=  -\psi ^{(-2)}(1+2 \sigma)-\psi ^{(-2)}(1-2 \sigma )+\sigma ^2 \log (t)+F^{\rm NS}_{\rm inst}(\sigma,t) ,\ee
where $\psi $ is the polygamma function,
and $ F^{\rm NS}_{\rm inst}$ is the instanton part of the NS free energy (or logarithm of $c\to\infty$ Virasoro conformal blocks). The first few terms read
\be F^{\rm NS}_{\rm inst}(\sigma,t)=-\frac{2 t}{-4\sigma^2+1}+\frac{t^2 \left(7 +20 \sigma^2\right)}{\left(-4 \sigma^2+1\right)^3 \left(-4 \sigma^2+4 \right)}+\mathcal{O}(t^3). \ee
Higher order terms can be computed by using combinatorics and Young diagrams,  we refer to \cite[Sec. 4.1]{Grassi:2019coc} for the details of the definition and a list of references.
It was shown in \cite{ggu} that  the two limits introduced above are in fact closely related by the Nakajima-Yoshioka blowup equations \cite{ny1} on $\mathbb{C}^2$. 

We have several blowup relations on $\mathbb{C}^2$.  One relation is \cite[Eq. (5.3)]{Nakajima:2003uh}
\begin{align} \label{eq:blowup c=1 D=0}
	 \sum_{n \in \mathbb{Z}+1/2}& \mathcal{Z}(a+n\epsilon_1,\epsilon_1,-\epsilon_1+\epsilon_2;z)\mathcal{Z}(a+n\epsilon_2,\epsilon_1-\epsilon_2,\epsilon_2;z)=0.	 	
\end{align} 
As in \cite{ggu}, we take the limit $\epsilon_2\rightarrow 0$ of  \eqref{eq:blowup c=1 D=0}  and we get
\begin{equation}\label{eqq}
\mathcal{T}_1(\sigma,\eta,t)=0, \quad \text{ for } \eta= \ri \partial_{\sigma} F^{\rm NS}(\sigma, t).
\end{equation} 
This means that if we look at the singularities matching condition \eqref{ttv} as an equation for $\eta$, the solution is
\be \eta= \ri \partial_{\sigma} F^{\rm NS}(\sigma, t). \ee
Moreover, if in addition we  impose \eqref{nor} we get
\be \label{nsmx} \ba  \partial_{\sigma} F^{\rm NS}(\sigma, t)&= 2 \ri \pi n, \quad n=1,2,\cdots.
\ea\ee
This is  the quantization condition proposed in \cite{ns}, where it was found that
the solutions $\{\sigma_n\}_{n>0}$ of \eqref{nsmx} are related to the spectrum $\{E_n\}_{n>0}$ of \eqref{mathieu} via the Matone relation \cite{matone,francisco,lmn,bkk-matone}
\be \label{4dmirrorpurex}E_n(t) = -{t}\partial_{t} F^{\rm NS}(\sigma_n, t).\ee
To reproduce Matone relation from the point of view of isomonodromic deformations we need another   blowup equation which takes the form of a differential bilinear relation and  reads \cite[eq. (6.14)]{ny1}
\be \ba	\sum_{n \in \mathbb{Z}}& \mathrm{D}^1_{\epsilon_1, \epsilon_2}\Big(\mathcal{Z}(a+n\epsilon_1,\epsilon_1,-\epsilon_1+\epsilon_2;t){ ,}\mathcal{Z}(a+n\epsilon_2,\epsilon_1-\epsilon_2,\epsilon_2;t)\Big)=0,  ~ \label{eq:blowup c=1 D=1}
\ea \ee
{where $\mathrm{D}^1_{\epsilon_1, \epsilon_2}$ was defined in \eqref{eq:Hirota operators} 
}. 
In  the limit $\epsilon_2\rightarrow 0$ this equation becomes
\begin{equation}\label{eqq2}
\left(t \partial_t\log \mathcal{T}_0(\sigma,\eta,t)\right)\Big|_{\eta=\ri \partial_\sigma F^{\rm NS}(\sigma, t)}=(t\partial_t F^{\rm NS}(\sigma, t)) .
\end{equation}
Note that the $t$ derivative on the l.h.s.~does not act on $\eta$.
Hence we have  an equivalence between Matone relation \eqref{4dmirrorpurex} and the Hamiltonian  \eqref{eeni} of Painlev\'e ${\rm III_3}$. This concludes the derivation of the NS quantization from the Kiev formula.

 \subsection{Weierstrass Potential} \label{butorus}

The starting point are Nakajima-Yoshioka blowup equations for 
Nekrasov partition function of  $\mathcal{N}=2^*$ \(SU(2)\) Seiberg-Witten theory in the four-dimensional $\Omega$ background. Such partition function is denoted by 
\be\label{zfull} \mathcal{Z}(a,\alpha;\epsilon_1,\epsilon_2|\mathfrak{q}).\ee
One can find the definition  in \eqref{full2s}.
For the purpose of this paper we are interested only in two limits of \eqref{zfull}.\footnote{In such limits we can absorb $\epsilon_1$ into a redefinition of parameters. So we set it to $\epsilon_1=1$.} In the self-dual limit we have
\be \mathcal{Z}(a,\alpha,\epsilon_1,\epsilon_2|\mathfrak{q})\xrightarrow{\epsilon_2 =- \epsilon_1=1, a=\sigma, \alpha=m}   \mathfrak{q}^{\sigma^2} (2 \pi)^{-m} \frac{\prod_{\epsilon'=\pm}G(1-m+2\epsilon'\sigma)}{\prod_{\epsilon'=\pm}G(1+2\epsilon' \sigma)}  {Z(\sigma,m,\mathfrak{q})},\ee
where $Z(\sigma,m,\mathfrak{q})$ is the $c=1$ conformal block on the torus as in \eqref{eq:dualNekrasov}. In the NS limit  we have
\be \label{NSl}
	\epsilon_2 \log \left(\mathcal{Z}(\sigma, m , 1,\epsilon_2| \mathfrak{q})\right) \xrightarrow{\epsilon_2 \to 0} {F}^{\rm NS}( \sigma,   m-{1\over 2},\mathfrak{q}),
\ee
where ${F}^{\rm NS}$ is defined in \eqref{ns2s}.
\subsubsection{Blowup relations}\label{ssec:blowup N=2*}

We first note that Nakajima-Yoshioka blowup relations for the four dimensional $\mathcal{N}=2^*$ theory  were not written explicitly in the literature. 
In this section we list the relevant relations that are used in the paper. Some of them have been worked out by one of us (MB) together with A.~Litvinov and A.~Shchechkin some time ago. The five-dimensional version of some of these equations was recently obtained in \cite[Sec.~3.1]{Gu:2019pqj}.  

We start with two algebraic blowup relations  on $\mathbb{C}^2$ 
\be\label{bua}\ba
	\frac{\theta_3(0| 2\tau)}{\varphi(\mathfrak{q})} 	\mathcal{Z}(a,\alpha;\epsilon_1,\epsilon_2|\mathfrak{q})=&\sum _{n \in \mathbb{Z}} \mathcal{Z}(a+n\epsilon_1,\alpha;\epsilon_1,\epsilon_2-\epsilon_1|\mathfrak{q}) \mathcal{Z}(a+n\epsilon_2,\alpha;\epsilon_1-\epsilon_2,\epsilon_2|\mathfrak{q}),
	\\
	\frac{\theta_2(0| 2\tau)}{\varphi(\mathfrak{q})} 	\mathcal{Z}(a,\alpha;\epsilon_1,\epsilon_2|\mathfrak{q})=&\sum _{n \in \mathbb{Z}+\frac{1}{2}} \mathcal{Z}(a+n\epsilon_1,\alpha;\epsilon_1,\epsilon_2-\epsilon_1|\mathfrak{q}) \mathcal{Z}(a+n\epsilon_2,\alpha;\epsilon_1-\epsilon_2,\epsilon_2|\mathfrak{q}).
\ea\ee
Using the symmetry \eqref{eq:alpha symm} we can obtain two more  equations:
\be\label{bua1}\ba
	\frac{\theta_3(0| 2\tau)}{\varphi(\mathfrak{q})} 	\mathcal{Z}(a,\alpha;\epsilon_1,\epsilon_2|\mathfrak{q})=&\sum _{n \in \mathbb{Z}} \mathcal{Z}(a+n\epsilon_1,\alpha-\epsilon_1;\epsilon_1,\epsilon_2-\epsilon_1|\mathfrak{q}) \mathcal{Z}(a+n\epsilon_2,\alpha-\epsilon_2;\epsilon_1-\epsilon_2,\epsilon_2|\mathfrak{q}),
	\\
	\frac{\theta_2(0| 2\tau)}{\varphi(\mathfrak{q})} 	\mathcal{Z}(a,\alpha;\epsilon_1,\epsilon_2|\mathfrak{q})=&\sum _{n \in \mathbb{Z}+\frac{1}{2}} \mathcal{Z}(a+n\epsilon_1,\alpha-\epsilon_1;\epsilon_1,\epsilon_2-\epsilon_1|\mathfrak{q}) \mathcal{Z}(a+n\epsilon_2,\alpha-\epsilon_2;\epsilon_1-\epsilon_2,\epsilon_2|\mathfrak{q}).
\ea\ee
There are also differential relations which are written in terms of the Hirota differential operators defined in \eqref{eq:Hirota operators}. The first order relations are 
\begin{multline}\label{bu3}
	\left( (\epsilon_1+\epsilon_2)\beta_0^{1,1}(\mathfrak{q}) +\alpha \beta_0^{1,2}(\mathfrak{q})  \right)\mathcal{Z}(a,\alpha;\epsilon_1,\epsilon_2|\mathfrak{q})\\=\sum _{n \in \mathbb{Z}} \mathrm{D}^{1}_{\epsilon_1,\epsilon_2}\big(\mathcal{Z}(a+n\epsilon_1,\alpha;\epsilon_1,\epsilon_2-\epsilon_1|\mathfrak{q}) { ,}\mathcal{Z}(a+n\epsilon_2,\alpha;\epsilon_1-\epsilon_2,\epsilon_2|\mathfrak{q})\big),
\end{multline}
\begin{multline}\label{bu4}
	\left( (\epsilon_1+\epsilon_2)\beta_1^{1,1}(\mathfrak{q}) +\alpha \beta_1^{1,2}(\mathfrak{q})  \right)\mathcal{Z}(a,\alpha;\epsilon_1,\epsilon_2|\mathfrak{q})\\=\sum _{n \in \mathbb{Z}+\frac{1}{2}} \mathrm{D}^{1}_{\epsilon_1,\epsilon_2}\big(\mathcal{Z}(a+n\epsilon_1,\alpha;\epsilon_1,\epsilon_2-\epsilon_1|\mathfrak{q}){ ,} \mathcal{Z}(a+n\epsilon_2,\alpha;\epsilon_1-\epsilon_2,\epsilon_2|\mathfrak{q})\big),
\end{multline}
where 
\begin{align}
		\beta_0^{1,1}(\mathfrak{q})=\partial_{\log \mathfrak{q}}\big(\frac{\theta_3(0|2\tau)}{\varphi(\mathfrak{q})} \big), \quad \beta_0^{1,2}(\mathfrak{q})=2 \frac{\partial_{\log \mathfrak{q}}\theta_3(0|2\tau)}{\varphi(\mathfrak{q})},
		\\
		\beta_1^{1,1}(\mathfrak{q})=\partial_{\log \mathfrak{q}}\big(\frac{\theta_2(0|2\tau)}{\varphi(\mathfrak{q})} \big), \quad \beta_1^{1,2}(\mathfrak{q})=2 \frac{\partial_{\log \mathfrak{q}}\theta_2(0|2\tau)}{\varphi(\mathfrak{q})}.
\end{align}
There are also second order differential relations which look rather cumbersome
\begin{multline}\label{eq:-1 2Hirota 1}
	\left( (\epsilon_1{+}\epsilon_2)^2 \beta_0^{2,1}(\mathfrak{q})+\alpha (\epsilon_1{+}\epsilon_2)  \beta_0^{2,2}(\mathfrak{q}) +\epsilon_1\epsilon_2 \beta_0^{2,3}(\mathfrak{q})+ \epsilon_1\epsilon_2 \beta_0^{2,4}(\mathfrak{q}) \partial_{\log \mathfrak{q}} \right)\mathcal{Z}(a,\alpha;\epsilon_1,\epsilon_2|\mathfrak{q})
	\\
	=\sum _{n \in \mathbb{Z}} \mathrm{D}^{2}_{\epsilon_1,\epsilon_2}\big(\mathcal{Z}(a+n\epsilon_1,\alpha;\epsilon_1,\epsilon_2-\epsilon_1|\mathfrak{q}) \mathcal{Z}(a+n\epsilon_2,\alpha;\epsilon_1-\epsilon_2,\epsilon_2|\mathfrak{q})\big),
\end{multline}
\begin{multline}\label{eq:-1 2Hirota 2}
	\left( (\epsilon_1{+}\epsilon_2)^2 \beta_1^{2,1}(\mathfrak{q})+\alpha (\epsilon_1{+}\epsilon_2)  \beta_1^{2,2}(\mathfrak{q}) +\epsilon_1\epsilon_2 \beta_1^{2,3}(\mathfrak{q})+ \epsilon_1\epsilon_2 \beta_1^{2,4}(\mathfrak{q}) \partial_{\log \mathfrak{q}} \right)\mathcal{Z}(a,\alpha;\epsilon_1,\epsilon_2|\mathfrak{q})
	\\
	=\sum _{n \in \mathbb{Z}+\frac12} \mathrm{D}^{2}_{\epsilon_1,\epsilon_2}\big(\mathcal{Z}(a+n\epsilon_1,\alpha;\epsilon_1,\epsilon_2-\epsilon_1|\mathfrak{q}) \mathcal{Z}(a+n\epsilon_2,\alpha;\epsilon_1-\epsilon_2,\epsilon_2|\mathfrak{q})\big),
\end{multline}
where 
\begin{align}
	\beta_0^{2,1}(\mathfrak{q})=\partial^2_{\log \mathfrak{q}}\frac{\theta_3(0|2\tau)}{\varphi(\mathfrak{q})},\; \beta_0^{2,2}(\mathfrak{q})=4\partial_{\log \mathfrak{q}}\frac{\partial_{\log \mathfrak{q}}\theta_3(0|2\tau)}{\varphi(\mathfrak{q})},\;
	\beta_0^{2,4}(\mathfrak{q})=-4 \frac{\partial_{\log \mathfrak{q}}\theta_3(0|2\tau)}{\varphi(\mathfrak{q})}
	\\
	\beta_0^{2,3}(\mathfrak{q})=\left(-\frac{1}{3}\frac{\partial_{\log \mathfrak{q}}\theta_3(0|2\tau)}{\varphi(\mathfrak{q})}-4\frac{\partial_{\log \mathfrak{q}}\theta_3(0, 2\tau) \partial_{\log \mathfrak{q}}\varphi(\mathfrak{q})}{\varphi(\mathfrak{q})^2}+\frac{4}{3}\frac{\partial^2_{\log \mathfrak{q}}\theta_3(0|2\tau)}{\varphi(\mathfrak{q})} \right),
	\\
	\beta_1^{2,1}(\mathfrak{q})=\partial^2_{\log \mathfrak{q}}\frac{\theta_2(0|2\tau)}{\varphi(\mathfrak{q})},\; \beta_1^{2,2}(\mathfrak{q})=4\partial_{\log \mathfrak{q}}\frac{\partial_{\log \mathfrak{q}}\theta_2(0|2\tau)}{\varphi(\mathfrak{q})},\; 
	\beta_1^{2,4}(\mathfrak{q})=-4 \frac{\partial_{\log \mathfrak{q}}\theta_2(0|2\tau)}{\varphi(\mathfrak{q})}
	\\
	\beta_1^{2,3}(\mathfrak{q})=\left(-\frac{1}{3}\frac{\partial_{\log \mathfrak{q}}\theta_2(0|2\tau)}{\varphi(\mathfrak{q})}-4\frac{\partial_{\log \mathfrak{q}}\theta_2(0, 2\tau) \partial_{\log \mathfrak{q}}\varphi(\mathfrak{q})}{\varphi(\mathfrak{q})^2}+\frac{4}{3}\frac{\partial^2_{\log \mathfrak{q}}\theta_2	(0|2\tau)}{\varphi(\mathfrak{q})} \right).	
\end{align}
We will not use the relations \eqref{eq:-1 2Hirota 1}, \eqref{eq:-1 2Hirota 2} to study the Nekrasov-Shatashvili quantization conditions, but they will be used later in Sec. \ref{Sec:-2blowup}.

We do not claim that this is the full list of blowup relations, this is just the list needed in this paper. Note also that, as far as we know, there is no rigorous proof of these relations. However we believe that this can be done either by using the geometric methods of \cite{ny1,Nakajima:2011} or by using the representation theory methods of \cite{Bershtein:2013oka}.

\subsubsection{From blowup to NS quantization conditions and spectrum}
By combining the relations \eqref{bua} we obtain 
\begin{multline}\label{bb1}
\theta_2(0|2\tau) \sum _{n \in \mathbb{Z}} \mathcal{Z}(a+n\epsilon_1,\alpha;\epsilon_1,\epsilon_2-\epsilon_1|\mathfrak{q}) \mathcal{Z}(a+n\epsilon_2,\alpha;\epsilon_1-\epsilon_2,\epsilon_2|\mathfrak{q})\\ -\theta_3(0|2\tau)\sum _{n \in \mathbb{Z}+\frac{1}{2}} \mathcal{Z}(a+n\epsilon_1,\alpha;\epsilon_1,\epsilon_2-\epsilon_1|\mathfrak{q}) \mathcal{Z}(a+n\epsilon_2,\alpha;\epsilon_1-\epsilon_2,\epsilon_2|\mathfrak{q})=0.
\end{multline}
Similarly from \eqref{bua1} we obtain
\begin{multline}\label{bb2}
\theta_2(0|2\tau) \sum _{n \in \mathbb{Z}} \mathcal{Z}(a+n\epsilon_1,\alpha;\epsilon_1,\epsilon_2-\epsilon_1|\mathfrak{q}) \mathcal{Z}(a+n\epsilon_2,\alpha{+\epsilon_1-\epsilon_2 };\epsilon_1-\epsilon_2,\epsilon_2|\mathfrak{q})\\ -\theta_3(0|2\tau)\sum _{n \in \mathbb{Z}+\frac{1}{2}} \mathcal{Z}(a+n\epsilon_1,\alpha;\epsilon_1,\epsilon_2-\epsilon_1|\mathfrak{q}) \mathcal{Z}(a+n\epsilon_2,\alpha{+\epsilon_1-\epsilon_2 };\epsilon_1-\epsilon_2,\epsilon_2|\mathfrak{q})=0.
\end{multline}
Taking the NS limit $a=\sigma$, $\alpha=m$, $\epsilon_1=1$, $\epsilon_2\to 0$ we are left with \footnote{The possibility of a relation between \eqref{eq:solutionTorus} and blowup equations has also been hypothesised by G.~Bonelli,  F.~Del Monte, A.~Tanzini in \cite{talkfabrizio}. }
\be\label{ttb}
	\theta_2(0|2\tau) Z_0^D(\sigma,m,\eta_\star^-,\tau )  -\theta_3(0|2\tau)  Z_{1/2}^D(\sigma,m,\eta_\star^-,\tau ) =0, \qquad \eta_\star^-=- \ri \partial_{\sigma}F^{\rm NS} ( \sigma,  m-{1\over 2}, \mathfrak{q})\, , 
\ee

\be\label{ttb2}
	\theta_2(0|2\tau) Z_0^D(\sigma,m,\eta_\star^+,\tau )  -\theta_3(0|2\tau)  Z_{1/2}^D(\sigma,m,\eta_\star^+,\tau ) =0, \qquad \eta_\star^+= -\ri \partial_{\sigma}F^{\rm NS} ( \sigma,  m+{1\over 2}, \mathfrak{q}).
\ee

In turn, this means that if we consider \eqref{eq:apparentSingularityCancellation}
as an equation for \(\eta\), then we have two solutions: $\eta_\star^-$ and $\eta_\star^+$. The solution $\eta_\star^-$
makes contact with the operator ${\rm O}_-$,  while  the solution $\eta_\star^+$ 
makes contact with the operator ${\rm O}_+$. This is in perfect agreement with what we discussed around \eqref{2o} and \eqref{eq:etaMinus}.

Let us now consider the operator corresponding to the case $\#2$   in Table \ref{tab:homologyClasses} (the other cases work analogously). From the above discussion it follows that the singularities matching condition and  the normalizability of the linear problem are equivalent to
\be\label{nsmx2}  \ba &\partial_{\sigma}F^{\rm NS} ( \sigma,   m\mp{1\over 2}, \mathfrak{q})=  2\pi \ri (n+1), \quad n=0,1,2,\cdots \quad \text{ if} \quad m<-1,  \\
& \partial_{\sigma}F^{\rm NS} ( \sigma,   -m\pm{1\over 2}, \mathfrak{q})=  2\pi \ri (n+1), \quad n=0,1,2,\cdots \quad \text{ if} \quad m>1. 
\ea \ee 
This is precisely the quantization condition proposed in \cite{ns,Nekrasov:2014yra} where it was found that
the solutions $\{\sigma_n\}_{n\geq0}$ of \eqref{nsmx2} are related to the spectrum $\{E_n\}_{n\geq 0}$ of $ {\rm O}_{\mp}$ via Matone relation 
\be \label{4dmirrorns}E_n(t) =  ( \log\mathfrak{q})^2  \left(\mathfrak{q}\partial_\mathfrak{q}  F^{\rm NS} (\sigma,m\mp{1\over 2},\mathfrak{q} )+{1\over 4 \pi^2}2 m(m\mp 1)\eta_1(\tau)\right).\ee
As in Sec.~\ref{sec:buMa}, to reproduce Matone relation in the context of isomonodromic deformations, we need another set of  blowup equations. These take the form of differential bilinear relations.
It follows from \eqref{bu3} and \eqref{bua} that 
\begin{multline}
	\left((\epsilon_1+\epsilon_2)
	\partial_{\log \mathfrak{q}}\big(\log \frac{\theta_3(0|2\tau)}{\varphi(\mathfrak{q})} \big)+2 \alpha \frac{\partial_{\log \mathfrak{q}}\theta_3(0|2\tau)}{\theta_3(0|2\tau)}
	\right)\times \\ \sum _{n \in \mathbb{Z}} \mathcal{Z}(a+n\epsilon_1,\alpha;\epsilon_1,\epsilon_2-\epsilon_1|\mathfrak{q}) \mathcal{Z}(a+n\epsilon_2,\alpha;\epsilon_1-\epsilon_2,\epsilon_2|\mathfrak{q}) \\ =\sum _{n \in \mathbb{Z}} \mathrm{D}^{1}_{\epsilon_1,\epsilon_2}\big(\mathcal{Z}(a+n\epsilon_1,\alpha;\epsilon_1,\epsilon_2-\epsilon_1|\mathfrak{q}) { ,}\mathcal{Z}(a+n\epsilon_2,\alpha;\epsilon_1-\epsilon_2,\epsilon_2|\mathfrak{q})\big)\, .
\end{multline}
Taking the NS  limit $a=\sigma$, $\alpha=m$, $\epsilon_1=1$, $\epsilon_2\to 0$ we get 
\begin{multline}
		\left(\partial_{\log \mathfrak{q}}\big(\log \frac{\theta_3(0|2\tau)}{\varphi(\mathfrak{q})} \big)+2 m \frac{\partial_{\log \mathfrak{q}}\theta_3(0|2\tau)}{\theta_3(0|2\tau)}
		\right)\mathfrak{q}^{1/24}Z_0^D(\sigma,m,\eta_\star^-,\tau) 
		\\
		=\partial_{\log \mathfrak{q}}  \left(\mathfrak{q}^{1/24}Z_0^D(\sigma,m,\eta_\star^-,\tau)\right)+ \mathfrak{q}^{1/24}Z_0^D(\sigma,m,\eta_\star^-,\tau)  \partial_{\log\mathfrak{q}}  F^{\rm NS}(\sigma,m-{1\over 2},\mathfrak{q} )\, ,
\end{multline}
where  we use \be \eta_\star^-=- \ri \partial_{\sigma}F^{\rm NS} ( \sigma,  m-{1\over 2}, \mathfrak{q}),\ee as in \eqref{ttb}. Using \eqref{eq:energyd} we obtain 
\begin{equation}
	\partial_{\log \mathfrak{q}}  \log Z_0^D(\sigma,m,\eta_\star^-,\tau)=
	-\frac{1}{4\pi^2}H_\star^--\partial_{\log \mathfrak{q}}\log \frac{\varphi(\mathfrak{q})}{\theta_3(0|2\tau)}-\frac1{24}+2 m \frac{\partial_{\log \mathfrak{q}}\theta_3(0|2\tau)}{\theta_3(0|2\tau)}
\end{equation}
and get
\begin{equation}
	{1\over 4 \pi^2}  H_\star^{-} =  \mathfrak{q}\partial_\mathfrak{q}  F^{\rm NS} (\sigma,m-{1\over 2},\mathfrak{q} ).
\end{equation}
The relation for $\eta_\star^+$ can be obtained similarly by using the symmetry \eqref{eq:alpha symm}. 
We have
\begin{equation}
	{1\over 4 \pi^2}  H_\star^{+} =  \mathfrak{q}\partial_\mathfrak{q}  F^{\rm NS} (\sigma,m+{1\over 2},\mathfrak{q} ).
\end{equation}
Hence we have a complete equivalence between the  Hamiltonian \eqref{esigma} and Matone relation \eqref{4dmirrorns}. This concludes the derivation of the NS quantization from the  Kiev formula on the torus.

As a final remark we note that, unlike in the example of modified Mathieu, to our knowledge there is no proof  of the NS quantization for the example of the quantum elliptic Calogero-Moser system\footnote{There are  however several  independent tests which have been performed (see for instance \cite[Sec.2]{Hatsuda_2018}, as well as  \cite{He:2011zk, Piatek:2011tp, Basar:2015xna, Beccaria:2016wop}, or also Appendix \ref{sec:nstest}). }. {Our derivation here is based on monodromy arguments, which are rigorous, and blowup relations, which we believe can be proven.}

\section{Bilinear relations on the torus} \label{Sec:-2blowup}

In this section we show that the isomonodromic equation \eqref{eq:Calogero} in $Q$ is equivalent to the bilinear relation \eqref{eq: bilin dual Nekrasov tildei} for the $\mathcal{T}$ function, more precisely for $Z_\epsilon^{D}$. Moreover, by using blowup equations, we demonstrate that \eqref{eq:dualNekrasov} indeed satisfies such bilinear relation. This provides an alternative proof for the work of \cite{Bonelli:2019boe}

\subsection{From blowup relations}
Let us first note that by substituting \eqref{eq:dualNekrasov} into \eqref{eq: bilin dual Nekrasov tildei} one gets a bilinear relation for the function $Z$. As already noted in a related context \cite{Bershtein:2014yia},
this type of relations cannot be a specialization of the $\IC^2$ blowup equations.
Usually such relations come from the so-called $(-2)$ or $\mathbb{C}^2/\mathbb{Z}_2$ blowup equations. In this terminology the relations from Sec.~\ref{ssec:blowup N=2*} are called $(-1)$ blowup equations.
The $(-2)$ blowup relations can be obtained using representation theory (see e.g. \cite{Bershtein:2014yia}) or algebraic geometry (see e.g. \cite{Bruzzo:2013daa}, \cite{Ohkawa:2018vow}) arguments.  There is a
transparent algebraic method to deduce them from the standard $(-1)$ blowup relations\footnote{We a grateful to H. Nakajima for explaining this idea and help with references.}. This method goes back to the  papers on Donaldson invariants \cite{Fintushel},\cite{Brussee}.  Recently this method was applied to the case of the pure theory in \cite{Shchechkin:2020ryb}. Here we apply it to the $\mathcal{N}=2^*$ case. 
As a result we get a first order differential relation 
	\begin{multline}\label{eq:-2 1Hirota}
	\sum _{2n \in \mathbb{Z}} \mathrm{D}^{1}_{2\epsilon_1,2\epsilon_2}\big(\mathcal{Z}(a+2n\epsilon_1,\alpha;2\epsilon_1,\epsilon_2-\epsilon_1|\mathfrak{q}) \mathcal{Z}(a+2n\epsilon_2,\alpha;\epsilon_1-\epsilon_2,2\epsilon_2|\mathfrak{q})\big) \\ =
	(\epsilon_1+\epsilon_2)\gamma_0(\mathfrak{q}) \sum _{2n \in \mathbb{Z}} \mathcal{Z}(a+2n\epsilon_1,\alpha;2\epsilon_1,\epsilon_2-\epsilon_1|\mathfrak{q}) \mathcal{Z}(a+2n\epsilon_2,\alpha;\epsilon_1-\epsilon_2,2\epsilon_2|\mathfrak{q})
	\end{multline}
and a second order differential relation
	\begin{multline}\label{eq:-2 2Hirota}
	\sum _{2n \in \mathbb{Z}} \mathrm{D}^{2}_{2\epsilon_1,2\epsilon_2}\big(\mathcal{Z}(a+2n\epsilon_1,\alpha;2\epsilon_1,\epsilon_2-\epsilon_1|\mathfrak{q}) \mathcal{Z}(a+2n\epsilon_2,\alpha;\epsilon_1-\epsilon_2,2\epsilon_2|\mathfrak{q})\big) \\ =
	\Big(\epsilon_1\epsilon_2 \gamma_1(\mathfrak{q})+\alpha^2 \gamma_2(\mathfrak{q})+(\epsilon_1+\epsilon_2)^2\gamma_3(\mathfrak{q})+\alpha(\epsilon_1+\epsilon_2)\gamma_4(\mathfrak{q})+\epsilon_1\epsilon_2 \gamma_5(\mathfrak{q}) \partial_{\log \mathfrak{q}}\Big) 	\\ \sum _{2n \in \mathbb{Z}} \mathcal{Z}(a+2n\epsilon_1,\alpha;2\epsilon_1,\epsilon_2-\epsilon_1|\mathfrak{q}) \mathcal{Z}(a+2n\epsilon_2,\alpha;\epsilon_1-\epsilon_2,2\epsilon_2|\mathfrak{q}),
	\end{multline}
where the Hirota differential operators were defined in  \eqref{eq:Hirota operators} and  we use 
	\begin{align*}
	\gamma_0(\mathfrak{q})&=\log'\left(\frac{\theta_3^2+\theta_2^2}{\varphi^2}\right)=2 \partial_{\log {\mathfrak{q}}} \log \left(\frac{\theta_3(0|\tau)}{\varphi(\mathfrak{q})}\right), 
	\\	
	\gamma_2(\mathfrak{q})&=-\gamma_4(\mathfrak{q})=-8 \frac{(\theta_3')^2+(\theta_2')^2} {\theta_3^2+\theta_2^2}=-4\big(\partial_{\log \mathfrak{q}}^2\log(\theta_3(0|\tau))\big), 
	\\	
	\gamma_5(\mathfrak{q})&=-8 \log' (\theta_3^2+\theta_2^2)=-16\big(\partial_{\log \mathfrak{q}} \log(\theta_3(0|\tau))\big),
	\\
	\gamma_1(\mathfrak{q})&= -8\frac{\theta_3^2\log''({\theta_3}/{\varphi})+\theta_2^2\log''({\theta_2}/{\varphi})}{\theta_2^2+\theta_3^2} 
	+8\log'\left(\frac{\theta_3^2+\theta_2^2}{\varphi^2} \right)  \log' (\theta_3^2+\theta_2^2)\\&=8\big(\partial_{\log \mathfrak{q}}^2\log\varphi(\mathfrak{q})\big)-32\big(\partial_{\log \mathfrak{q}}\log \varphi(\mathfrak{q})\big) \big(\partial_{\log \mathfrak{q}}\log(\theta_3(0|\tau))\big)+16 \big(\partial_{\log \mathfrak{q}} \log \theta_3(0|\tau)\big)^2, 
	\\	
	\gamma_3(\mathfrak{q})&=\frac{2 (3\varphi'\varphi'{-}\varphi \varphi'')}{\varphi^2}-\frac{16\varphi'( \theta_3\theta_3'{+}\theta_2\theta_2')}{\varphi(\theta_3^2{+}\theta_2^2)}
	-\frac{18 (\theta_3'\theta_3'{+}\theta_2'\theta_2')+10(\theta_3\theta_3''{+}\theta_2\theta_2'')- (\theta_3\theta_3'{+}\theta_2\theta_2')}{3(\theta_3^2{+}\theta_2^2)}\, .
\end{align*}
 with the understanding that $\theta_i=\theta_i(0|2\tau)$, $\varphi=\varphi(\mathfrak{q})$, and $'$ means that the derivatives are taken with respect to $\log \mathfrak{q}$.
We give a detailed proof of the relation \eqref{eq:-2 1Hirota} in App. \ref{app:proof -2}. The proof of the relation \eqref{eq:-2 2Hirota} is based on the same ideas, but involves more cumbersome calculations. 
After the substitution  $\epsilon_2=-\epsilon_1=\frac12$, $a={\sigma}$, $ \alpha={m}$,  equation \eqref{eq:-2 2Hirota} leads to the following bilinear relation for the self-dual Nekrasov function $\mathcal{Z}(\sigma,m,-1,1|\mathfrak{q})$
\begin{multline}\label{eq:Hirota selfdual}
	\sum _{2n \in \mathbb{Z}} \mathrm{D}^{2}_{-1,1}\big(\mathcal{Z}(\sigma-n,m,-1,1|\mathfrak{q}) \mathcal{Z}(\sigma+n,m,-1,1|\mathfrak{q})\big) \\ =
	\Big(-\frac14\gamma_1(\mathfrak{q})+m^2 \gamma_2(\mathfrak{q})-\frac14 \gamma_5(\mathfrak{q}) \partial_{\log \mathfrak{q}}\Big)  \sum _{2n \in \mathbb{Z}} \big(\mathcal{Z}(\sigma-n,m,{-}1,1|\mathfrak{q}) \mathcal{Z}(\sigma+n,m,{-}1,1|\mathfrak{q})\big). 
\end{multline}
This is equivalent to 
\begin{multline}\label{eq: bilin dual Nekrasov}
	\mathrm{D}^{2}_{-1,1}({Z}_{0}^D,{Z}_{0}^D)+\mathrm{D}^{2}_{-1,1}({Z}_{1/2}^D,{Z}_{1/2}^D) 
	\\
	= \Big(-\frac{1}4(\gamma_1(\mathfrak{q})-\frac1{12}\gamma_5(\mathfrak{q}))+m^2 \gamma_2(\mathfrak{q})-\frac{1}4 \gamma_5(\mathfrak{q}) \partial_{\log \mathfrak{q}}\Big) \big({Z}_{0}^D {Z}_{0}^D+{Z}_{1/2}^D {Z}_{1/2}^D\big).
\end{multline}
Indeed, by comparing terms with given $\eta$ exponents (say $\re^{\ri k \eta}$ in \eqref{eq: bilin dual Nekrasov}) we obtain \eqref{eq:Hirota selfdual}. This is the same argument as in \cite[Sec.~4.2]{Bershtein:2014yia}.
Using the notation \eqref{eq: dual Nekrasov tilde} and the relation $\mathrm{D}^2_{-1,1}(F,F)=2 F^2 \big(\partial_{\log \mathfrak{q}}^2 \log F\big)$ we get
	\begin{multline}\label{eq: bilin dual Nekrasov tilde}
	(\tilde{{Z}}_{0}^D)^2 \partial_{\log \mathfrak{q}}^2 \log \tilde{{Z}}_{0}^D+ 	(\tilde{{Z}}_{1/2}^D)^2 \partial_{\log \mathfrak{q}}^2 \log \tilde{{Z}}_{1/2}^D
	\\= 
	\Big(-2\big(\partial_{\log \mathfrak{q}} \log(\theta_3(0|\tau))\big)^2-2m^2 \big(\partial_{\log \mathfrak{q}}^2\log(\theta_3(0|\tau))\big)+2\big(\partial_{\log \mathfrak{q}} \log(\theta_3(0|\tau))\big) \partial_{\log \mathfrak{q}}\Big) \\ \big(\tilde{{Z}}_{0}^D \tilde{{Z}}_{0}^D+\tilde{{Z}}_{1/2}^D \tilde{{Z}}_{1/2}^D\big),
\end{multline}
which is precisely equation \eqref{eq: bilin dual Nekrasov tildei}. 
To have a better intuition on the meaning of \eqref{eq: bilin dual Nekrasov tilde} let us consider two particular cases. 
\newline	

\noindent{\bf Example 1.} Let $m=0$. Then we have $Z(\sigma,m,\mathfrak{q})=1/\varphi(\mathfrak{q})$ and get (cf. formulas \eqref{eq:Q for m=0} and \eqref{simple})
\begin{equation}
	\tilde{Z}_{0}^D = \mathfrak{q}^{\sigma^2}\theta_3(\frac{\eta}{2\pi}+2  \sigma \tau |2\tau),\quad  \tilde{Z}_{1/2}^D = \mathfrak{q}^{\sigma^2}\theta_2(\frac{\eta}{2\pi}+2  \sigma \tau|2\tau).
\end{equation}
It is convenient to reduce everything to theta functions with modular parameter $\tau$ by using\begin{equation}\label{eq:2tau=tau 1}
	\theta_3(z|2\tau)^2+\theta_2(z|2\tau)^2=\theta_3(z|\tau)\theta_3(0|\tau).
\end{equation}
We have
\begin{multline}\label{eq:2tau=tau 2}
	\theta_3(z|2\tau)^2\, \big(\partial_{\log \mathfrak{q}}^2 \log \theta_3(z|2\tau)\big) +\theta_2(z|2\tau)^2\, \big(\partial_{\log \mathfrak{q}}^2 \log \theta_2(z|2\tau)\big)
	\\
	=2\big(\partial_{\log \mathfrak{q}}\theta_3(z|\tau)\big)\,\big(\partial_{\log \mathfrak{q}} \theta_3(0|\tau)\big),
\end{multline}
\begin{equation}\label{eq:2tau=tau 3}
	\theta_3(z|2\tau)^2\, \big(\partial_{2 \pi \ri z}^2 \log \theta_3(z|2\tau)\big) +\theta_2(z|2\tau)^2\, \big(\partial_{2 \pi \ri z}^2 \log \theta_2(z|2\tau)\big)=\theta_3(z|\tau)\,\big(\partial_{\log \mathfrak{q}} \theta_3(0|\tau)\big),
\end{equation}
\begin{multline}\label{eq:2tau=tau 4}
	\theta_3(z|2\tau)^2\, \big(\partial_{\log \mathfrak{q}}\partial_{2 \pi \ri z} \log \theta_3(z|2\tau)\big) +\theta_2(z|2\tau)^2\, \big(\partial_{\log \mathfrak{q}}\partial_{2 \pi \ri z} \log \theta_2(z|2\tau)\big)
	\\
		=\big(\partial_{2 \pi \i z}\theta_3(z|\tau)\big) \,\big(\partial_{\log \mathfrak{q}} \theta_3(0|\tau)\big),
\end{multline}
\begin{equation}\label{eq:2tau=tau 5}
	\theta_3(z|2\tau)^2\, \big(\partial_{2 \pi \ri z} \log \theta_3(z|2\tau)\big) +\theta_2(z|2\tau)^2\, \big(\partial_{2 \pi \ri z} \log \theta_2(z|2\tau)\big)=\big(\partial_{2 \pi \ri z} \theta_3(z|\tau)\big)\,\theta_3(0|\tau)\, .
\end{equation}
These relations can be easily proven by using the definition of theta function as power series in $\mathfrak{q}$. We present the details of the calculation for \eqref{eq:2tau=tau 2} and \eqref{eq:2tau=tau 4} in App. \ref{app:2tau=tau}. The other relations are similar.
By using \eqref{eq:2tau=tau 1}, we can write the relation \eqref{eq: bilin dual Nekrasov tilde} at $m=0$ as 
\begin{multline}
	\mathfrak{q}^{2\sigma^2}\Big(\theta_3(\frac{\eta}{2\pi}{+}2  \sigma \tau |2\tau)^2\, \partial_{\log \mathfrak{q}}^2\theta_3(\frac{\eta}{2\pi}{+}2  \sigma \tau |2\tau)+ \theta_2(\frac{\eta}{2\pi}{+}2  \sigma \tau |2\tau)^2\, \partial_{\log \mathfrak{q}}^2\theta_2(\frac{\eta}{2\pi}{+}2  \sigma \tau |2\tau)\Big)
	\\=2 \frac{\partial_{\log \mathfrak{q}}\theta_3(0|\tau)}{\theta_3(0|\tau)} \left(\partial_{\log \mathfrak{q}}-\frac{\partial_{\log \mathfrak{q}}\theta_3(0|\tau)}{\theta_3(0|\tau)}\right)\left(\mathfrak{q}^{2\sigma^2}\theta_3(\frac{\eta}{2\pi}{+}2  \sigma \tau|\tau)\theta_3(0|\tau)\right).
\end{multline}
This equality can be easily proven from \eqref{eq:2tau=tau 2}, \eqref{eq:2tau=tau 3}, \eqref{eq:2tau=tau 4}.
\newline 

\noindent{\bf Example 2.}
Consider the limit $m \rightarrow \infty$, $\mathfrak{q} m^4 \rightarrow t$. In this limit ${Z}_{\epsilon}^D \rightarrow  \mathcal{T}_{2\epsilon}$, where $\mathcal{T}_{2\epsilon}$ is the Painlev\'e $\rm III_3$ tau function given in \eqref{taup3}.
The equation \eqref{eq: bilin dual Nekrasov} becomes  the Toda equation in the form \cite[Prop 2.3]{Bershtein:2016uov}
\begin{equation}
	\mathrm{D}^2_{-1,1}(\mathcal{T}_{0},\mathcal{T}_{0})+\mathrm{D}^2_{-1,1}(\mathcal{T}_{1},\mathcal{T}_{1})=2t^{1/2}(\mathcal{T}_{0}^2+\mathcal{T}_{1}^2).	
\end{equation}


\subsection{From isomonodromic deformations}
In this part we deduce the bilinear equation \eqref{eq: bilin dual Nekrasov tilde} from the isomonodromic equations \eqref{pdef},\eqref{eq:Calogero} and the formula for the tau function \eqref{eq:tauFunction}.
The logic of the calculation is the same as in the example $m=0$ above but instead of the simple formula for $Q$ given in \eqref{eq:Q for m=0}, we have to use the formulas \eqref{pdef},\eqref{eq:Calogero}. It is convenient to rewrite them as
\begin{equation}
\partial_{\log \mathfrak{q}}(2\pi \ri Q)=\frac{p}{2 \pi \ri},\quad \partial_{\log \mathfrak{q}}\frac{p}{2 \pi \ri}=-m^2 \partial_1^3 \log \theta_1(2Q|\tau).
\end{equation}
In this section we use the following short notation for the derivatives of theta functions: \be \partial_1\theta_j(z,\tau)=\partial_{2 \pi \ri z}\theta_j(z,\tau),\quad\partial_2\theta_j(z,\tau)=\partial_{2 \pi \ri \tau}\theta_j(z,\tau)\,. \ee
After these preparations we compute the left and the right sides of  equation \eqref{eq: bilin dual Nekrasov tilde}. We have
\begin{equation}\label{left}
\begin{aligned}
	{\rm LHS}=&\tilde{{Z}}_0^D\tilde{{Z}}_0^D \partial_{\log \mathfrak{q}}^2\log \tilde{{Z}}_{0}^D+\tilde{{Z}}_{1/2}^D\tilde{{Z}}_{1/2}^D \partial_{\log \mathfrak{q}}\log \tilde{{Z}}_{1/2}^D
	\\ 
	=&\mathcal{T}^2  \big(\theta_3(2Q |2\tau)^2+ \theta_2(2Q|2\tau)^2\big) \partial_{\log \mathfrak{q}}^2\log \mathcal{T}
	\\
	&+\mathcal{T}^2\Big(\theta_3(2Q |2\tau)^2\,\partial_{\log \mathfrak{q}}^2\log (\theta_3(2Q |2\tau)+\theta_2(2Q |2\tau)^2\,\partial_{\log \mathfrak{q}}^2\log (\theta_2(2Q |2\tau)\Big)
	\\
	=&\mathcal{T}^2 \Big( m^2\theta_3(2Q |\tau) \,\theta_3(0|\tau)\,  \partial_2 \partial_1^2 \log \theta_1(2Q |\tau)-m^2\partial_1\theta_3(2Q|\tau)\,\theta_3(0|\tau)\, \partial_1^3 \log \theta_1(2Q|\tau)
	\\
	&+2 \partial_2\theta_3(2Q|\tau)\,\theta_3(0|\tau)-\frac{p^2}{\pi^2}\theta_3(2Q,\tau)\,\partial_2\theta_3(0|\tau)+\frac{2p}{\pi \ri} \partial_1\theta_3(2Q|\tau)\,\partial_2\theta_3(0|\tau)\Big),
\end{aligned}
\end{equation}
	where we used the relations \eqref{hdef},\eqref{eq:1} for the tau functions and the relations \eqref{eq:2tau=tau 2}-\eqref{eq:2tau=tau 5} for the theta functions. On the other side we have
\begin{equation}\label{right}
\begin{aligned}
	\rm{RHS}=&2\Big(\partial_2\log(\theta_3(0|\tau))\, \Big(\partial_{\log \mathfrak{q}}-\partial_2\log(\theta_3(0|\tau))\Big)-m^2 \partial_2^2\log(\theta_3(0|\tau))\Big)\big(\tilde{{Z}}_{0}^D \tilde{{Z}}_{0}^D+\tilde{{Z}}_{1/2}^D \tilde{{Z}}_{1/2}^D\big)
	\\
	=&2\Big(\partial_2\log(\theta_3(0|\tau))\, \Big(\partial_{\log \mathfrak{q}}-\partial_2\log(\theta_3(0|\tau))\Big)-m^2 \partial_2^2\log(\theta_3(0|\tau))\Big) \big(\mathcal{T}^2\theta_3(2Q |\tau)\theta_3(0|\tau)\big)
	\\
	=&2\theta_3(0|\tau) \Big(\partial_2\log(\theta_3(0|\tau)\,\partial_{\log \mathfrak{q}}-m^2 \partial_2^2\log(\theta_3(0|\tau)))\Big) \big(\mathcal{T}^2\theta_3(2Q |\tau)\big)
	\\
	=&\mathcal{T}^2\Big(m^2\theta_3(2Q |\tau)\Big( -2 \theta_3(0|\tau)\,\partial_2^2\log\theta_3(0|\tau)+4 \partial_1^2 \log \theta_1(2Q |\tau)\, \partial_2\theta_3(0|\tau)
	\Big) 
	\\
	&+2 \partial_2\theta_3(0|\tau)\,\partial_2\theta_3(2Q |\tau)- \frac{p^2}{\pi ^2}  \theta_3(2Q |\tau)\, \partial_2 \theta_3(0|\tau)+\frac{2p}{\pi \ri} \partial_2 \theta_3(0|\tau) \, \partial_1 \theta_3(2Q|\tau) \Big), 
\end{aligned}
\end{equation}
where we used \eqref{hdef} \eqref{eq:1} for the tau function and \eqref{eq:2tau=tau 1} for the theta functions.
	The last three terms in \eqref{left} agree with the last three terms in \eqref{right}. The  equality of the two other terms is equivalent to a theta function identity
	\begin{multline}\label{thetaid}
		\theta_3(2Q |\tau)\Big( -2 \theta_3(0|\tau)\,\partial_2^2\log\theta_3(0|\tau)+4 \partial_1^2 \log \theta_1(2Q |\tau) \, \partial_2\theta_3(0|\tau)	 \Big)  
		\\ = \theta_3(0|\tau)\Big( \theta_3(2Q |\tau)\,\partial_2 \partial_1^2 \log \theta_1(2Q |\tau)-\partial_1 \theta_3(2Q |\tau)\, \partial_1^3 \log\theta_1(2Q |\tau)
		\Big)	
	\end{multline}
	which we are going to prove now.
	 After some simple algebra \eqref{thetaid} reduces to a relation $F(z,\tau)=0$, where (here all derivatives are taken with respect to $2 \pi \ri z$)
	\begin{multline}
		F(z,\tau)=\theta_1(z |\tau)^3\theta_3(z |\tau)\Big(\theta_3^{(4)}(0 |\tau)\theta_3(0 |\tau)-\theta_3^{(2)}(0 |\tau)^2\Big)
		\\
		-4\theta_1(z |\tau)\Big(\theta_1^{(2)}(z |\tau)\theta_1(z |\tau)-\theta_1^{(1)}(z |\tau)^2\Big)\theta_3(z |\tau)\theta^{(2)}_3(0 |\tau)\theta_3(0 |\tau)
		\\
		-2\Big(\theta_1^{(3)}(z |\tau)\theta_1(z |\tau)^2-3\theta_1^{(2)}(z |\tau)\theta_1^{(1)}(z |\tau)\theta_1(z |\tau)+2\theta_1^{(1)}(z |\tau)^3\Big)\theta_3^{(1)}(z |\tau)\theta_3(0 |\tau)^2
		\\
		+\Big(\theta_1^{(4)}(z |\tau)\theta_1(z |\tau)^2 +\theta_1^{(2)}(z |\tau)^2\theta_1(z |\tau)-2\theta_1^{(3)}(z |\tau)\theta_1^{(1)}(z |\tau)\theta_1(z |\tau)+2\theta_1^{(2)}(z |\tau)\theta_1^{(1)}(z |\tau)^2\Big)\theta_3(z |\tau)\theta_3(0 |\tau)^2.
\end{multline}
Using the power series expression for $F(z,\tau)$ one can deduce the following modular properties
\begin{equation}
	F(z+\tau,\tau)=-\re^{-4\pi \ri \tau}\re^{-8\pi \ri z}F(z,\tau),\quad F(z+1,\tau)=-F(z,\tau).
\end{equation}
The function $F(z,\tau)$ does not have poles. Therefore it should have exactly 4 zeroes in the fundamental domain otherwise $F =0$ (as we will see). Clearly $F$ is an odd function $F(z,\tau)=-F(-z,\tau)$, and it is easy to see that $F^{(1)}(0,\tau)=0$. Hence $F$ has a zero of order at least 3 at $z=0$. Moreover one can check that $F(\frac{\tau+1}2,\tau)=0$, $F(\frac{1}2,\tau)=0$. Hence we must have $F=0$.
Actually the only nontrivial check is $F(\frac{1}2,\tau)=0$. It reduces to the identity 
\begin{multline}
	\Big(\theta_2^{(4)}(0 |\tau)\theta_2(0 |\tau)-\theta_2^{(2)}(0 |\tau)^2\Big)\theta_3(0 |\tau)^2+	\theta_2(0 |\tau)^2\Big(\theta_3^{(4)}(0 |\tau)\theta_3(0 |\tau)-\theta_3^{(2)}(0 |\tau)^2\Big)\\ -4\theta_2(0 |\tau)\theta^{(2)}_2(0 |\tau)\theta_3(0 |\tau)\theta^{(2)}_3(0 |\tau)=0.
\end{multline}
This identity can be proven directly using the $\mathfrak{q}$ series expansion for the $\theta$ functions.
\newline

\noindent \textbf{Remark.} Equation \eqref{eq: bilin dual Nekrasov tilde} can be viewed as a system of two second order differential equations. Hence its general solution depends on four constants of integration. But there is a simple two-parametric set of transformations of the form 
\begin{equation}
	\tilde{Z}_{\epsilon}\mapsto C_1 \exp^{C_2\int \theta_3(0|\tau)^4 d\tau} 	\tilde{Z}_{\epsilon}\,, \quad \epsilon=0,1/2,
\end{equation}
which preserves equation \eqref{eq: bilin dual Nekrasov tilde}. These transformations preserve the ratio $Z_0/Z_{1/2}$, hence the equation for $Q$ (which follows \eqref{eq:tauFunction})
\begin{equation}\label{qdefi}
	\frac{\theta_3(2Q|2\tau)}{\theta_2(2Q|2\tau)}=\frac{\tilde{Z}^D_{0}(\sigma,m,\eta,\tau)}{\tilde{Z}^D_{1/2}(\sigma,m,\eta,\tau)}
\end{equation}
will depend only on two constants of integration. In terms of the Ansatz \eqref{eq:dualNekrasov} these constants are $\sigma,\eta$. As we explained at the beginning of Sec.~\ref{sec:toro}, the equation \eqref{qdefi} determines $Q$ essentially uniquely. In addition we proved that for any solution of  the isomonodromic deformation equation \eqref{eq:Calogero}, the corresponding functions $\tilde{Z}_{0}, \tilde{Z}_{1/2}$ satisfy \eqref{eq: bilin dual Nekrasov tilde}. Since a generic solution of \eqref{eq:Calogero} depends on two parameters, we get a correspondence.  
Therefore any generic solution $\tilde{Z}_0, \tilde{Z}_{1/2}$ of the equation \eqref{eq: bilin dual Nekrasov tilde} determines $Q$ which solves the isomonodromic deformation equation \eqref{eq:Calogero}.

\section{Blowup equations from regularized action functional}\label{functional}

In this section we derive the $\epsilon_2\to 0$ limit of blowup equations from the regularized action functional.  This was first understood in the example of the pure $\cN=2, SU(2)$ SW theory in \cite{Gavrylenko:2020gjb}, which was also inspired by the works of \cite{Litvinov:2013sxa,Lukyanov:2011wd}.

\subsection{Definition and derivatives of the action functional}

The Lagrangian of the non-autonomous classical Calogero-Moser system is given by
\begin{equation}
\label{eq:11}
\mathcal{L}= (2\pi \ri\partial_{\tau} Q)^2+m^2\wp(2Q|\tau)+2m^2\eta_1(\tau).
\end{equation}
Now we study its asymptotics in two limits: \(\tau\to \ri\infty\) and \(\tau\to\tau_{\star}\).
We know that in the second limit the Hamiltonian is finite, so the leading singularity is given by \(2m^2\wp(2Q|\tau)\approx \frac{m^2}{2Q^2}\).
By using equation \eqref{eq:tauExpansion} we get
\begin{equation}
\label{eq:21}
\mathcal{L}= \pm \frac{\pi \ri m}{\tau-\tau_{\star}}+\mathcal{O}(1).
\end{equation}
In the other limit \(\tau\to \ri\infty\) the leading asymptotics is given by the derivative term, the constant term of \(\wp(2Q|\tau)\) from \eqref{eq:wpSin} cancels with the asymptotics of \(\eta_1(\tau)\) \eqref{eq:etaInf}:
\begin{equation}
\label{eq:22}
\mathcal{L}= (2\pi \ri \sigma)^2+\mathcal{O}(e^{2\pi\ri\tau}).
\end{equation}
We need to subtract both these  asymptotics and define the regularized Lagrangian\footnote{We also subtract a finite part. This is needed to get the \eqref{eq:StauDer} and \eqref{eq:72} in the current form.}:
\begin{equation}
\label{eq:Ltilde}
\tilde{\mathcal{L}}^{\mp}=\mathcal{L}\mp \frac{\pi \ri m}{\tau-\tau_{\star}}\pm \frac{\pi \ri m}{\tau}-(2\pi \ri\sigma)^2.
\end{equation}
Now we define the regularized action functional:
\begin{equation}\begin{gathered}
\label{eq:Stilde}
\tilde{S}^{\mp}(\sigma,m,\tau_{\star})=
\int\limits^{\tau_{\star}}_{\ri\infty}\tilde{\mathcal{L}}^{\mp}\frac{\mathrm{d}\tau}{2\pi \ri} =
\\=
\frac1{2\pi \ri}\int\limits^{\tau_{\star}}_{\ri\infty} \left(\left( (2\pi \ri\partial_{\tau} Q)^2+m^2\left(\wp(2Q|\tau)+2\eta_1(\tau)\right) \right)\mathrm{d}\tau \mp \pi \ri m\, \mathrm{d}\log \frac{\tau-\tau_{\star}}{\tau} - (2\pi \ri \sigma)^2\mathrm{d}\tau\right).
\end{gathered}\end{equation}

The \(\tau_{\star}\) derivative is:
\begin{equation}
\label{eq:28}
\partial_{\tau_{\star}}\tilde{S}^{\mp}(\sigma,m,\tau_{\star})=\frac{\tilde{\mathcal{L}}^{\mp}(\tau_{\star})}{2\pi \ri}+\int\limits^{\tau_{\star}}_{\ri\infty}\mathrm{d} \left( 4\pi \ri\partial_{\tau}Q\partial_{\tau_{\star}}Q\pm \frac{m/2}{\tau-\tau_{\star}} \right),
\end{equation}
where we used the equation of motion \eqref{eq:Calogero}.
For this computation we need the expansions of \(Q\) in both limits.
The expansion around \(\tau_{\star}\) is given by \eqref{eq:tauExpansion}:
\begin{equation}
\label{eq:82}
Q\approx \sqrt{\pm \frac{m}{2\pi \ri}(\tau-\tau_{\star})} \left( 1\pm \frac{H^{\mp}_{\star}+2m^2\eta_1(\tau_{\star})}{4\pi \ri m}(\tau-\tau_{\star}) \right).
\end{equation}
The expansion around \(\ri\infty\) is given by \eqref{eq:etaBeta}: \(Q\approx \sigma\tau+\beta\), where
\begin{equation}
\label{eq:88}
\beta=\frac{\eta}{4\pi}+\frac1{2\pi \ri}\log \frac{\Gamma(-m+2\sigma)\Gamma(1-2\sigma)}{\Gamma(1-m-2\sigma)\Gamma(2\sigma)}=:\frac{\eta}{4\pi}+\frac{\phi(\sigma,m)}{2\pi \ri}.
\end{equation}
We now compute  the value of the regularized Lagrangian at the upper limit:
\begin{equation}
\label{eq:81}
\tilde{\mathcal{L}}^{\mp}(\tau_{\star})\approx \frac12 \left( H^{\mp}_{\star}+2m^2\eta_1(\tau) \right)+2m^2\eta_1(\tau)\pm\frac{\pi \ri m}{\tau_{\star}}-(2\pi \ri \sigma)^2.
\end{equation}
Other useful expressions are the expansions around \(\tau_{\star}\)
\begin{equation}
\label{eq:83}
4\pi \ri\partial_{\tau}Q\partial_{\alpha}Q\approx
\mp \frac{\frac{m}{2} \partial_{\alpha}\tau_{\star}}{\tau-\tau_{\star}}-3 \frac{\partial_{\alpha}\tau_{\star}(H_{\star}^{\mp}+2m^2\eta_1(\tau))}{4\pi \ri},
\end{equation}
and around \(\ri\infty\):
\begin{equation}
\label{eq:84}
4\pi \ri\partial_{\tau}Q\partial_{\alpha}Q\approx 4\pi \ri \sigma (\tau\partial_{\alpha}\sigma + \partial_{\alpha}\beta),
\end{equation}
where \(\alpha\) is either \(\sigma\) or \(\tau_{\star}\).
Using these relations we finally find
\begin{equation}
\label{eq:StauDer}
2\pi \ri\partial_{\tau_{\star}}\tilde{S}^{\mp}(\sigma,m,\tau_{\star})=-H^{\mp}_{\star}\pm\frac{\pi \ri m}{\tau_{\star}}-(2\pi \ri \sigma)^2-2\pi \ri \partial_{\tau_{\star}}\left(4\pi \ri\sigma \beta\right).
\end{equation}
In the same way we also compute the \(\sigma\)-derivative:
\begin{equation}\begin{gathered}
\label{eq:SsigmaDer}
\partial_{\sigma}\tilde{S}^{\mp}(\sigma,m,\tau_{\star})=\int\limits^{\tau_{\star}}_{\ri\infty}\mathrm{d} \left( 4\pi \ri\partial_{\tau}Q\partial_{\sigma}Q - 4\pi \ri\tau \sigma \right)=
-4\pi \ri\sigma\partial_{\sigma}\beta-4\pi \ri\sigma\tau_{\star}=
\\=
\partial_{\sigma} \left( -4\pi \ri\sigma\beta -2\pi \ri\sigma^2\tau_{\star}\right)+4\pi \ri\beta.
\end{gathered}\end{equation}
Now we consider the following equality:
\begin{equation}\begin{gathered}
\label{eq:72}
\tilde{S}^{\mp}-m\partial_m \tilde{S}^{\mp}=
\\=
\frac1{2\pi i}\int\limits^{\tau_{\star}}_{\ri\infty}
\left( (2\pi \ri\partial_{\tau} Q)^2-m^2\left(\wp(2Q|\tau)+2\eta_1(\tau)\right) -(2\pi \ri\sigma)^2\right)\mathrm{d}\tau-
\mathrm{d} \left( 8(\pi \ri)^2\partial_{\tau}Q\partial_mQ \right)=
\\=
\int\limits^{\tau_{\star}}_{\ri\infty}\mathrm{d} \left(\log \mathcal{T}-2\pi \ri\sigma^2\tau - 4\pi \ri\partial_{\tau}Q\, m\partial_mQ \right)
\end{gathered}\end{equation}
To complete this computation we need to know the asymptotics of \(\mathcal{T}\) at \(+\ri\infty\).
This can be found from \eqref{eq:tauFunction} and \eqref{eq:dualNekrasov}:
\begin{equation}
\label{eq:86}
\mathcal{T}\approx \re^{2\pi \ri\tau\sigma^2} \prod_{\epsilon=\pm1}\frac{G(1-m+2\epsilon\sigma)}{G(1+2\epsilon\sigma)}.
\end{equation}
Therefore
\begin{equation}
\label{eq:Sblowup}
\exp\left(\tilde{S}^{\mp}-m\partial_m\tilde{S}^{\mp}-m\partial_m\left(4\pi \ri\sigma \beta\right)\right)=
\prod_{\epsilon=\pm1}\frac{G(1+2\epsilon\sigma)}{G(1-m+2\epsilon\sigma)}\re^{-2\pi \ri\tau_{\star}\sigma^2}
\mathcal{T}(\sigma,m,\eta^{\mp},\tau_{\star}).
\end{equation}

We introduce the new function
\begin{equation}
\label{eq:Sshift}
{\mathcal{S}}^{\mp}=\tilde{S}^{\mp}\mp m/2\log \tau_{\star}+2\pi \ri\sigma^2\tau_{\star}+4\pi \ri\sigma\beta+\varphi(\sigma,m)
\end{equation}
with some function \(\varphi(\sigma,m)\), which will be chosen later.
By using the \({\mathcal{S}}\) identities \eqref{eq:StauDer}, \eqref{eq:SsigmaDer}, and \eqref{eq:Sblowup} we get
\begin{equation}\begin{gathered}
\label{eq:85}
2\pi \ri\partial_{\tau_{\star}}{\mathcal{S}}^{\mp}=-H_{\star}^{\mp},\\
\partial_{\sigma}{\mathcal{S}}^{\mp}=\ri \eta +2\phi(\sigma,m) + \partial_{\sigma}\varphi(\sigma,m),\\
\exp({\mathcal{S}}^{\mp}-m\partial_m {\mathcal{S}}^{\mp} -\ri\sigma\eta-2\sigma\phi(\sigma,m) -\varphi(\sigma,m)+m\partial_m\varphi(\sigma,m) )=\\
=\prod_{\epsilon=\pm1}\frac{G(1+2\epsilon\sigma)}{G(1-m+2\epsilon\sigma)}\mathcal{T}(\sigma,m,\eta^{\mp},\tau_{\star}).
\end{gathered}\end{equation}
We would like to cancel some unwanted terms, namely, to find  \(\varphi(\sigma,m)\) such that
\begin{equation}\begin{gathered}
\label{eq:90}
\partial_{\sigma}\varphi(\sigma,m)=-2\phi(\sigma,m),\\
\varphi(\sigma,m)-m\partial_m \varphi(\sigma,m)=-2\sigma\phi(\sigma,m)+\ldots,
\end{gathered}\end{equation}
where ``\(\ldots\)'' stands for the logarithm of the Barnes functions and is almost completely defined by the first equation (\(\sigma\)-derivative).
We can solve the first equation by integration:
\begin{equation}\begin{gathered}
\label{eq:91}
\varphi(\sigma,m)=
\varphi_0(m)-\int^{\sigma}\mathrm{d}(2\sigma)\log \frac{\Gamma(-m+2\sigma)\Gamma(1-2\sigma)}{\Gamma(1-m-2\sigma)\Gamma(2\sigma)}=
\\=
\log\frac{G(-m+2\sigma)G(1-m-2\sigma)}{G(1-2\sigma)G(2\sigma)}
-2\sigma\log\Gamma(1-2\sigma)+(2\sigma-1)\log\Gamma(2\sigma)-
\\-
(2\sigma-m-1)\log\Gamma(-m+2\sigma)+(2\sigma+m)\log\Gamma(1-m-2\sigma)+m^2+m\log 2\pi+\varphi_0(m).
\end{gathered}\end{equation}
Here we used the following identity:
\begin{equation}
\label{eq:92}
\int^x \mathrm{d}x \log \Gamma(x)=\frac{x(1-x)}{2}+\frac{x}{2}\log 2\pi+(x-1)\log\Gamma(x)-\log G(x).
\end{equation}
We now take \(\varphi_0(m)=-m^2+am\), where \(a\) is an arbitrary constant,
which can be fixed after identification of \(\mathcal{S}\) with the properly normalized conformal block.
In this way we get
\begin{equation}
\label{eq:94}
\varphi(\sigma,m)-m\partial_m\varphi(\sigma,m)+2\sigma\phi(\sigma,m)=
\log \frac{G(1-m+2\sigma)G(1-m-2\sigma)}{G(1-2\sigma)G(1+2\sigma)}.
\end{equation}
After choosing \(\varphi(\sigma,m)\) in such a way we have
\begin{equation}\begin{split}
\label{eq:actionBlowup}
2\pi \ri\partial_{\tau_{\star}}{\mathcal{S}}^{\mp}(\sigma,m,\tau_{\star})&=-H^{\mp}_{\star}(\sigma,m,\tau_{\star}),\\
\partial_{\sigma}{\mathcal{S}}^{\mp}(\sigma,m,\tau_{\star})&=\ri\eta^{\mp}(\sigma,m,\tau_{\star}),\\
\exp({\mathcal{S}}^{\mp}-m\partial_m {\mathcal{S}}^{\mp} -\sigma\partial_{\sigma}\mathcal{S}^{\mp} )&=\mathcal{T}(\sigma,m,\eta^{\mp},\tau_{\star}).
\end{split}\end{equation}

\subsection{Relation to classical conformal blocks}

\paragraph{Classical conformal blocks and BPZ equations.}
Following \cite{Litvinov:2013sxa, Gavrylenko:2020gjb} $\mathcal{S}$ should be identifies with the $c=\infty$ conformal blocks (or classical conformal blocks, or NS free energy), and therefore \eqref{eq:85} reproduces the $\epsilon_2\to 0$ limit of blowup equations used in Sec.~\ref{butorus}.
Following \cite{Maruyoshi:2010iu,Fateev:2009aw,Alday:2009fs,Drukker:2009id,Fateev:2009me}, we start from consideration of the correlators with heavy degenerate field \(\phi_{(1,2)}(w)\), light degenerate field \(\phi_{(2,1)}(w)\), and energy-momentum tensor \(T(z)\).
We define
\begin{equation}\begin{split}
\label{eq:32}
G(\sigma,m,\tau)=&\tr_{\Delta(\sigma)} \mathcal{R} \left(\mathfrak{q}^{L_0-\frac{c}{24}} V_{\Delta(m)}(0)\right),\\
G_T(\sigma,m,z,\tau)=&\tr_{\Delta(\sigma)} \mathcal{R} \left(\mathfrak{q}^{L_0-\frac{c}{24}} V_{\Delta(m)}(0) T(z)\right),\\
G_{h}(\sigma,m,w,\tau)=&\tr_{\Delta(\sigma)} \mathcal{R} \left(\mathfrak{q}^{L_0-\frac{c}{24}} V_{\Delta(m)}(0) \phi_{(1,2)}(w)\right),\\
G_{l}(\sigma,m,w,\tau)=&\tr_{\Delta(\sigma)} \mathcal{R} \left(\mathfrak{q}^{L_0-\frac{c}{24}} V_{\Delta(m)}(0) \phi_{(2,1)}(w)\right),\\
G_{T,h}(\sigma,m,z,w,\tau)=&\tr_{\Delta(\sigma)} \mathcal{R} \left(\mathfrak{q}^{L_0-\frac{c}{24}} V_{\Delta(m)}(0) T(z)\phi_{(1,2)}(w)\right),
\end{split}\end{equation}
where \(\mathcal{R}\) denotes the cyclic ordering on a cylinder (analog of the radial ordering), and we used the following parameterization of \(\Delta\):
\begin{equation}
\label{eq:deltaM}
\Delta(m)=\frac1{4}(b+b^{-1})^2-m^2/b^2 = b^{-2}(1/4-m^2) + \mathcal{O}(1).
\end{equation}
Now we use the OPE with degenerate field:
\begin{equation}
\label{eq:10}
T(z)\phi_{(1,2)}(w)=\frac{\Delta_{(1,2)}\phi_{(1,2)}(w)}{(z-w)^2}+\frac{\partial_w\phi_{(1,2)}(w)}{z-w}+(\mathcal{L}_{-2}\phi_{(1,2)})(w)+\ldots 
\end{equation}
Using the explicit form of the null-vector and the formula for \(\Delta_{(1,2)}\) we rewrite it as
\begin{equation}
\label{eq:OPEdeg}
T(z)\phi_{(1,2)}(w)=\frac{-\frac{2b^2+3}{4b^2}\phi_{(1,2)}(w)}{(z-w)^2}+\frac{\partial_w\phi_{(1,2)}(w)}{z-w}-b^2\partial^2_w\phi_{(1,2)}(w)+\ldots 
\end{equation}
Another OPE is
\begin{equation}
\label{eq:79}
T(z)V_{\Delta(m)}(0)=\frac{\Delta(m) V_{\Delta(m)}(0)}{z^2}+\frac{\partial V_{\Delta(m)}(0)}{z}+\ldots.
\end{equation}
Combining these OPE's together, and using the fact that correlator depends only on difference of coordinates, we finally write
\begin{equation}\begin{gathered}
\label{eq:GTh1}
G_{T,h}(\sigma,m,z,w,\tau)= I(\sigma,m,w,\tau)+\frac{2b^2+3}{4b^2} \left(\log \theta_1(z-w)\right)'' G_h(\sigma,m,w,\tau)+\\
+\left( \log \frac{\theta_1(z-w)}{\theta_1(z)} \right)'\partial_w G_h(\sigma,m,w,\tau)
-\Delta(m) \left(\log \theta_1(z)\right)''G_h(\sigma,m,w,\tau).
\end{gathered}\end{equation}
To derive this formula we first wrote explicitly the globally defined functions \(\left(\log \theta_1(z-w)\right)''\), \(\left(\log \theta_1(z)\right)''\), \(\left( \log \frac{\theta_1(z-w)}{\theta_1(z)} \right)'\), which are fixed up to constants by their singular behavior.
Coefficients in front of these functions are dictated by the OPE's with \(T(z)\) (conformal Ward identities).
These terms have vanishing \(A\)-cycle integral in the variable \(z\).
The constant term \(I(\sigma,m,w,\tau)\) is not fixed by the singular parts of the OPE's.
We can find it in two different ways, which will give a non-trivial equation \eqref{eq:BPZGh} on \(G_h(\sigma,m,w,\tau)\). This is the analog of the BPZ equation \cite{BELAVIN1984333} on the torus, see  \cite{EGUCHI1987308}.
   
On one side,
\begin{equation}
\label{eq:96}
\oint_A T(z)\mathrm{d}z = (2\pi \ri)^2\left( L_0-\frac{c}{24}\right),
\end{equation}
therefore 
\begin{equation}
\label{eq:97}
I(\sigma,m,w,\tau)=2\pi \ri \partial_{\tau}G_h(\sigma,m,w,\tau).
\end{equation}
On the other side, we have not used the regular part of \eqref{eq:OPEdeg} yet.
To do this first rewrite \eqref{eq:GTh1} in a more suitable form using \eqref{wpdef} and \eqref{zdef}:
\begin{multline}
\label{eq:87}
G_{T,h}(\sigma,m,z,w,\tau)=-\frac{2b^2+3}{4b^2} \left( \wp(z-w|\tau)+2\eta_1(\tau) \right) G_h(\sigma,m,w,\tau)+\\
+\left(\zeta(z-w|\tau)-\zeta(z|\tau)+2\eta_1(\tau)w\right)\partial_w G_h(\sigma,m,w,\tau) +\\
+\Delta(m) \left(\wp(z|\tau)+2\eta_1(\tau)\right)G_h(\sigma,m,w,\tau)
+ 2\pi \ri \partial_{\tau}G_h(\sigma,m,w,\tau).
\end{multline}
The regular part at \(z=w\) is:
\begin{multline}
\label{eq:BPZGh}
-b^2\partial_w^2G_h(\sigma,m,w,\tau)=-\frac{2b^2+3}{2b^2}\eta_1(\tau)G_h(\sigma,m,w,\tau)+\left( 2\eta_1(\tau)w-\zeta(w|\tau) \right)\partial_wG_h(\sigma,m,w,\tau)+\\
 +\Delta(m)\left(\wp(w|\tau)+2\eta_1(\tau)\right)G_h(\sigma,m,w,\tau)
+ 2\pi \ri \partial_{\tau}G_h(\sigma,m,w,\tau).
\end{multline}
To get rid of the first derivative we redefine 
\cite{Maruyoshi:2010iu,Fateev:2009aw}
\begin{equation}
\label{eq:99}
G_h(\sigma,m,w,\tau)=\theta_1(w|\tau)^{\frac1{2b^2}}\eta(\tau)^{-1-\frac{3}{2b^2}}\tilde{G}_h(\sigma,m,w,\tau)
\end{equation}
and then get
\begin{multline}
\label{eq:101}
2\pi \ri\partial_{\tau}\tilde{G}_h(\sigma,m,w,\tau)+b^2\partial_w^2\tilde{G}_h(\sigma,m,w,\tau)+\\
+
\left(\frac12 \frac{\partial_w^2\theta_1(w|\tau)}{\theta_1(w|\tau)}+\left(\frac1{4b^2}+ (m^2-1/4) b^{-2}+\mathcal{O}(1)\right)\partial_w^2\log \theta_1(w|\tau)\right)\tilde{G}_h(\sigma,m,w,\tau)=0,
\end{multline}
where we substituted the \(b\to 0\) expansion of \(\Delta(m)\) \eqref{eq:deltaM}.
It is convenient to make the following Ansatz
\begin{equation}\begin{gathered}
\label{eq:100}
G_h(\sigma,m,w,\tau)=\re^{b^{-2}f_h(\sigma,m,w,\tau)+\mathcal{O}(1)},\qquad
\tilde{G}_h(\sigma,m,w,\tau)=\re^{b^{-2}\tilde{f}_h(\sigma,m,w,\tau)+\mathcal{O}(1)},\\
G(\sigma,m,\tau)=\re^{b^{-2}f(\sigma,m,\tau)+\mathcal{O}(1)},
\end{gathered}\end{equation}
where \(f\)'s are called classical conformal blocks.
Using the leading order of \eqref{eq:100} one gets:
\begin{equation}
\label{eq:102}
2\pi \ri\partial_{\tau}\tilde{f}_h(\sigma,m,w,\tau)+(\partial_w \tilde{f}_h(\sigma,m,w,\tau))^2-m^2\left( \wp(w|\tau)+2\eta_1(\tau) \right)=0.
\end{equation}
\paragraph{Identification with the action functional.}
Equation \eqref{eq:102} is nothing but the Hamilton-Jacobi equation for a system with  Hamiltonian given by
\begin{equation}
\label{eq:105}
H(p,w,\tau)=p^2-m^2(\wp(w|\tau)+2\eta_1(\tau)).
\end{equation}
The equations of motion are \(2\pi \ri\partial_{\tau}w=2p\), \(2\pi \ri\partial_{\tau}p=m^2\wp'(w|\tau)\).
If we define
\begin{equation}
\label{eq:106}
w=2Q
\end{equation}
we recover \eqref{pdef} and \eqref{eq:Calogero}.
Moreover if we evaluate \(Q\) on the solution of the equations of motions, we have
\begin{equation}
\label{eq:CBregularization}
\tilde{f}_h\left(\sigma,m,2Q(\tau_2),\tau_2\right)-\tilde{f}_h\left(\sigma,m,2Q(\tau_1),\tau_1\right)
=\int_{\tau_1}^{\tau_2}\frac{\mathrm{d}\tau}{2\pi \ri}\left( (2\pi \ri\partial_{\tau}Q)^2+m^2(\wp(2Q|\tau)+2\eta_1(\tau)) \right).
\end{equation}
To get a non-trivial statement we consider the limit \(\tau_1\to \ri\infty\) and \(\tau_2\to \tau_{\star}\).
First we look at the \(\tau_1\to \ri\infty\) limit.
In this region \(Q\approx\sigma\tau_1+\beta\) \eqref{eq:etaBeta}.
Hence
\begin{equation}
\label{eq:108}
G_h(\sigma,m,2Q(\tau_1),\tau_1)\approx \re^{-\pi \ri\tau_1 \frac{c}{12}}\tr_{\Delta(\sigma)} \left(V_{\Delta(m)}(0) \re^{2\pi \ri(2\sigma\tau_1+2\beta)L_0} \phi_{(1,2)}(0) \re^{2\pi \ri(\tau_1-2\sigma\tau_1-2\beta)L_0}\right).
\end{equation}
Taking the logarithm we get in the leading order
\begin{multline}
\label{eq:109}
f_h(\sigma,m,2Q(\tau_1),\tau_1)\approx 2\pi \ri(2\sigma\tau_1+2\beta)\left(\frac1{4}-(\sigma-\frac1{2})^2\right)+\\
+2\pi \ri(\tau_1-2\sigma\tau_1-2\beta)(\frac1{4}-\sigma^2)-\frac{\pi \ri\tau_1}2
+ c(\sigma,m,\sigma-\frac12)+c(\sigma-\frac12,1,\sigma)=\\
=2\pi \ri \tau_1\sigma^2+4\pi \ri\sigma\beta-\pi \ri(\sigma\tau_1+\beta)+c(\sigma,m,\sigma-\frac12)+c(\sigma-\frac12,1,\sigma),
\end{multline}
where \(c\)'s are NS limits of the 3-point functions: \(\log C(\alpha_1,\alpha_2,\alpha_3)=b^{-2}c(\alpha_1,\alpha_2,\alpha_3)+\mathcal{O}(1)\).

Now we switch to \(\tilde{f}_h(\sigma,m,2Q(\tau_1),\tau_1)=f_h(\sigma,m,2Q(\tau_1),\tau_1)-\frac12\log\theta_1(2Q(\tau_1)|\tau_1)+\frac{3}{2}\log\eta(\tau_1)\). By using  the asymptotic \be \theta_1(2Q(\tau_1)|\tau_1)\approx \ri\re^{\pi \ri\tau_1/4}\re^{-2\pi \ri(\sigma\tau_1+\beta)} \ee we get:
\begin{equation}
\label{eq:tau1asymptotics}
\tilde{f}_h(\sigma,m,2Q(\tau_1),\tau_1)\approx -\pi \ri/4+2\pi \ri\tau_1\sigma^2+4\pi \sigma\beta + c(\sigma,m,\sigma-\frac12)+c(\sigma-\frac12,1,\sigma).
\end{equation}

Let us now focus on the \(\tau_2\to\tau_{\star}\) limit.
Here we use the OPE
\begin{multline}
\label{eq:111}
V_{\Delta(m)}(0)\phi_{(1,2)}(2Q)=C(m+\frac12,1,m)(-2Q)^{\Delta(m+\frac12)-\Delta(1)-\Delta(m)} V_{\Delta(m+\frac12)}(0)+\\
+C(m-\frac12,1,m)(-2Q)^{\Delta(m-\frac12)-\Delta(1)-\Delta(m)} V_{\Delta(m-\frac12)}(0) + \ldots 
\end{multline}
 In the \(b\to 0\) limit we can neglect the sub-leading term. However for different signs of \(m\), different terms will dominate. We will again describe these two possibilities by using \(\pm\) sign:
\begin{equation}
\label{eq:112}
V_{\Delta(m)}(0)\phi_{(1,2)}(2Q)=C(m\mp\frac12,1,m)(-2Q)^{b^{-2}(\pm m+\frac12)} V_{\Delta(m\mp\frac12)}(0)+ \ldots 
\end{equation}
Using this OPE we write the asymptotics of \(\tilde{f}_h\):
\begin{multline}
\label{eq:tau2asymptotics}
\tilde{f}_h(\sigma,m,2Q(\tau_2),\tau_2)\approx
\left(\frac12\pm m\right)\ri\pi\pm m \log 2Q - \frac12 \log \theta_1'(0|\tau_{\star})+\frac{3}{2}\log\eta(\tau_{\star})+
\\
+c(m\mp \frac12,1,m)+f(\sigma,m-\frac12,\tau_{\star})=\\
=
\left(\frac12\pm m\right)\ri\pi\pm m \log \left(\sqrt{\frac{2m}{\pi}}\re^{\mp \frac{\ri\pi}{4}}\sqrt{\tau_2-\tau_{\star}}\right) - \frac12 \log \theta_1'(0|\tau_{\star})+\frac{3}{2}\log\eta(\tau_{\star})+
\\
+c(m\mp \frac12,1,m)+f(\sigma,m-\frac12,\tau_{\star})=\\
=\left(\frac12\pm m\right)\ri\pi \pm \frac{m}{2}\log \frac{2m}{\pi} - \frac{\ri\pi m}{4}\pm \frac{m}{2}\log(\tau_2-\tau_{\star})
+c(m\mp \frac12,1,m)+f(\sigma,m\mp \frac12,\tau_{\star}),
\end{multline}
where we used
\be G(\sigma,m,\tau)=\re^{b^{-2}f(\sigma,m,\tau)+\mathcal{O}(1)}.\ee
Combining together \eqref{eq:CBregularization}, \eqref{eq:tau1asymptotics}, \eqref{eq:tau2asymptotics} we get the following equality:
\begin{multline}
\label{eq:104}
f(\sigma,m\mp \frac12,\tau_{\star})=\lim_{\substack{\tau_1\to \ri\infty\\\tau_2\to \tau_{\star}}} \int_{\tau_1}^{\tau_2}\frac{\mathrm{d}\tau}{2\pi \ri}\left( (2\pi \ri\partial_{\tau}Q)^2+m^2(\wp(2Q|\tau)+2\eta_1(\tau)) \right)+\\
+
2\pi \ri\tau_1\sigma^2+4\pi \sigma\beta + c(\sigma,m,\sigma-\frac12)+c(\sigma-\frac12,1,\sigma)-\pi \ri/4-\\
-\left(\frac12\pm m\right)\ri\pi \mp \frac{m}{2}\log \frac{2m}{\pi} + \frac{\ri\pi m}{4}\mp \frac{m}{2}\log(\tau_2-\tau_{\star})-c(m\mp \frac12,1,m).
\end{multline}
Now we would like to compare \eqref{eq:104} with  the regularized and redefined action \(\mathcal{S}\). 
To do that we  combine  \eqref{eq:Stilde} and \eqref{eq:Sshift} and rewrite \(\mathcal{S}\) as: \begin{multline}
\label{eq:23}
\mathcal{S}^{\mp}(\sigma,m,\tau_{\star})=\lim_{\substack{\tau_1\to \ri\infty\\\tau_2\to \tau_{\star}}} \int_{\tau_1}^{\tau_2}\frac{\mathrm{d}\tau}{2\pi \ri}\left( (2\pi \ri\partial_{\tau}Q)^2+m^2(\wp(2Q|\tau)+2\eta_1(\tau)) \right)+\\
+2\pi \ri \sigma^2(\tau_1-\tau_\star)\mp \frac{m}{2}\log(\tau_2-\tau_{\star})\pm \frac{m}{2}\log \tau_{\star}+\\
+
2\pi \ri\sigma^2\tau_{\star}+4\pi \ri\sigma\beta\mp\frac{m}{2}\log \tau_{\star}+\varphi(\sigma,m)=\\
=
\lim_{\substack{\tau_1\to \ri\infty\\\tau_2\to \tau_{\star}}} \int_{\tau_1}^{\tau_2}\frac{\mathrm{d}\tau}{2\pi \ri}\left( (2\pi \ri\partial_{\tau}Q)^2+m^2(\wp(2Q|\tau)+2\eta_1(\tau)) \right)+\\
+2\pi \ri \sigma^2\tau_1+4\pi \ri\sigma\beta\mp \frac{m}{2}\log(\tau_2-\tau_{\star})+\varphi(\sigma,m).
\end{multline}
By comparing the above expressions we get  the following identification:
\begin{multline}
\label{eq:27}
f(\sigma,m\mp \frac12,\tau_{\star})=\mathcal{S}^{\mp}(\sigma,m,\tau_{\star})+
 c(\sigma,m,\sigma-\frac12)+c(\sigma-\frac12,1,\sigma)-c(m\mp \frac12,1,m)-\\
-\varphi(\sigma,m)
+ \frac{\ri\pi (m-1)}{4}-\left(\frac12\pm m\right)\ri\pi \mp \frac{m}{2}\log \frac{2m}{\pi} .
\end{multline}
This proves that the regularized action is equals to the classical conformal block up to some possible \(\tau\)-independent constant.
Together with \eqref{eq:actionBlowup} this gives an additional  proof of the classical/\(c=1\) blowup relations.

\paragraph{Energy from classical conformal blocks.}
Though we already know that the energy for the spectral problem can be described by \(-H_{\star}^{\mp}=2\pi \ri\partial_\tau f(\sigma,m\mp \frac12,\tau)\), we can see how this fact follows directly from CFT.
To do this we consider the equivalent of equation \eqref{eq:BPZGh} for the light degenerate field. More precisely we have \(\phi_{(2,1)}(w)\):
\begin{multline}
\label{eq:BPZGh2}
-b^{-2}\partial_w^2G_l(\sigma,m,w,\tau)=-\frac{2+3b^2}{2}\eta_1(\tau)G_l(\sigma,m,w,\tau)+\left( 2\eta_1(\tau)w-\zeta(w|\tau) \right)\partial_wG_l(\sigma,m,w,\tau)+\\
+\Delta(m) \left(\wp(w|\tau)+2\eta_1(\tau)\right)G_l(\sigma,m,w,\tau)
+ 2\pi \ri \partial_{\tau}G_l(\sigma,m,w,\tau).
\end{multline}
As before we use the Ansatz 
\begin{equation}
\label{eq:95}
G_l(\sigma,m,w,z)=\psi(\sigma,m,w,\tau)\re^{b^{-2} f(\sigma,m,\tau)+\mathcal{O}(1)},
\end{equation}
where \(\mathcal{O}(1)\) is a function of \(\sigma,m,\tau\) only. In  the \(b\to 0\) limit we get \begin{equation}
\label{eq:98}
\left(\partial^2_w-(m^2-1/4) \left( \wp(w|\tau)+2\eta_1(\tau) \right)+2\pi \ri\partial_{\tau}f(\sigma,m,\tau)\right)\psi(\sigma,m,w,\tau)=0.
\end{equation}
By shifting \(m\) we end up with
\begin{multline}
\label{eq:89}
\left(-\partial^2_w+m(m\mp 1)\wp(w|\tau)\right)\psi(\sigma,m\mp \frac12,w,\tau)=\\
=\left(2\pi \ri\partial_{\tau}f(\sigma,m\mp \frac12,\tau)-2m(m\mp 1)\eta_1(\tau)\right)\psi(\sigma,m\mp \frac12,w,\tau),
\end{multline}
in agreement with \cite{Maruyoshi:2010iu,Fateev:2009aw}. 
One can also add that to get the solution of the general \(2\times 2\) system, or equivalently the solution of the equation with apparent singularity \eqref{eq:scalarEquation}, \eqref{eq:ellipticPotential-0}, one should also insert heavy degenerate field \(\phi_{(1,2)}(2Q)\) into \(G_l\).
It has \((-1)\) monodromy with \(\phi_{(2,1)}\), so its insertion will give precisely apparent singularity.

Now using the AGT relation between conformal blocks and Nekrasov partition functions \cite{agt} we can identify classical conformal block with the NS limit of Nekrasov partition function:
\begin{equation}
\label{eq:107}
f(\sigma,m\mp \frac12,\tau)=F^{\rm NS}(\sigma,m\mp \frac12,\re^{2\pi \ri\tau}).
\end{equation}

\section{Other examples and generalizations}\label{sec:others}

In Sec.~\ref{sec:p3}, \ref{sec:toro}  we illustrated in detail the example of isomonodromic deformations on the torus and the one of Painlev\'e $\rm III_3$.
In this section we briefly comment on other examples, even though we do not spell out all the details. 
We limit ourselves to list the quantum operators 
corresponding to the other Painlev\'e equations. 
These operators coincide with the quantum Seiberg-Witten (SW) curves of the gauge theories underlying Painlev\'e equations,  in agreement with several existing results in the literature that we will discuss below. Some of these operators have  been studied recently in \cite{Lisovyy:2021bkm}.
According to the procedure spelled out in Sec.~\ref{main},  the exact spectrum of such operators should  be obtained by  imposing vanishing of some combination of tau functions with suitable normalizability conditions on the monodromy parameters of the associated linear system.   \newline

{\bf  Painlev\'e I.} The associated linear problem is defined by  the following Lax matrix (see for instance \cite[eq. (2.2)]{Lisovyy_2017})
\be\label{linearP1}	A(z,t)=	\left(
\begin{array}{cc}
 -p & q^2+q z+\frac{t}{2}+z^2 \\
 4 z-4 q & p \\
\end{array}
\right).\ee
The compatibility conditions are 
\be\label{compp1}\ba 
&{\rd q \over \rd t}=p , \\
&{\rd p\over \rd t}=6 q^2+t ,
\ea \ee
leading to the Painlev\'e I equation
\be \label{p1}{\rd^2 q \over \rd^2 t}=6 q^2+t .\ee
The Hamiltonian is
\be H_0=\frac{p^2}{2}-2 q^3-q t .\ee
It is also useful to define 
\be H_1=2H_0+{p\over q} .\ee
Moreover, since \be A_{12}=q^2+q z+\frac{t}{2}+z^2,  \ee we have  auxiliary poles unless $q=\infty$. This is our requirement for singularities matching and it gives the following
operator
\be \label{up1}-\partial_z^2 + 2 t z+4 z^3 +H_1 \ee
in agreement with expectations from \cite{dm1, dm2,nok2}.  It is also straightforward to see that the  singularities matching condition imposes vanishing of the Painlev\'e I tau function.
The condition of normalizability instead is more subtle and will not be addressed in this work. 
Notice that, as expected, the operator \eqref{up1} is the one arising in the quantization of the Seiberg-Witten curve corresponding to the ${\mathcal H}_0$ Argyres-Douglas theory.  It is also well known that the  quantization condition for this potential involves the ${\mathcal H}_0$  NS free energy (see for instance \cite{ito-shu,gg-qm}). It should be possible to relate such NS type quantization to the vanishing of the Painlev\'e I tau function. For that we would need some kind of  blowup equations for Argyres-Douglas theories which, at present, are not known.  Alternatively one can try to develop a functional approach similar to \cite{Gavrylenko:2020gjb}, see also Sec.~\ref{functional}.  This is  under investigation and will appear in \cite{olegetal}.
\newline

{\bf  Painlev\'e II.} The  linear problem is obtained from the following Lax matrix (see for instance \cite[eq. (3.14)]{blmst})
\be A(z,t)=\left(
\begin{array}{cc}
 p+\frac{t}{2}+z^2 & u (z-q) \\
 -\frac{2 (\theta +p q+p z)}{u} & -p-\frac{t}{2}-z^2 \\
\end{array}
\right).\ee
The associated compatibility condition leads to the Painlev\'e II equation
\be {\rd^2 q \over \rd^2 t}=2 q^3+q t+{1\over 2}-\theta .\ee
The Hamiltonian is
\be H_0=\frac{p^2}{2}+p q^2+\frac{p t}{2}+\theta  q.\ee
We also define \be H_1=2 H_0-2 q.\ee
Since
\be A_{12}=u(z-q),\ee
we have singularities matching if $u=\infty$ (or $q=\infty$). It is easy to show that at this special point the relevant operator reads
\be \label{quadratic}-\partial_z^2+ t z^2+z^4-2 \theta  z+\frac{t^2}{4}+ H_1. \ee
As expected, this is the operator  arising in the quantization of the Seiberg-Witten curve to the $\mathcal{H}_1$ Argyres-Douglas theory. 
It is also straightforward to see that condition of singularities matching imposes the vanishing of the Painlev\'e II tau function as a quantization condition for the potential \eqref{quadratic}.
As before, the condition of normalizability instead is more subtle and will not be addressed in this work. Likewise we do not know blowup equations that would link the vanishing of tau function to the  NS quantization.  Hence it would be nice to develop a functional approach to this problem as done in \cite{Gavrylenko:2020gjb} for Painlev\'e $\rm III_3$ and in Sec.~\ref{functional} for the torus. \newline

{\bf  Painlev\'e IV.} The  linear problem is obtained from the following Lax matrix (see for instance \cite[eq. (3.36)]{blmst})
\be A(z,t)=\left(
\begin{array}{cc}
 \frac{\theta_0-p q}{z}+t+z & u \left(1-\frac{q}{2 z}\right) \\
 \frac{2 (-\theta_0-\theta_{\infty}+p q)}{u}+\frac{2 p (p q-2 \theta_0)}{u z} & -\frac{\theta_0-p q}{z}-t-z \\
\end{array}
\right). \ee
The compatibility condition leads to the Painlev\'e IV equation:
\be {\rd^2 q \over \rd^2 t}=   \left({\rd q \over \rd t}\right)^2{1\over 2 q} + {3q^3\over 2} + 4 t q^2 + 2q(t^2 - 2\theta_{\infty} + 1) - \dfrac{8\theta_0^2}{q}. \ee
The Painlev\'e IV Hamiltonian is defined as
\be H_0=2 p^2 q-p \left(4 \theta_0+q^2+2 q t\right)+q (\theta_0+\theta_{\infty}) .\ee
We also introduce
\be H_1=H_0-2p +t + 2 t \theta_0. \ee
 After imposing the singularities matching condition we obtain
the following operator in agreement with \cite{Masoero_2018}
\be -\partial_z^2 -2 \theta_{\infty}+t^2+2t z+{H_1\over z}+\frac{\theta_0^2-\frac{1}{4}}{z^2}+z^2 .\ee
This is the operator appearing in quantization of SW curve to $\mathcal{H}_2$ Argyres-Douglas theory.
 \newline

{\bf  Painlev\'e $\rm III_2$.}  We follow \cite{saito}. The relevant Lax matrix is
\be A(z,t)=\left(
\begin{array}{cc}
 -\frac{2 p+q^2 (-t)-\theta_{\infty} q}{2 z^2}-\frac{t}{2}-\frac{\theta_{\infty}}{2 z} & -\frac{4 p^2-4 p q^2 t+q^4 t^2-\theta_{\infty}^2 q^2-4 q}{4 q^2 z}-\frac{\left(-2 p+q^2 t+\theta_{\infty} q\right)^2}{4 q z^2} \\
 \frac{q}{z^2}-\frac{1}{z} & \frac{2 p+q^2 (-t)-\theta_{\infty} q}{2 z^2}+\frac{t}{2}+\frac{\theta_{\infty}}{2 z}. ~~ \\
\end{array}
\right),\ee
leading to the Painlev\'e $\rm III_2$ equation
\be {\rd^2 q\over \rd^2 t}={1\over q}\left({\rd q\over \rd t}\right)^2-{\rd q\over \rd t}{1\over t}+{q^2(1+\theta_{\infty} )\over t}+q^3-{2\over t^2}.\ee
The Hamiltonian is
\be H_0=-\frac{p^2}{q^2 t}+\frac{q^2 t}{4}+\frac{1}{2} (\theta_{\infty}+1) q+\frac{1}{q t}. \ee
It is useful to define
\be H_1= \frac{1}{2} (q t-\theta_{\infty})-H_0 t .\ee
Singularities matching leads to 
\be -\partial_z^2+\frac{t^2}{4}+t \frac{\frac{\theta_{\infty} }{2}-1}{z}+\frac{1}{z^3}+{H_1\over z^2}.\ee
This is precisely the operator corresponding to the  quantization of the $N_f=1$ SW curve, see for instance \cite{zen11,ito2017}. \newline

{\bf  Painlev\'e $\rm III_1$.}   The relevant Lax matrix is (see for instance \cite[eq. (A.23)]{blmst}
\be A(z,t)= \left(
\begin{array}{cc}
 \frac{(2 p-1) \sqrt{t}}{2 z^2}+\frac{\sqrt{t}}{2}-\frac{\theta_{*}}{z} & -\frac{p q u}{z}-\frac{p \sqrt{t} u}{z^2} \\
 \frac{2 (\theta_{*}+\theta_{\star})-2 p^2 q-4 \theta_{*} p+2 p q}{2 p u z}+\frac{(p-1) \sqrt{t}}{u z^2} & -\frac{(2 p-1) \sqrt{t}}{2 z^2}-\frac{\sqrt{t}}{2}+\frac{\theta_{*}}{z} \\
\end{array}
\right), \ee
leading to the Painlev\'e $\rm III_1$ equation
\be
{\rd ^2 q\over \rd ^2t} = \left({\rd  q \over \rd t}\right)^2{1\over q}- {\rd  q\over \rd t}{1\over t} + {q^3\over t^2} + {2 q^2 \theta_{\star}\over t^2} 
+ \dfrac{1-2\theta_*}{t} - \dfrac{1}{q} .
\ee
The Hamiltonian is
\be H_0=\theta_{*}^2+p^2 q^2-p q^2+2 \theta_{*} p q+p t-q (\theta_{*}+\theta_{\star})-\frac{t}{2} \, ,\ee
and we define
\be H_1=H_0+q- p q - \theta_{*}.  \ee
We have  singularities matching if $p=\infty$, $q=0$, such that $p q=\infty$ and $u$ finite. After some algebra this leads to the following operator
\be-\partial_z^2+ \frac{H_1}{z^2}+\frac{t}{4 z^4}-\frac{\theta_{\star} \sqrt{t}}{z^3}-\frac{\theta_{*} \sqrt{t}}{z}+\frac{t}{4}, \ee
which is the quantum  SW curve with $N_f=2$, see for instance \cite{zen11,ito2017}.\newline

{\bf  Painlev\'e $V$.}    The relevant Lax matrix is (see for instance \cite[Sec.~4.3]{saito})
\be A(z,t)=\left(
\begin{array}{cc}
 \frac{2 p+(q-z) (\theta_{\infty}+t (q+z-1))}{2 (z-1) z} & \frac{(q-1) \left(\left(p+\frac{1}{2} q (\theta_{\infty}+(q-1) t)\right)^2-\frac{\theta_0^2}{4}\right)}{q z}+\frac{\frac{\theta_1^2}{4}-\left(p+\frac{1}{2} (q-1) (\theta_{\infty}+q t)\right)^2}{z-1} \\
 \frac{q}{z-q z}+\frac{1}{z-1} & -\frac{2 p+(q-z) (\theta_{\infty}+t (q+z-1))}{2 (z-1) z} \\
\end{array}
\right).\ee
The compatibility condition leads to the Painlev\'e  $V$ equation:
\be {\rd ^2 q\over \rd ^2 t}= {2q-1\over 2(q-1)q} \left({\rd  q\over \rd  t}\right)^2-{1\over t} {\rd q\over \rd  t}+{(q-1)q(2 q t -t +2 \theta_{\infty}-2)\over 2 t}+{\theta_0^2\over 2 q t^2}+{\theta_1^2\over 2 (q-1)t^2}.\ee
The Hamiltonian is
\be H_0=-\frac{p^2}{(q-1) (q t)}-\frac{\theta_0^2}{(4 q) t}+\frac{\theta_1^2}{(q-1) (4 t)}+\frac{1}{4} q (2 \theta_{\infty}+q t-t-2), \ee
we also define 
\be H_1=t H_0 +\frac{\theta_0^2}{4}+\frac{\theta_1^2}{4}+\frac{\theta_{\infty}}{2}+\frac{q t}{2}-\frac{\theta_{\infty} t}{2}+\frac{t}{2}-\frac{1}{2}.\ee
Singularities matching leads to confluent Heun equation
\be \ba -\partial_z^2+{1\over 4 z^2 (z-1)^2}&\left(\theta_{0}^2+t^2 z^4-1 +z \left(-2 \theta_{0}^2+4 H_1+2 (\theta_{\infty}-2) t+2\right)\right.\\
&\left. + z^2 \left(\theta_{0}^2+\theta_{1}^2-4 H_1+t^2-4 (\theta_{\infty}-2) t-2\right)+z^3\left(2 (\theta_{\infty}-2) t-2 t^2\right) \right)~,\\
\ea\ee 
which is the operator appearing in quantization of  $N_f=3$ SW curve, see for instance \cite{zen11,ito2017}.
We  note that this curve also plays a role in the study of black hole quasinormal modes.
 By adapting the  procedure illustrated in Sec.~\ref{main} to potentials with  resonance  eigenstates, one should be able to reproduce \cite{daCunha:2015ana, CarneirodaCunha:2019tia}. In addition by using Sec.~\ref{sec:ns} one should be able to 
 provide a more direct link between \cite{daCunha:2015ana, CarneirodaCunha:2019tia} and \cite{zen11, Aminov:2020yma}. It would also be interesting to further investigate the connection with the Rabi model \cite{daCunha:2015npa} by using the NS approach.
 \newline

{\bf  Painlev\'e VI.} 
The relevant Lax matrix in this case is quite complicated. We write it as
\be A(z,t)=\left(
\begin{array}{cc}
A_{11}(z) & A_{12}(z) \\
 A_{21}(z) & -A_{11}(z) \\
\end{array}
\right),\ee
where 
\be \label{a12h} A_{12}(z)= -\frac{(t-1) t (z-q)}{(z-1) z (t-z)}. \ee
The expressions for $A_{11}$ and $A_{21}$ take the following forms
\be A_{21}={1\over 64 \theta_\infty ^2 (t-1)^2 t^2 (t-q)^2 (q-1)^2 q^2} \Big (\sum_{i=0}^7 u_i (t,z,p) q ^i\Big),\ee
\be\ba A_{11}={1\over 8 \theta_\infty  (z-1) z (t-q) (q-1) q (t-z)}\Big (\sum_{i=0}^7 s_i(t,z,p) q ^i\Big).\ea \ee
 The expressions for  $u_i (t,z,p)$ and $s_i (t,z,p)$ are quite cumbersome, hence we do not write them explicitly. 
Compatibly condition for this system is
\be \ba
{\rd\over \rd t}q=&-\frac{2}{(t-1) t}\left(-p q^3+p q^2 t+p q^2+p q (-t)\right), \\
 {\rd\over \rd t}p=&{1\over 4t}\left(\frac{4 \theta_1^2}{(q-1)^2}+\frac{(2 \theta_\infty +1)^2}{t-1}\right)+\frac{1-4 \theta_t^2}{4(t-q)^2}-\frac{4 \left(P(t)^2 q^2 \left(3 q^2-2 (t+1) q+t\right)+\theta_0^2 t\right)}{4(t-1) t q^2},\ea\ee
leading to Painlev\'e VI
\be\label{p6}\ba  {\rd^2 q \over \rd^2 t}= & \frac{1}{2} \left(\frac{1}{q-t}+\frac{1}{q}+\frac{1}{q-1}\right)\left({\rd q \over \rd t}\right)^2-\left(\frac{1}{q-t}+\frac{1}{t}+\frac{1}{t-1}\right){\rd q \over \rd t}\\
&+\frac{(q-1) q (q-t)}{2 (t-1)^2 t^2}\left((2 \theta_\infty +1)^2-\frac{4 \theta_0^2 t}{q^2}+\frac{4 \theta_1^2 (t-1)}{(q-1)^2}-\frac{\left(4 \theta_t^2-1\right) (t-1) t}{(t-q)^2}\right).\ea\ee
From \eqref{a12h} is easy to see that 
the singularities matching condition leads to the Heun operator appearing in the quantization of  the $N_f=4$ SW curve. This example was  studied  in \cite{Litvinov:2013sxa, Lencses:2017dgf,Kashani-Poor:2013oza, Barragan-Amado:2018pxh,Anselmo:2018zre,Novaes:2018fry, Barragan-Amado:2020pad, Dubrovin_2018,1805497,Jeong:2020uxz}. In particular the self-dual approach to the Heun equation was studied in great details in \cite{Barragan-Amado:2018pxh,Anselmo:2018zre,Novaes:2018fry, Barragan-Amado:2020pad}.
\newline

{\bf  Some further comments and generalisations:} 

\begin{itemize}

\item 
In this work we took the approach of studying the spectral properties of quantum mechanical operators by using the knowledge about isomonodromic deformations. However, one can read our result by taking the inverse logic and, in line with \cite{nok},  use the spectral properties of quantum  operators to study the distribution of movable poles in solutions of second order nonlinear ODEs arising as compatibility conditions of isomonodromic deformations.

\item  All the examples listed above correspond to  isomonodromic problems associated to $2\times 2$ linear system. Nevertheless, a similar story is expected to hold also for the  higher rank situation. In this case some related properties for the corresponding tau function(s) and generalisation of the Kiev formula can be found in \cite{Gavrylenko:2015wla, Gavrylenko:2018ckn, Bershtein:2018srt,Gavrylenko:2016moe,Bonelli:2017ptp, Bonelli:2019yjd,Gavrylenko:2015cea}.

\item We also note that recently a new class of nonlinear eigenvalue problems has been related to a set of generalized Painlev\'e equations \cite{Bender_2019}. 
 It would be interesting to study these  problems and their stability/instability notion within our gauge theoretic framework.

\item In this work we studied  in detail  examples of operators with confining potential.  However our  formalism can also be applied  straightforwardly to study the band structure of periodic potentials, including the band edges and the corresponding energy splitting. These results will appear somewhere else. It would also be interesting to extend our analysis to the study of potentials which admit a spectrum of resonance modes.  

\item Another set of generalised problems which it would be interesting to investigate are these connected to q-deformed Painlev\'e equations and five-dimensional gauge theories \cite{Bershtein:2016aef,Bonelli:2017gdk,Bershtein:2018srt, Matsuhira_2019,jimbo2017cft,Bonelli:2020dcp,Nosaka:2020tyv, Moriyama:2021mux}. In this case the relevant quantum spectral problems are the ones associated to relativistic quantum integrable systems. 
 
For example, we know that the NS quantization condition does not extend directly to the five-dimensional/relativistic setup. In particular, to compute the exact spectrum of relativistic integrable systems one needs to supply the naive NS quantization \cite{ns,acdkv} with additional non-perturbative corrections \cite{Grassi:2014zfa}. Nevertheless, if  we think of the four dimensional quantization as the vanishing of  Painlev\'e tau functions, then this fact extends directly to the five dimensional/relativistic integrable system setup. Indeed it was found in \cite{Bonelli:2017gdk} that the zeroes of the tau functions for q-Painlev\'e  compute the exact spectrum of relativistic integrable systems.  
Hence thinking of the quantization condition as vanishing of (q-) Painlev\'e tau functions provides a unifying framework for both relativistic and non-relativistic quantum systems.  From that perspective it would be interesting to understand how the quantum mirror map is realised on the q-Painlev\'e side. 

Recent interesting related work in this direction is also  \cite{Noumi_2020}.

\item In the  Painlev\'e $\rm III_3$ example we have an intriguing bridge between the following two operators. On one side we have the modified Mathieu
  \be \partial_x^2 -\sqrt{t} \left(\re^x+\re^{-x}\right), \ee
  and on the other side we have a "dual" Fermi gas operator, which reads  \cite[eq (1.3)]{Bonelli:2016idi}
    \be \re^{{4 t^{1/4} } \cosh( x)}{ \left(\re^{{\ri\over 2} \partial_x}+ \re^{-{\ri\over 2} \partial_x} \right)}\re^{{4 t^{1/4} } \cosh( x)} .\ee
In particular, the spectral properties of both operators are encoded in the isomonodromic deformation equations of the linear system \eqref{linear}. For example, the quantization condition of both operators can be expressed as vanishing of the Painlev\'e $\rm III_3$ tau function.
 This provided a concrete link between the results of \cite{WMT,zamo} and \cite{gil}, which was also generalised to the q-deformed/five-dimensional framework, see \cite{Bonelli:2016idi} and \cite{Bonelli:2017gdk} for more details \footnote{Similarly, relations between Kiev formulas and tt* equations are also discussed in \cite{Bonelli:2021rrg} and forthcoming publications by the same Authors.}.
It would be very interesting to find such "dual" operator for other Painlev\'e equations by using the geometrical guideline coming from the TS/ST correspondence \cite{Grassi:2014zfa}. This could provide some  concrete realisation of ideas presented in \cite{Fock:1999ae} \footnote{We thank N.~Nekrasov for discussions on this point and bringing our attention to this reference.}.

\end{itemize}

\section*{Acknowledgements}
We would like to thank Giulio Bonelli, Bruno Carniero da Cunha, Joao Cavalcante, Fabrizio Del Monte, Jie Gu, Yasuyuki Hatsuda, Oleg Lisovyy, Hiraku Nakajima, Nikita Nekrasov,  Anton Shchechkin, Alessandro Tanzini for useful discussions and correspondence. We  are  especially  grateful  to  Anton  Shchechkin  for a careful reading of the  preliminary version of the paper and many critical remarks. AG would like to thank Skoltech and the HSE Moscow for the kind hospitality during fall 2019 when this project started.  The work of AG is partially supported by the Fonds National Suisse, Grant No. 185723  
and by the NCCR "The Mathematics of Physics" (SwissMAP).
The work was partially carried out in Skolkovo Institute of Science and Technology under financial support of Russian Science Foundation within grant 19-11-00275 (PG).
\appendix

\section{Conventions for elliptic functions}\label{conv}

For the elliptic functions we use the same conventions and definitions as  in Appendix A of \cite{Bonelli:2019boe}.
We always use 
\be \mathfrak{q}=\re^{2\pi \ri\tau}.\ee
The conventions for Jacobi theta functions are
\begin{equation}\begin{split}
\label{eq:thetaFunctions}
\theta_1(z |\tau)&=-\ri \sum_{n\in \IZ}(-1)^n \re^{\ri \pi \tau(n+{1\over 2})^2}\re^{2\pi \ri z (n+{1\over 2})}, \\
\theta_2(z |\tau)&=\sum_{n\in \IZ} \re^{\ri \pi \tau(n+{1\over 2})^2}\re^{2\pi \ri z (n+{1\over 2})},  \\
\theta_3(z |\tau)&=\sum_{n\in \IZ} \re^{\ri \pi \tau n^2}\re^{2\pi \ri z n}, \\
 \theta_4(z |\tau)&=\sum_{n\in \IZ} (-1)^n \re^{\ri \pi \tau n^2}\re^{2\pi \ri z n}.\\
\end{split}\end{equation}
There is also a useful infinite product representation for \(\theta_1\):
\begin{equation}
\label{eq:theta1product}
\theta_1(z|\tau)=2\mathfrak{q}^{\frac1{4}}\sin\pi z\prod_{k=1}^{\infty}(1-\mathfrak{q}^k)(1-\mathfrak{q}^ke^{2\pi \ri z})(1-\mathfrak{q}^ke^{-2\pi \ri z}).
\end{equation}
%
The Dedekind $\eta$ is defined as
\be \label{deddek} \eta(\tau)=\mathfrak{q}^{1/24}\prod_{n\geq 1}(1-\mathfrak{q}^n),\ee
and  satisfies 
\begin{equation}
\label{eq:24}
\eta(-1/\tau)=\sqrt{-\ri\tau}\eta(\tau).
\end{equation}
We also define
\begin{align}\label{eq:phiDef}
\varphi(\mathfrak{q})=\prod_{k=1}^\infty (1-\mathfrak{q}^k)=\mathfrak{q}^{-1/24}\eta(\tau). ~
\end{align}
One also has a relation
\begin{equation}
\label{eq:thetaprime}
\left.\partial_z\theta_1(z|\tau)\right|_{z=0}=2\pi \eta(\tau)^3.
\end{equation}
The Weierstrass $\wp$ function is
\begin{equation}
\label{wpdef} \wp(z|\tau)=-\partial_z^2 \log \theta_1(z|\tau)-2\eta_1(\tau),
\end{equation}
where
\begin{equation}\label{eta1def}
\eta_1(\tau)=-2\pi \ri \partial_{\tau}\log\eta(\tau)=
-{1\over 6}\left({\partial_z^3\theta_1(z|\tau)\over \partial_z\theta_1(z|\tau)}\right)\Big|_{z=0}.
\end{equation}
We also use
\begin{equation}\label{zdef} \zeta(z|\tau)= \partial_z \log \theta_1(z|\tau)+2z \eta_1(\tau).
\end{equation}
We now list some useful  representations of the Weierstrass function. 
The first one is \begin{equation}
\label{eq:wpSum}
\wp(z|\tau)=\frac1{z^2}+\sum_{(n,k)\neq (0,0)}\left( \frac1{(z-k\tau-n)^2}-\frac1{(k\tau+n)^2} \right).
\end{equation}
This representation makes the modular transformation obvious:
\begin{equation}
\label{eq:wpModular}
\wp(z|\tau)=\tau^{-2}\wp(z/\tau|-1/\tau).
\end{equation}
If we take \eqref{eq:wpSum} and we perform the sum  in the \(n\)-direction we get another representation:
\begin{equation}
\label{eq:wpSin}
\wp(z|\tau)=\frac{\pi^2}{\sin^2\pi z}-\frac{\pi^2}{3}+\sum_{k\neq 0}\left( \frac{\pi^2}{\sin^2\pi(z-k\tau)}-\frac{\pi^2}{\sin^2(\pi k\tau)} \right).
\end{equation}
From \eqref{eq:wpSin} one can easily get the expansion in the limit \(\tau\to \ri\infty\):
\begin{equation}
\label{eq:wpZero}
\wp(z|\tau)=\frac{\pi^2}{\sin^2\pi z}-\frac{\pi^2}{3}+16\pi^2\re^{2\pi i\tau}\sin^2\pi z+\mathcal{O}(\re^{4\pi i\tau}).
\end{equation}
Another option is to compute the sum in \eqref{eq:wpSum} along the \(k\)-direction:
\begin{equation}
\label{eq:wpSinh}
\wp(z|\tau)=\frac{\pi^2/\tau^2}{\sin^2 \frac{\pi z}{\tau}}-\frac{\pi^2}{3\tau^2}+\sum_{n\neq 0}\left( \frac{\pi^2/\tau^2}{\sin^2 \frac{\pi(z-n)}{\tau}}-\frac{\pi^2/\tau^2}{\sin^2 \frac{\pi n}{\tau}} \right).
\end{equation}
There are also some useful expansions of \(\eta_1(\tau)\). For example
\begin{equation}
\label{eq:etaInf}
\eta_1(\tau)=\frac{\pi^2}{6}\left( 1-24\sum_{n=1}^{\infty}\sigma_1(n)\re^{2\pi i n\tau} \right),
\end{equation}
where \(\sigma_1(n)\) is sum of all divisors of \(n\):
\begin{equation}
\label{eq:26}
\sigma_1(n)=\sum_{d|n}d.
\end{equation}
Another useful expansion can be obtained by using the modular transformation of the Dedekind function:
\begin{equation}
\label{eq:etaZero}
\eta_1(\tau)=\frac{\pi \ri}{\tau}+\frac1{\tau^2}\eta_1(-1/\tau)=\frac{\pi^2}{6\tau^2}+\frac{\pi \ri}{\tau}-\frac{4\pi^2}{\tau^2}\sum_{n=1}^{\infty}\sigma_1(n)\re^{-2\pi \ri/\tau}.
\end{equation}

\section{Perturbative study of quantum mechanical potentials}\label{sec:nstest}

The results presented in this Appendix are not new and can be found in various textbook, as well as in  \cite[Sec.~2]{Hatsuda_2018} where they also discuss them in relation to gauge theory. 
We added this Appendix in order to perform another verification of the computations done in the main part of the paper, and also to study several limiting cases in more detail.

\subsection{Perturbed P\"oschl--Teller potential}\label{b1}

Here we study the quantum mechanical problem of the Weierstrass potential by considering its expansion \eqref{eq:wpZero} in the limit \(\tau\to \ri\infty\).
The corresponding quantum mechanical Hamiltonian is
\begin{equation}\begin{gathered}
\label{eq:15}
\mathrm{O}_-=-\partial_x^2+m(m-1)\wp(x|\tau)=\\
=-\partial_x^2+\frac{\pi^2m(m-1)}{\sin^2\pi x}-\frac{\pi^2}{3}m(m-1)+ 16\pi^2m(m-1)\re^{2\pi \ri\tau}\sin^2\pi x+ \mathcal{O}(\re^{4\pi \ri\tau}).
\end{gathered}\end{equation}
The first term is the well-known trigonometric P\"oschl--Teller potential.
Its eigenfunctions can be constructed explicitly, for example, using supersymmetric quantum mechanics (we take \(m>1\) so that the eigenfunctions are well defined for $x \in  [0,1]$):
\begin{equation}
\label{eq:16}
|\Psi_k\rangle=(\partial_x+\pi m\cot \pi x)(\partial_x+\pi (m+1)\cot \pi x)\cdots (\partial_x+\pi (m+k-1)\cot \pi x)\left( \sin \pi x \right)^{m+k}.
\end{equation}
The eigenvalue equation has the form
\begin{equation}
\label{eq:17}
\left( -\partial_x^2+\frac{m(m-1)\pi^2}{\sin^2\pi x} \right)|\Psi_k\rangle=
\pi^2(m+k)^2|\Psi_k\rangle.
\end{equation}
Now we compute the first order correction to the energy (we computed it for the first few levels and then guessed the general form):
\begin{equation}
\label{eq:18}
\frac{\langle \Psi_k|\sin^2\pi x|\Psi_k\rangle}{\langle \Psi_k|\Psi_k\rangle}
=\frac12+\frac{m(m-1)}{2((m+k)^2-1)}.
\end{equation}
Hence\begin{multline}
\label{eq:energyAcycle}
E_k=\frac{\langle \Psi_k|\hat{H}|\Psi_k\rangle}{\langle \Psi_k|\Psi_k\rangle}+\mathcal{O}(\re^{4\pi \ri\tau})=\\
=-\frac{\pi^2}{3}m(m-1)+\pi^2(m+k)^2+8\pi^2\re^{2\pi \ri\tau}
\left( 1+ \frac{m(m-1)}{(m+k)^2-1}\right)+\mathcal{O}(\re^{4\pi \ri\tau}),
\end{multline} 
see also \cite[formula (2.36)]{Hatsuda_2018}.
This is in perfect agreement with the gauge theory computation, see \eqref{eq:70}.
 
We notice that this formula is applicable also for \(\tau\to -\frac12+\ri\infty\).
In this case the potential is still real, and all the formulas for the energy can be applied directly.
The only difference is that now \(\mathfrak{q}=\re^{2\pi \ri\tau}\) will be a small negative real number.

\subsection{Distant potential walls approximation}
We now switch to another approximation. We  consider the operator
\be \mathrm{O}_-=-\partial_x^2 +m (m-1)\wp (x, \widehat \tau),\ee
where \be x\in [0,1],\quad {\widehat{\tau}}\in \ri \IR_+, \quad m>1\, . \ee
We introduce here \(\widehat{\tau}\) to distinguish it from the modular parameter in the conformal blocks, which will be \(\tau=-1/\widehat{\tau}\).
We want to study the limit \be\label{tau0l} -\ri{\widehat{\tau}}\to 0\, .\ee
In this regime the potential takes the form of an  infinite collection of  walls  \be \frac{\pi^2m(m-1)/{\widehat{\tau}}^2}{\sin^2 \frac{\pi (x-n)}{\widehat{\tau}}} \ee
located at integer points \(n\in \mathbb{Z}\), see \eqref{eq:wpSinh}.
Each wall decays exponentially at a distance of the order \(-\ri{\widehat{\tau}}\).
We could naively rescale the $x$ coordinate as \be y=\pi x/(-\ri\widehat{\tau})\ee 
and  say that the limiting potential is \(\frac{m(m-1)}{\sinh^2 y}\). However, such procedure produces some inconvenient  artifacts like a continuous spectrum (it was discrete before the limit).
In fact the correct way to solve this problem is to keep also the second neighboring wall before taking the limit \eqref{tau0l}\footnote{In this limit the second wall goes to infinity in the \(y\)-coordinate.}.
This procedure will give us a result which is valid up to exponentially small  corrections of the form \(\mathcal{O}(\re^{-2 \pi \ri/{\widehat\tau}})\).

We proceed as follows. We first consider the scattering on the single potential wall and then glue the two  scattered wave functions corresponding to the two neighboring walls.  
 The equation for the \(n=0\) wall is (hyperbolic P\"oschl--Teller)
 \begin{equation}
\label{eq:25}
-\partial_x^2\Psi+\frac{\pi^2 m(m-1)/{\widehat{\tau}}^2}{\sin^2 \frac{\pi x}{{\widehat{\tau}}}}\Psi=\frac{\pi^2 \kappa^2}{{\widehat{\tau}}^2}\Psi,
\end{equation}
where \(\kappa\) parameterizes the energy.
Two independent solutions of this equation are 
\begin{equation}
\label{eq:76}
\Psi^{(1)}(x)=\left(2\ri \sin \frac{\pi x}{{\widehat{\tau}}} \right)^m {}_2F_1(m+\kappa,m-\kappa,m+\frac12,\sin^2 \frac{\pi x}{2{\widehat{\tau}}}),
\end{equation}

\begin{equation}
\label{eq:77}
\Psi^{(2)}(x)=\left( \sin \frac{\pi x}{{\widehat{\tau}}} \right)^m \left( \sin \frac{\pi x}{2{\widehat{\tau}}} \right)^{1-2m} {}_2F_1(\frac12+\kappa,\frac12-\kappa,\frac12-m,\sin^2 \frac{\pi x}{2{\widehat{\tau}}}).
\end{equation}
We see that if \(m>1\) only the first one is regular as \(x\to 0\).  So in order to study the scattering on the potential wall  it is sufficient to consider \(\Psi^{(1)}(x)\).
We also need to re-expand \(\Psi^{(1)}(x)\) in the limit \eqref{tau0l}.
This can be done by using  the formula \eqref{eq:FhypReexpansion}. We have
\begin{multline}
\label{eq:37}
\Psi^{(1)}(x)\approx \frac{\Gamma(-2\kappa)\Gamma(m+1/2)}{\Gamma(m-\kappa)\Gamma(1/2-\kappa)}\left(-\sin^2 \frac{\pi x}{2{\widehat{\tau}}}\right)^{-m-\kappa}\left(2\ri \sin \frac{\pi x}{{\widehat{\tau}}} \right)^m+
\\+
\frac{\Gamma(2\kappa)\Gamma(m+1/2)}{\Gamma(m+\kappa)\Gamma(1/2+\kappa)}\left(-\sin^2 \frac{\pi x}{2{\widehat{\tau}}}\right)^{-m+\kappa}\left(2\ri \sin \frac{\pi x}{{\widehat{\tau}}} \right)^m\approx
\\\approx
\frac{\Gamma(-2\kappa)\Gamma(m+1/2)}{\Gamma(m-\kappa)\Gamma(1/2-\kappa)} 2^{2m+2\kappa} \re^{-\kappa\frac{\ri\pi x}{{\widehat{\tau}}}} +
\frac{\Gamma(2\kappa)\Gamma(m+1/2)}{\Gamma(m+\kappa)\Gamma(1/2+\kappa)} 2^{2m-2\kappa} \re^{\kappa\frac{\ri\pi x}{{\widehat{\tau}}}}+\mathcal{O}(e^{-\pi\ri x/\widehat{\tau}}) \,.
\end{multline}
Likewise the wave function coming from the potential wall at \(x=1\) should have the form
\begin{multline}
\label{eq:78}
\Psi^{(\tilde{1})}=\Psi^{(1)}(1-x)\approx
\frac{\Gamma(-2\kappa)\Gamma(m+1/2)}{\Gamma(m-\kappa)\Gamma(1/2-\kappa)} 2^{2m+2\kappa} \re^{-\frac{\kappa \ri\pi}{{\widehat{\tau}}}} \re^{\kappa\frac{\ri\pi x}{{\widehat{\tau}}}} + \\ +
\frac{\Gamma(2\kappa)\Gamma(m+1/2)}{\Gamma(m+\kappa)\Gamma(1/2+\kappa)} 2^{2m-2\kappa} \re^{\frac{\kappa \ri\pi}{{\widehat{\tau}}}} \re^{-\kappa\frac{\ri\pi x}{{\widehat{\tau}}}}+\mathcal{O}(e^{-\pi\ri (1-x)/\widehat{\tau}}).
\end{multline}
One should have \(\Psi^{(1)}(x)\approx  \pm \Psi^{(\tilde{1})}(x)\), because the potential is symmetric.
This gives us the following relation:
\begin{equation}
\label{eq:29}
\re^{\pi \ri \kappa/\widehat{\tau}} \frac{\Gamma(2\kappa)\Gamma(1/2-\kappa)\Gamma(m-\kappa)}{\Gamma(-2\kappa)\Gamma(1/2+\kappa)\Gamma(m+\kappa)}2^{-4\kappa} \approx \pm 1.
\end{equation}
This can be simplified using Legendre duplication formula
\begin{equation}
\label{eq:80}
\Gamma(2\kappa)=\frac{2^{2\kappa-1}}{\sqrt{\pi}}\Gamma(\kappa)\Gamma(1/2+\kappa).
\end{equation}
We get
\begin{equation}
\re^{\pi \ri\kappa/{\widehat{\tau}}}\frac{\Gamma(1+\kappa)\Gamma(m-\kappa)}{\Gamma(1-\kappa)\Gamma(m+\kappa)}\approx\mp1.
\end{equation}
We can also rewrite it in the logarithmic form:
\begin{equation}
\label{eq:wallsQuantization}
\pi \ri k \approx \frac{\pi \ri\kappa}{{\widehat{\tau}}} + \log \frac{\Gamma(1+\kappa)\Gamma(m-\kappa)}{\Gamma(1-\kappa)\Gamma(m+\kappa)}\, , \quad k \in \IZ\,. 
\end{equation}
 This is in perfect agreement with the gauge theory computation. Indeed this example is  
case \(\#\) 2 in Table \ref{tab:homologyClasses}. Therefore the quantization condition coming from the gauge theory \eqref{eq:etaQuantization}, \eqref{eq:etaMinus}  give us
\begin{equation}\begin{gathered}
\label{eq:19}
\re^{\ri \pi (k+1)}=
-\mathfrak{q}^{-\sigma}\frac{\Gamma(m-2\sigma)\Gamma(2\sigma)}{\Gamma(m+2\sigma)\Gamma(-2\sigma)}\re^{ \partial_\sigma F^{\rm NS}_{\rm inst}(\sigma,m-1/2,\mathfrak{q})/2}.
\end{gathered}\end{equation}
Taking the logarithm of this equation we get
\begin{equation}
\label{eq:71}
\ri\pi k=-2\ri\pi\tau\sigma + \log \frac{\Gamma(m-2\sigma)\Gamma(2\sigma)}{\Gamma(m+2\sigma)\Gamma(-2\sigma)}+ {1\over 2}\partial_\sigma F^{\rm NS}_{\rm inst}(\sigma,m-1/2,\mathfrak{q}).
\end{equation}
We see that this expression coincides with the approximate quantum mechanical quantization condition \eqref{eq:wallsQuantization} after identification 
\be 2\sigma=\kappa ,\quad {\widehat{\tau}}= -\frac{1}{\tau}\,. 
\ee 
The difference between these two expressions is given by the derivative of  the conformal block, which is exponentially small i.e.~\(\mathcal{O}(\re^{-2 \pi \ri /\hat \tau}) \).
On the quantum mechanical side this corresponds to the ``interaction'' between the potential walls.

We can also try to analyze the spectrum in the limit \({\widehat{\tau}}\to 0\).
To do this we expand the gamma functions around \(\sigma=0\) with \(\pi>\arg \sigma \geq0\):
\begin{equation}
\label{eq:73}
\log \frac{\Gamma(m-2\sigma)\Gamma(2\sigma)}{\Gamma(m+2\sigma)\Gamma(-2\sigma)}=-\ri \pi -4 \sigma  (\psi ^{(0)}(m)+\gamma_{\rm Euler} )+\frac{8}{3} \sigma ^3 (\psi ^{(2)}(1)-\psi ^{(2)}(m))+O\left(\sigma ^5\right),
\end{equation}
where $\psi$ is the polygamma function.
Now we can find \(\sigma\) in a form of a double series expansion
\begin{equation}\begin{gathered}
\label{eq:74}
\sigma=\frac{-k-1}{2 \tau }-\frac{\ri (k+1) (\psi ^{(0)}(m)+\gamma_{\rm Euler}  )}{\pi  \tau ^2}+\frac{2 (k+1) (\psi ^{(0)}(m)+\gamma_{\rm Euler}  )^2}{\pi ^2 \tau ^3}+\mathcal{O}(\tau^{-4})+\mathcal{O}(\mathfrak{q}).
\end{gathered}\end{equation}
By substituting this expression into the formula for the energy of the one-wall problem we get\footnote{The first term comes from the $\pi^2/3$ in \eqref{eq:wpSinh}.}
\be E\approx -\frac{\pi^3}{3}m(m-1)+\pi^2 \tau^2 \kappa^2. \ee
Hence
\begin{equation}
\label{eq:75}
E=-\frac{\pi^3}{3}m(m-1)+\pi^2(k+1)^2+\frac{4\ri\pi (k+1)^2}{\tau}(\gamma_{\rm Euler}+\psi ^{(0)}(m))+\ldots.
\end{equation}
At the leading order \eqref{eq:75} coincides with the spectrum of two  infinite walls  potentials.
Corrections of order \(\tau^{-1}\) and higher come from the fact that the potential walls have a non-zero width.
Other exponentially small corrections of order \(\mathcal{O}(\mathfrak{q})\) come from all the other terms in the expansion \eqref{eq:wpSinh}
and also from the fact that we used the one-wall wave functions for the two-wall problem.

We also notice that a similar computation can be done for the case \(\# 6\) in Table \ref{tab:homologyClasses}.
In this case the limiting quantum-mechanical potential  consists of potential walls \(\frac{m(m-1)}{\sinh^2y}\)  with potential wells \(-\frac{m(m-1)}{\cosh^2y}\) between them.

\section{Computation of monodromies}
\label{app:monodromies}
To make the exposition self-consistent we report in this Appendix
 the explicit computations of monodromies in terms of  \(\eta\) and \(\sigma\). These results were obtained  in \cite{Bonelli:2019boe}.  We assume everywhere that \(2\sigma\notin \mathbb{Z}\) and the same for \(m\).

\subsection{\(B\)-cycle monodromy \(M_{B}\)}
The fundamental solution of the linear system \eqref{eq:torusSystem} has the following properties 
\begin{equation}\label{eq:T and M}
	Y(z+1)=\mathrm{T}_AY(z)M_A ,\qquad Y(z+\tau)=\mathrm{T}_BY(z)M_B .
\end{equation}
Here the matrices the \(\mathrm{T}_A,\mathrm{T}_B\) are called twists, they encode nontrivial shifts of the matrix $A(z)$. Geometrically they encodes the fact that we have connection in nontrivial holomorphic bundle on $\IT ^2$.
It follows from the formula \eqref{At} that 
\begin{equation}\label{eq:Twists}
\mathrm{T}_A =1, \quad \mathrm{T}_B=\re^{2\pi \ri Q\sigma_3},
\end{equation}
where \(\sigma_3\) denotes the Pauli matrix. 

The matrices $M_A, M_B$ are called monodromy matrices. In this section we find their explicit expressions for an appropriate choice of \(Y(z)\) (see \cite[eq. (D.32)]{Bonelli:2019boe}), the reader in hurry can skip the derivation and go to the answers \eqref{eq:MB}, \eqref{eq:MBtilde} ,\eqref{eq:MA}. We will assume through this derivation that $\sigma, m$ are generic, then expect possible singularities in these parameters.

Since the monodromies do not depend on \(\tau\) and \(z\), we can consider the system \eqref{eq:linearSystem} in the limit
\be \tau\to \ri\infty, \quad {\rm Re}(\tau)=0. \ee
Moreover, we  take \(z\) in the vicinities of four points, \(\frac12-\frac{\tau}{2}\), \(\frac12+\frac{\tau}{2}\), \(-\frac12+\frac{\tau}{2}\), \(-\frac12-\frac{\tau}{2}\).
\begin{figure}[h!]
\begin{center}
\begin{tikzpicture}[scale=1]

\begin{scope}

\clip(-3.5,-4) rectangle (3.5,4);

\path[pattern color=green!15!white, pattern=north west lines, decoration=snake,decorate](-5,1.3) rectangle (5,5.5);
\path[pattern color=red!15!white, pattern=north west lines, decoration=snake,decorate](-5,-1.2) rectangle (5,-5.5);
\path[pattern color=blue!15!white, pattern=north east lines, decoration=snake,decorate](-5,-2.3) rectangle (5,1.8);

\path[draw=gray, decoration=snake,decorate](-5,1.3) rectangle (5,5.5);
\path[draw=gray, decoration=snake,decorate](-5,-1.2) rectangle (5,-5.5);
\path[draw=gray, decoration=snake,decorate](-5,-2.3) rectangle (5,1.8);

\end{scope}

\foreach \j in {-3,0,3}
  {
  \foreach \i in {-3,-2,-1,0,1,2,3}{
  \draw[fill,radius=0.07](\i,\j) circle;
  }};
\node at (-0.2,-0.2){\(0\)};
\node at (1-0.2,-0.2){\(1\)};
\node at (-0.2,3-0.2){\(\tau\)};
\node at (1-0.2,3-0.2){\(\tau+1\)};


\node [label={[xshift=-0.25cm, yshift=-0.6cm]\(z_4\)}] [label={[xshift=0.25cm, yshift=-0.4cm]\(z_0\)}] (p0) at (0.5,-1.5){};
\node [label={[xshift=0.25cm, yshift=-0.4cm]\(z_1\)}] (p1) at (0.5,1.5){};
\node [label={[xshift=-0.25cm, yshift=-0.4cm]\(z_2\)}] (p2) at (-0.5,1.5){};
\node [label={[xshift=-0.25cm, yshift=-0.4cm]\(z_3\)}] (p3) at (-0.5,-1.5){};

\draw[->,blue] (p0)--(p1);
\draw[->,blue] (p1)--(p2);
\draw[->,blue] (p2)--(p3);
\draw[->,blue] (p3)--(p0);

\draw[radius=0.05] (p0) circle;
\draw[radius=0.05] (p1) circle;
\draw[radius=0.05] (p2) circle;
\draw[radius=0.05] (p3) circle;

\node at (0,-1.3){\color{blue}\(A\)};
\node at (0.3,0.5){\color{blue}\(B\)};
\node at (0.1,1.7){\color{blue}\(A^{-1}\)};
\node at (-0.15,-0.7){\color{blue}\(B^{-1}\)};


\end{tikzpicture}
\end{center}
\caption{\label{fig:linSystemDomains} Degenerate limit of the linear system.}
\end{figure}
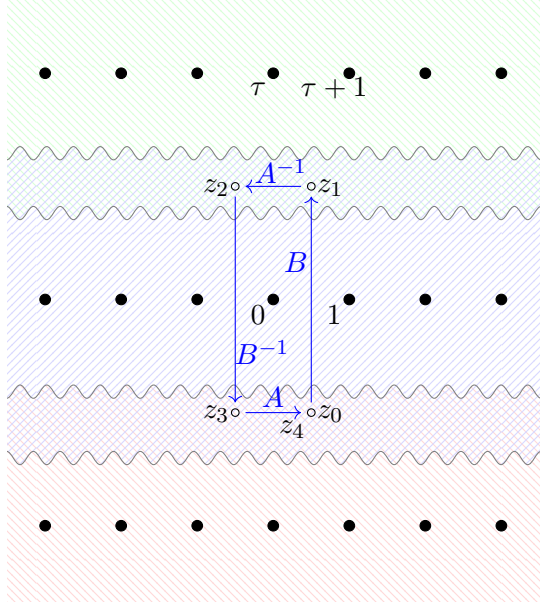
As a first step we find the asymptotics of \(Q\) in this regime.
By using \eqref{eq:solutionTorus} we get the following expansion:
\begin{equation}\begin{gathered}\begin{split}
\label{eq:9}
\re^{\pi \ri\tau/2}(\re^{2\pi \ri Q}+\re^{-2\pi \ri Q})
&\frac{\prod_{\epsilon=\pm}G(1-m+2\epsilon \sigma)}{\prod_{\epsilon=\pm}G(1+2\epsilon \sigma)}\re^{2\pi \ri\sigma^2}\approx
\\\approx
&\frac{\prod_{\epsilon=\pm}G(1-m+\epsilon(2\sigma+1))}{\prod_{\epsilon=\pm}G(1+\epsilon(2\sigma+1))}\re^{\ri \eta/2}\re^{2\pi \ri\tau(\sigma+1/2)^2}+
\\+
&\frac{\prod_{\epsilon=\pm}G(1-m+\epsilon(2\sigma-1))}{\prod_{\epsilon=\pm}G(1+\epsilon(2\sigma-1))}\re^{-\ri \eta/2}\re^{2\pi \ri\tau(\sigma-1/2)^2}.
\end{split}\end{gathered}\end{equation}
The leading term has the following form:
\begin{equation}
\label{eq:etaBeta}
\re^{2\pi \ri Q}\approx \re^{2\pi \ri\tau\sigma+\ri \eta/2}
\frac{\Gamma(-m+2\sigma)\Gamma(1-2\sigma)}{\Gamma(1-m-2\sigma)\Gamma(2\sigma)} =:\re^{2\pi \ri(\sigma\tau+\beta)}.
\end{equation}
Hence we have \(Q\approx \sigma\tau+\beta\).
The corresponding momentum is \(p\approx 2\pi \ri\sigma\).
 Here we are focusing on the upper sign in \eqref{eq:momentumSingular}, but this does not matter for the computation of the monodromies.

To compute the limit of the Lax matrix we first analyze the element \(12\) of \eqref{eq:A12}. Using \eqref{eq:theta1product} we have
\begin{multline}
\label{eq:A12expansion}
A_{12}(z)=m\frac{(\re^{\pi \ri(z-2Q)}-\re^{-\pi \ri(z-2Q)})\prod_{n=1}^{\infty}(1-\re^{2\pi \ri n\tau} \re^{2\pi \ri(z-2Q)})(1-\re^{2\pi \ri n\tau} \re^{-2\pi \ri(z-2Q)})}{(\re^{\pi \ri z}-\re^{-\pi \ri z})\prod_{n=1}^{\infty}(1-\re^{2\pi \ri n\tau} \re^{2\pi \ri z})(1-\re^{2\pi \ri n\tau} \re^{-2\pi \ri z})}\times
\\\times
\frac{\prod_{n=1}^{\infty}(1-\re^{2\pi \ri n\tau} )^2}{(\re^{2\pi \ri Q }-\re^{-2\pi \ri Q})\prod_{n=1}^{\infty}(1-\re^{2\pi \ri n\tau} \re^{4\pi \ri Q})(1-\re^{2\pi \ri n\tau} \re^{-4\pi \ri Q})}\times 2\pi \ri.
\end{multline}
It is convenient to consider 
\begin{equation}
\label{eq:Resigma}
\mathrm{Re} \sigma\in(0,1/4),
\end{equation}
and then perform analytic continuation to all values of \(\sigma\).
We also consider \(z\) in the following region containing the fundamental domain:
\be  {\rm Im} z\in (-\frac{1}{2}{\rm Im}\tau,\frac{1}{2}{\rm Im}\tau). \ee
Under such assumptions we can just drop the infinite products in \eqref{eq:A12expansion} and write
\begin{equation}\begin{gathered}
\label{eq:31}
A_{12}(z)\approx \frac{2\pi \ri m(\re^{\pi \ri(z-2Q)}-\re^{-\pi \ri(z-2Q)})}{(\re^{\pi \ri z}-\re^{-\pi \ri z})(\re^{2\pi \ri Q }-\re^{-2\pi \ri Q})}=
\frac{-2\pi \ri m}{\re^{2\pi \ri z}-1}+\frac{2\pi \ri m}{\re^{4\pi \ri Q}-1}\approx
\frac{-2\pi \ri m\re^{2\pi \ri z}}{\re^{2\pi \ri z}-1}.
\end{gathered}\end{equation}
The same procedure can be done for \(A_{21}(z)\):
\begin{equation}
\label{eq:34}
A_{21}(z)\approx
\frac{-2\pi \ri m}{\re^{2\pi \ri z}-1}+\frac{2\pi \ri m}{\re^{-4\pi \ri Q}-1}\approx
\frac{-2\pi \ri m}{\re^{2\pi \ri z}-1}.
\end{equation}
Therefore in our approximation the connection matrix has the form
\begin{equation}
\label{eq:35}
A(z)\approx 
2\pi \ri\begin{pmatrix}
\sigma & -\frac{m\re^{2\pi \ri z}}{\re^{2\pi \ri z}-1}\\
-\frac{m}{\re^{2\pi \ri z}-1} & -\sigma
\end{pmatrix}.
\end{equation}
The solution of the linear system \eqref{eq:torusSystem} becomes
\begin{multline}
\label{eq:Yhyp0}
Y(z)\approx(1-\re^{2\pi \ri z})^m \times
\\\times
\begin{pmatrix}
{}_2F_1(m,m+2\sigma,2\sigma,\re^{2\pi \ri z}) &
\frac{-m \re^{2\pi \ri z}}{2\sigma-1} {}_2F_1(1+m,1+m-2\sigma,2-2\sigma,\re^{2\pi \ri z})\\
\frac{m}{2\sigma} {}_2F_1(1+m,m+2\sigma,1+2\sigma,\re^{2\pi \ri z}) &
{}_2F_1(m,1+m-2\sigma,1-2\sigma,\re^{2\pi \ri z})
\end{pmatrix}\times
\\\times
\operatorname{diag}\left((-\re^{2\pi \ri z})^{\sigma},(-\re^{2\pi \ri z})^{-\sigma}\right).
\end{multline}
We now compare this solution in the vicinities of the two  following points. The first one is
\be z=\frac12+\frac{\tau}{2}  \ee
which corresponds to \(\re^{2\pi \ri z}\to -0\), so in this case \eqref{eq:Yhyp0} can be used.
The second point is
\be  z=\frac12-\frac{\tau}{2}\, . \ee
In this case \(\re^{2\pi \ri z}\to -\infty\), so one has to perform analytic continuation of \(Y(z)\) along the straight line from \(\frac12+\frac{\tau}{2}\) to \(\frac12-\frac{\tau}{2}\).
This can be done by using the standard formula for hypergeometric function:
\begin{multline}
\label{eq:FhypReexpansion}
{}_2F_1(a,b,c,x)=\frac{\Gamma(b-a)\Gamma(c)}{\Gamma(b)\Gamma(c-a)} (-x)^{-a} {}_2F_1(a,a-c+1,a-b+1,x^{-1})+
\\+
\frac{\Gamma(a-b)\Gamma(c)}{\Gamma(a)\Gamma(c-b)} (-x)^{-b} {}_2F_1(b-c+1,b,b-a+1,x^{-1}).
\end{multline}
The result of this analytic continuation \(Y_{cont}(z)\) is given by the following explicit formula
\begin{multline}
\label{eq:33}
Y_{cont}(z)\approx(1-\re^{-2\pi \ri z})^m\times
\\\times
\begin{pmatrix}
{}_2F_1(h,1+m-2\sigma,1-2\sigma,\re^{-2\pi \ri z}) &
\frac{m}{2\sigma}{}_2F_1(1+m,m+2\sigma,1+2\sigma,\re^{-2\pi \ri z})\\
\frac{-m\re^{-2\pi \ri z}}{2\sigma-1}{}_2F_1(1+m,1+m-2\sigma,2-2\sigma,\re^{-2\pi \ri z}) &
{}_2F_1(m,m+2\sigma,2\sigma,\re^{-2\pi \ri z}) 
\end{pmatrix}\times
\\\times
\operatorname{diag}\left((-\re^{2\pi \ri z})^{\sigma},(-\re^{2\pi \ri z})^{-\sigma}\right)
\begin{pmatrix}
\frac{\Gamma(2\sigma)^2}{\Gamma(2\sigma-m)\Gamma(2\sigma+m)} &
-\frac{\sin\pi m}{\sin 2\pi\sigma}\\
\frac{\sin\pi m}{\sin  2\pi\sigma} &
\frac{\Gamma(1-2\sigma)^2}{\Gamma(1-m-2\sigma)\Gamma(1+m-2\sigma)}
\end{pmatrix}.
\end{multline}
Now we rewrite the \(B\)-cycle monodromy equation 
\eqref{eq:T and M} as
\begin{equation}\label{eq:38}
	M_B^{(0)}\approx
	Y_{cont}(z)^{-1}\operatorname{diag} \left( \re^{-2\pi \ri\sigma\tau}\re^{-2\pi \ri\beta}, \re^{2\pi \ri\sigma\tau}\re^{2\pi \ri\beta} \right)Y(z+\tau).
\end{equation}
We now use this formula for \(z\) near  \(\frac12-\frac{\tau}{2}\).
In this region \(\re^{2\pi \ri z}\) is large and \(\re^{2\pi \ri(z+\tau)}\) is small.
Hence we get
\begin{multline}
\label{eq:40}
M_B^{(0)}\approx
\begin{pmatrix}
\frac{\Gamma(1-2\sigma)^2}{\Gamma(1-m-2\sigma)\Gamma(1+m-2\sigma)} &
\frac{\sin\pi m}{\sin 2\pi \sigma}\\
-\frac{\sin\pi m}{\sin 2\pi \sigma} & 
\frac{\Gamma(2\sigma)^2}{\Gamma(-m+2\sigma)\Gamma(m+2\sigma)}
\end{pmatrix}
\begin{pmatrix}
1 & \frac{m}{2\sigma}(-\re^{2\pi \ri z})^{-2\sigma}\\
0 & 1
\end{pmatrix}\times
\\\times
\operatorname{diag} \left( \re^{-2\pi \ri\beta},\re^{2\pi \ri\beta} \right)
\begin{pmatrix}
1 & 0\\
(-\re^{2\pi \ri(z+\tau)})^{2\sigma} & 1
\end{pmatrix}.
\end{multline}
In this computation we already neglected the terms of order \(\re^{-2\pi \ri z}\) and \(\re^{2\pi \ri(z+\tau)}\). Therefore,  given \eqref{eq:Resigma},  we can also neglect the terms of order \((\re^{2\pi \ri z})^{-2\sigma}\).
Finally,  by using the definition of \(\beta\) from \eqref{eq:etaBeta} we obtain
\begin{equation}
\label{eq:12}
M_B^{(0)}=
\begin{pmatrix}
\frac{\sin\pi(2\sigma-m)}{\sin2\pi\sigma}\re^{-\ri \eta/2} &
\frac{\sin\pi m}{\sin2\pi\sigma}
\frac{\Gamma(1-2\sigma)\Gamma(2\sigma-m)}{\Gamma(1-m-2\sigma)\Gamma(2\sigma)}\re^{-\ri \eta/2}\\
-\frac{\sin\pi m}{\sin2\pi\sigma}
\frac{\Gamma(1-m-2\sigma)\Gamma(2\sigma)}{\Gamma(1-2\sigma)\Gamma(2\sigma-m)}\re^{\ri \eta/2} & 
\frac{\sin\pi(2\sigma+m)}{\sin2\pi\sigma}\re^{\ri \eta/2}
\end{pmatrix}.
\end{equation}
To get rid of the factors with gamma functions and the two \(\re^{\pm \ri \eta/2}\) in the out of diagonal elements, we
 can perform conjugation by a diagonal matrix. We get \begin{equation}
\label{eq:MB}
M_B=
\begin{pmatrix}
\frac{\sin\pi(2\sigma-m)}{\sin2\pi\sigma}\re^{-\ri \eta/2} &
\frac{\sin\pi m}{\sin2\pi\sigma}\\
-\frac{\sin\pi m}{\sin2\pi\sigma}& 
\frac{\sin\pi(2\sigma+m)}{\sin2\pi\sigma}\re^{\ri \eta/2}
\end{pmatrix}.
\end{equation}
This matrix can also be written in terms of \(\tilde{\eta}\) defined by
\begin{equation}
\label{eq:etaTildeApp}
\re^{\ri\tilde{\eta}/2}=\frac{\sin\pi(2\sigma+m)}{\sin\pi(2\sigma-m)}\re^{\ri \eta/2}.
\end{equation}
We have
\begin{equation}
\label{eq:MBtilde}
M_B=
\begin{pmatrix}
\frac{\sin\pi(2\sigma+m)}{\sin2\pi\sigma}\re^{-\ri\tilde{\eta}/2} &
\frac{\sin\pi m}{\sin2\pi\sigma}\\
-\frac{\sin\pi m}{\sin2\pi\sigma}& 
\frac{\sin\pi(2\sigma-m)}{\sin2\pi\sigma}\re^{\ri\tilde{\eta}/2}
\end{pmatrix}.
\end{equation}
These two representations give different monodromies in the dangerous cases when \be \label{danger}\sin\pi(2\sigma\pm m)=0\,,\ee
which will be explained in Appendix \ref{app:TraceCoordinates}.
\subsection{Other monodromies: \(M_{A}\), \(M_{0}\)}

Now we compute the remaining monodromies  \(M_{A}\), \(M_{0}\).
We take the starting point  to be \(z_0\) in the vicinity  of \(\frac12-\frac{\tau}{2}\). Using expression \eqref{eq:Yhyp0} for Y(z) we get 
\begin{equation}\label{eq:MA}
	M_A=
		\begin{pmatrix}
			\re^{2\pi \ri\sigma} & 0\\
			0 & \re^{-2\pi \ri\sigma}
		\end{pmatrix}.
\end{equation}
Now we compute the analytic continuation around the singular point as in Fig.~\ref{fig:linSystemDomains}:
\begin{equation}\begin{gathered}
\label{eq:42}
Y(z_1)=Y(z_0+\tau)=\mathrm{T}_BY(z_0)M_B,\\
Y(z_2)=Y(z_1-1)=\mathrm{T}_BY(z_0{ -}1)M_B=T_BY(z_0)M_A^{-1}M_B,\\
Y(z_3)=Y(z_2-\tau)=\mathrm{T}_BY(z_0-\tau)M_A^{-1}M_B=Y(z_0)M_B^{-1}M_A^{-1}M_B,\\
Y(z_4)=Y(z_3+1)=Y(z_0+1)M_B^{-1}M_A^{-1}M_B=Y(z_0)M_AM_B^{-1}M_A^{-1}M_B=Y(z_0)M_0,
\end{gathered}\end{equation}
where \(T_B=\re^{2\pi \ri Q\sigma_3}\).
Thus
\begin{equation}
\label{eq:43}
M_0=M_AM_B^{-1}M_A^{-1}M_B.
\end{equation}
We see from this relation that the monodromies corresponding to consecutive pieces of the path should be written from the right to the left.

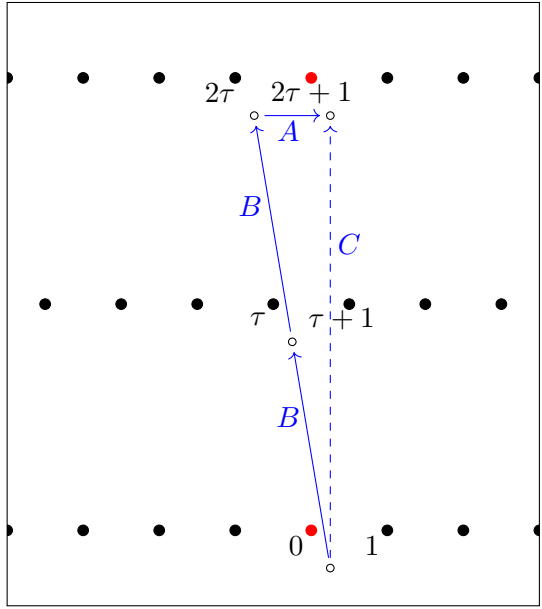
\begin{figure}[h!]
\begin{center}
\begin{tikzpicture}[scale=1]

\clip[draw](-3.5,-4) rectangle (3.5,4);

\tikzmath{int \i,\j;
for \j in {-3,0,3}{
  for \i in {-4,-3,-2,-1,0,1,2,3,4}{
    if (\i-\j/6-1/2)==0
      then {{\draw[fill,red,radius=0.07](\i-\j/6,\j) circle;};}
      else {{\draw[fill,radius=0.07](\i-\j/6,\j) circle;};};
  };};
}

\node at (-0.2,-0.2){\(\tau\)};
\node at (1-0.1,-0.2){\(\tau+1\)};
\node at (-0.7,3-0.2){\(2\tau\)};
\node at (1-0.5,3-0.2){\(2\tau+1\)};
\node at (0.3,-3-0.2){\(0\)};
\node at (1+0.3,-3-0.2){\(1\)};

\node  (p0) at (0.75,-3-0.5){};
\node  (p1) at (0.25,-0.5){};
\node  (p2) at (-0.25,2.5){};
\node  (p3) at (0.75,2.5){};

\draw[->,blue] (p0)--(p1);
\draw[->,blue] (p1)--(p2);
\draw[->,blue] (p2)--(p3);
\draw[->,dashed,blue] (p0)--(p3);

\draw[radius=0.05] (p0) circle;
\draw[radius=0.05] (p1) circle;
\draw[radius=0.05] (p2) circle;
\draw[radius=0.05] (p3) circle;

\node at (0.2,2.3){\color{blue}\(A\)};
\node at (-0.3,1.3){\color{blue}\(B\)};
\node at (0.2,-1.5){\color{blue}\(B\)};
\node at (1,0.8){\color{blue}\(C\)};


\end{tikzpicture}
\end{center}
\caption{\label{fig:linSystemTwisted} Monodromy \(M_C\).}
\end{figure}

Another monodromy which we also need is the one over the straight line \(C\) connecting the point \(0\) with \(2\tau+1\), see Fig.~\ref{fig:linSystemTwisted}.
It corresponds to the self-adjoint operator living on \(C\) when \(\tau=-\frac12+\ri\mathfrak{t}\).
Looking at Fig.~\ref{fig:linSystemTwisted} we conclude that 
\begin{equation}
\label{eq:44}
M_C=M_AM_B^2.
\end{equation}

\subsection{Trace coordinates}
\label{app:TraceCoordinates}

Using the formulas \eqref{eq:MB}, \eqref{eq:MBtilde}, \eqref{eq:MA} we will consider \((\sigma, \eta)\) or \((\sigma, \tilde{\eta})\) as coordinates on the moduli space of monodromy data on torus \footnote{To be precise, \(\eta\) and 	\(\sigma\) are not actual coordinates.
They are defined up to simultaneous sign inversion \((\eta,\sigma)\sim(-\eta,-\sigma)\), and also up to integer shifts.
}. This monodromy manifold can be described in terms of traces of some products of matrices
\begin{equation}
	\label{eq:39}
	p_{AB}=\tr M_AM_B,\quad p_A=\tr M_A,\quad p_B=\tr M_B,\quad p_0=\tr M_0.
\end{equation}
Here \(p_0=2\cos 2 \pi m\) is considered as a fixed parameter. By using the explicit expressions above we easily get the following formulas:
\begin{equation} 	\label{eq:pParameterization}
	\begin{gathered}
		p_{AB}-\re^{2\pi i\sigma}p_B=-2\ri\sin\pi(2\sigma+m)\re^{\ri \frac{\eta}{2}}=-2\ri\sin\pi(2\sigma-m)\re^{\ri \frac{\tilde{\eta}}{2}},\\
		p_{AB}-\re^{-2\pi i\sigma}p_B=2\ri\sin\pi(2\sigma-m)\re^{-\ri \frac{\eta}{2}}=2\ri\sin\pi(2\sigma+m)\re^{-\ri\frac{\tilde{\eta}}{2}},
	\end{gathered}
\end{equation}
By multiplying the above expressions we get the following relation
\begin{equation}
\label{eq:14}
\left(p_{AB}-\re^{2\pi \ri \sigma}p_B \right)\left(p_{AB}-\re^{-2\pi \ri \sigma}p_B \right)=p_0-p_A^2+2,
\end{equation}
or after simplification
\begin{equation}
\label{eq:51}
p_{AB}^2+p_A^2+p_B^2-p_A\,p_B\,p_{AB}=p_0+2.
\end{equation}

This equation defines a surface in \(\mathbb{C}^3\) with coordinates \(p_A,p_B,p_{AB}\). This surface is the monodromy manifold in our case. 

Now we can look to the dangerous points \eqref{danger}. In terms of trace coordinates this equations reads \(p_A^2=p_0+2\). Hence it defines two points on the monodromy manifold 
\begin{equation} \label{sgnpab}
	p_{AB}=\re^{\pm 2\pi \ri\sigma}p_B\,.
\end{equation}
It follows from equations \eqref{eq:pParameterization} that these two points belong to two different charts, one for finite \(\eta\) and another for finite \(\tilde{\eta}\). For example for $\sin (2\sigma+m)=0$ we have finite \(\eta\) for \("+"\) sign in \eqref{sgnpab} and finite \(\eta\) for \("-"\) sign in \eqref{sgnpab}.

The consequence of these considerations is that for generic \(m\) the corresponding monodromy manifold (without the very bad points \(2\sigma\in \mathbb{Z}\) ) can be covered by two charts where either \(\eta\) or \(\tilde{\eta}\) is finite.

\subsection{Diagonalization of \(M_{0}\)}
\label{app:M0diag}

To study  normalizability of the solution to the linear system \eqref{eq:torusSystem} we will need to study its asymptotics around \(0\).
To do this it is very convenient  to work  in a basis where \(M_0\) is diagonal.
There are two possible diagonalizations corresponding to two different charts, where either \(\eta\) or \(\tilde{\eta}\) are kept finite.
These diagonalizations will be denoted by \(^{(I)}\) and \(^{(II)}\), respectively:
\begin{equation}
\label{eq:52}
M_{\nu}^{(I,II)}=\left(T^{(I,II)}\right)^{-1}M_{\nu}T^{(I,II)}.
\end{equation}
The first diagonalization of \(M_0\) is done by the matrix
\begin{equation}
\label{eq:53}
T^{(I)}=
\begin{pmatrix}
\re^{\ri \frac{\eta}{2}+2\pi \ri\sigma}\frac{\sin\pi(2\sigma+m)}{\sin2\pi\sigma\cos\pi m}&
1\\
-\frac{\sin\pi(2\sigma-m)}{\sin2\pi\sigma\cos\pi m} &
\re^{-\ri \frac{\eta}{2}-2\pi \ri\sigma}
\end{pmatrix}.
\end{equation}
It is always non-degenerate, since \(\det T^{(I)}=2\).
The corresponding conjugated monodromy matrices are

\begin{equation}
\label{eq:49}
M_0^{(I)}=M_0^{(II)}=
\begin{pmatrix}
\re^{2\pi \ri m} & 0\\
0 & \re^{-2\pi \ri m}
\end{pmatrix},
\end{equation}

\begin{equation}
\label{eq:48}
M_A^{(I)}=
\begin{pmatrix}
\re^{\pi \ri m}\frac{\cos2\pi\sigma}{\cos\pi m} &
\ri\re^{-\ri \frac{\eta}{2}-2\pi \ri\sigma}\sin2\pi\sigma\\
\frac{\ri\sin\pi(2\sigma+m)\sin\pi(2\sigma-m)}{\re^{-\ri \frac{\eta}{2}-2\pi \ri\sigma}\sin2\pi\sigma\cos^2\pi m} &
\re^{-\pi \ri m}\frac{\cos2\pi\sigma}{\cos\pi m}
\end{pmatrix},
\end{equation}

\begin{equation}
\label{eq:47}
M_B^{(I)}=
\begin{pmatrix}
\frac{\re^{\ri\frac{\eta}{2}}\sin\pi(2\sigma+m)+\re^{-\ri \frac{\eta}{2}}\sin\pi(2\sigma-m)}{2\re^{-\pi \ri m}\cos\pi m \sin 2\pi\sigma} &
\frac{\re^{-\ri \eta}-1}{2\re^{\pi \ri(2\sigma+m)}} \\
\frac{\sin^2\pi(2\sigma-m)-\re^{\ri \eta}\sin^2\pi(2\sigma+m)}{2\re^{-\pi \ri(2\sigma+m)}\cos^2\pi m\sin^22\pi\sigma} & 
\frac{\re^{\ri \frac{\eta}{2}}\sin\pi(2\sigma+m)+\re^{-\ri \frac{\eta}{2}}\sin\pi(2\sigma-m)}{2\re^{\ri\pi m}\cos\pi m \sin2\pi\sigma}
\end{pmatrix}.
\end{equation}
The matrix \(M_C\) has a cumbersome expression, so we present here only its off-diagonal entries:
\begin{equation}
\label{eq:57}
\left(M^{(I)}_C\right)_{12}=
\frac{\left(\cos\pi(\sigma-\frac{m}{2})-\re^{\ri \eta-2\pi \ri\sigma}\cos\pi(\sigma+\frac{m}{2})\right) \left( \re^{\ri \eta-2\pi \ri\sigma}\sin\pi(\sigma+\frac{m}{2})+\sin\pi(\sigma-\frac{m}{2}) \right)}{\re^{3\ri \frac{\eta}{2}+\pi \ri m}\sin2\pi\sigma},
\end{equation}

\begin{multline}
\label{eq:58}
\left(M^{(I)}_C\right)_{21}=
\frac{\left( \re^{\ri \eta-2\pi \ri\sigma}\sin\pi(2\sigma+m)\cos\pi(\sigma+\frac{m}{2})+\sin\pi(2\sigma-m)\cos\pi(\sigma-\frac{m}{2}) \right)}{\re^{\ri \frac{\eta}{2}-\pi \ri m-4\pi \ri\sigma}\sin^32\pi\sigma\cos^2\pi m}\times
\\\times
\left( \sin\pi(2\sigma-m)\sin\pi(\sigma-\frac{m}{2})-\re^{\ri \eta-2\pi \ri\sigma}\sin\pi(2\sigma+m)\sin\pi(\sigma+\frac{m}{2}) \right)
\end{multline}
The second diagonalization of \(M_0\) is done by the matrix 

\begin{equation}
\label{eq:54}
T^{(II)}=
\begin{pmatrix}
1 & 
\re^{\ri \frac{\tilde{\eta}}{2}+2\pi \ri\sigma}\frac{\sin\pi(2\sigma-m)}{\sin2\pi\sigma\cos\pi m}\\
-\re^{-\ri \frac{\tilde{\eta}}{2}-2\pi \ri\sigma} & 
\frac{\sin(2\sigma+m)}{\sin2\pi\sigma\cos\pi m}
\end{pmatrix}\, .
\end{equation}
The conjugated monodromy matrices are
\begin{equation}
\label{eq:55}
M_A^{(II)}=
\begin{pmatrix}
\re^{\ri\pi m}\frac{\cos2\pi\sigma}{\cos\pi m} &
\frac{\ri \sin\pi(2\sigma+m)\sin\pi(2\sigma-m)}{\re^{-\ri \frac{\tilde{\eta}}{2}-2\pi \ri\sigma}\sin2\pi\sigma\cos^2\pi m}\\
\ri \re^{-\ri \frac{\tilde{\eta}}{2}-2\pi \ri\sigma}\sin2\pi\sigma&
\re^{-\pi \ri m}\frac{\cos2\pi\sigma}{\cos\pi m}
\end{pmatrix},
\end{equation}

\begin{equation}
\label{eq:56}
M_B^{(II)}=
\begin{pmatrix}
\frac{\re^{\ri \frac{\tilde{\eta}}{2}}\sin\pi(2\sigma-m)+\re^{-\ri \frac{\tilde{\eta}}{2}}\sin\pi(2\sigma+m)}{2\re^{-\pi \ri m}\sin2\pi\sigma\cos\pi m} &
\frac{\sin^2\pi(2\sigma+m)-\re^{\ri \tilde{\eta}}\sin^2\pi(2\sigma-m)}{2\re^{\pi \ri(m-2\sigma)}\sin^22\pi\sigma\cos^2\pi m}\\
\frac{\re^{-\ri\tilde{\eta}}-1}{2\re^{\pi \ri(2\sigma-m)}} & 
\frac{\re^{\ri \frac{\tilde{\eta}}{2}}\sin\pi(2\sigma-m)+\re^{-\ri \frac{\tilde{\eta}}{2}}\sin\pi(2\sigma+m)}{2\re^{\pi \ri m}\sin2\pi\sigma\cos\pi m}
\end{pmatrix},
\end{equation}

\begin{multline}
\label{eq:60}
\left(M^{(II)}_C\right)_{12}=
\frac{\left( \re^{\ri \tilde{\eta}-2\pi \ri\sigma}\sin\pi(2\sigma-m)\cos\pi(\sigma-\frac{m}{2})+\sin\pi(2\sigma+m)\cos\pi(\sigma+\frac{m}{2}) \right)}{\re^{\ri \frac{\tilde{\eta}}{2}+\pi \ri m-4\pi \ri\sigma}\sin^32\pi\sigma\cos^2\pi m}\times
\\\times
\left( \sin\pi(2\sigma+m)\sin\pi(\sigma+\frac{m}{2})-\re^{\ri \tilde{\eta}-2\pi \ri\sigma}\sin\pi(2\sigma-m)\sin\pi(\sigma-\frac{m}{2}) \right),
\end{multline}

\begin{equation}
\label{eq:59}
\left( M_C^{(II)} \right)_{21}=
\frac{\left(\cos\pi(\sigma+\frac{m}{2})-\re^{\ri \tilde{\eta}-2\pi i\sigma}\cos\pi(\sigma-\frac{m}{2})\right) \left( \re^{\ri \tilde{\eta}-2\pi \ri\sigma}\sin\pi(\sigma-\frac{m}{2})+\sin\pi(\sigma+\frac{m}{2}) \right)}{\re^{3\ri \frac{\tilde{\eta}}{2}-\pi \ri m}\sin2\pi\sigma}.
\end{equation}

\section{Conformal blocks on the torus: conventions}\label{torusconvention}

	The torus conformal blocks almost coincide with the Nekrasov partition functions for $\mathcal{N}=2^*$  four-dimensional theory. Hence here we write the formulas for Nekrasov functions. We follow the notations of \cite{Bershtein:2018zcz}, or rather adopt the notations from loc.~cit.~since there is no $\mathcal{N}=2^*$ there. 
	
	The  Nekrasov function is a product 
	\begin{equation}\label{full2s}
	\mathcal{Z}(a,\alpha;\epsilon_1,\epsilon_2|\mathfrak{q})=\mathcal{Z}_{\mathrm{cl}}\mathcal{Z}_{\mathrm{1-loop}}\cdot
	\varphi(\mathfrak{q})^{1-2\frac{\alpha(\epsilon_1+\epsilon_2-\alpha)}{\epsilon_1\epsilon_2}} \mathcal{Z}^{U(2)}_{\mathrm{inst}}(a,\alpha;\epsilon_1,\epsilon_2|\mathfrak{q})\, ,
	\end{equation}
	where 
	\begin{align}
	\mathcal{Z}_{\mathrm{cl}}(a;\epsilon_1,\epsilon_2|\mathfrak{q})&= \mathfrak{q}^{-a^2/\epsilon_1\epsilon_2},
	\\ 
	\mathcal{Z}_{\mathrm{1-loop}}(a,\alpha;\epsilon_1,\epsilon_2)&=\exp\left(\gamma_{\epsilon_1,\epsilon_2}(2a-\alpha;1)+\gamma_{\epsilon_1,\epsilon_2}(-2a-\alpha;1)-\gamma_{\epsilon_1,\epsilon_2}(2a;1)-\gamma_{\epsilon_1,\epsilon_2}(-2a;1)\right),
	\end{align}
	with \cite[App.~E]{Nakajima:2003uh}
	\be \gamma_{\epsilon_1, \epsilon_2}(x, \Lambda)={\rd \over \rd s}{ \Lambda^s\over \Gamma(s)}\int_0^{\infty}{\rd t \over t}t^s {\re^{-t x} \over (\re^{\epsilon_1 t}-1)(\re^{\epsilon_2 t}-1)}\Big |_{s=0}, \quad {\rm Re}(x)>0.\ee	%
	The function $\mathcal{Z}^{U(2)}_{\mathrm{inst}}(a,\alpha;\epsilon_1,\epsilon_2|\mathfrak{q})$ is defined as a sum over partitions: 
	\begin{equation}
		\begin{aligned}
		\mathcal{Z}_{inst}(a,\alpha;\epsilon_1,\epsilon_2|\Lambda)&=\sum_{\lambda^{(1)},\lambda^{(2)}}\frac{\prod_{i,j=1}^2\mathsf{N}_{\lambda^{(i)},\lambda^{(j)}}(\alpha+a_i-a_j;\epsilon_1,\epsilon_2)}
		{\prod_{i,j=1}^2\mathsf{N}_{\lambda^{(i)},\lambda^{(j)}}(a_i-a_j;\epsilon_1,\epsilon_2)} \mathfrak{q}^{|\lambda^{(1)}|+|\lambda^{(2)}|}, 
		\\
		\mathsf{N}_{\lambda,\mu}(a;\epsilon_1,\epsilon_2)&=\prod_{s\in\mathbb{\lambda}}(a-\epsilon_2(a_{\mu}(s)+1)+\epsilon_1l_{\lambda}(s))\prod_{s\in\mathbb{\mu}}(a+\epsilon_2a_{\lambda}(s)-\epsilon_1(l_{\mu}(s)+1)),
\end{aligned}
\end{equation}
	where $\lambda^{(1)}, \lambda^{(2)}$ are partitions and $|\lambda|=\sum\lambda_j$ denotes the number of boxes. We also use $a_{\lambda}(s), l_{\lambda}(s)$ to denote the lengths of arms and legs for the box $s$ in the Young diagram corresponding to the partition $\lambda$. The parameters $a_1$ and $a_2$ satisfy $a_1=a, a_2=-a$.  The first terms of these functions are 
	\[	
		\mathcal{Z}^{U(2)}_{\mathrm{inst}}(a,\alpha;\epsilon_1,\epsilon_2|\mathfrak{q})=1+\mathfrak{q} \frac{2 (\alpha -\epsilon_1) (\alpha -\epsilon_2) \left(-4 a^2+\alpha ^2-\alpha  \epsilon_1-\alpha  \epsilon_2+\epsilon_1^2+2 \epsilon_1 \epsilon_2+\epsilon_2^2\right)}{\epsilon_1 \epsilon_2 (-2 a+\epsilon_1+\epsilon_2) (2 a+\epsilon_1+\epsilon_2)}+\mathcal{O}(\mathfrak{q}^2).
	\]
	It is sometimes convenient to factor out the $U(1)$ part of the partition function by defining
	\begin{equation}
		\mathcal{Z}^{SU(2)}_{\mathrm{inst}}(a,\alpha;\epsilon_1,\epsilon_2|\mathfrak{q})=\varphi(\mathfrak{q})^{1-2\frac{\alpha(\epsilon_1+\epsilon_2-\alpha)}{\epsilon_1\epsilon_2}} \mathcal{Z}^{U(2)}_{\mathrm{inst}}(a,\alpha;\epsilon_1,\epsilon_2|\mathfrak{q}).
	\end{equation}
	On the CFT side this is the transition from the sum of Virasoro and Heisenberg to Virasoro algebra. The first few terms of this function are 
	\[
		\mathcal{Z}^{SU(2)}_{\mathrm{inst}}(a,\alpha;\epsilon_1,\epsilon_2|\mathfrak{q})=
		1+\mathfrak{q}\left(1-\frac{2 \alpha  (\epsilon_1-\alpha ) 	(\epsilon_2-\alpha ) (-\alpha +\epsilon_1+\epsilon_2)}{\epsilon_1 \epsilon_2 (-2 a+\epsilon_1+\epsilon_2) (2 a+\epsilon_1+\epsilon_2)}\right)+\mathcal{O}(\mathfrak{q}^2).
	\]
The function $\mathcal{Z}^{U(2)}_{\mathrm{inst}}$ has the following reflection symmetry 	\begin{align}
		\mathcal{Z}^{U(2)}_{\mathrm{inst}}(a,\epsilon_1+\epsilon_2-\alpha;\epsilon_1,\epsilon_2|\mathfrak{q})=\mathcal{Z}^{U(2)}_{\mathrm{inst}}(a,\alpha;\epsilon_1,\epsilon_2|\mathfrak{q})~.
	\end{align}
	This follows from the relation 
	\begin{equation}
		\mathsf{N}_{\lambda,\mu}(2a+\alpha;\epsilon_1,\epsilon_2)=(-1)^{|\lambda|+|\mu|}\mathsf{N}_{\mu,\lambda}(\epsilon_1+\epsilon_2-2a-\alpha;\epsilon_1,\epsilon_2)
	\end{equation}
	The classical part $\mathcal{Z}_{\mathrm{cl}}(a;\epsilon_1,\epsilon_2|\mathfrak{q})$ also posses such symmetry, but not $\mathcal{Z}_{\mathrm{1-loop}}(a;\epsilon_1,\epsilon_2|\mathfrak{q})$. Hence we have 	%
	\be\ba \label{eq:alpha symm}
		 \mathcal{Z}(a,\epsilon_1+\epsilon_2-\alpha;\epsilon_1,\epsilon_2|\mathfrak{q}) = C^{\mathrm{refl}}_{\epsilon_1,\epsilon_2}(a,\alpha)\mathcal{Z}(a,\alpha;\epsilon_1,\epsilon_2|\mathfrak{q}) 
	\ea\ee
	where
	\begin{multline}
		C^{\mathrm{refl}}_{\epsilon_1,\epsilon_2}(a,\alpha)=\exp\Big(\gamma_{\epsilon_1,\epsilon_2}(2a{+}\alpha{-}\epsilon_1{-}\epsilon_2;1) + \gamma_{\epsilon_1,\epsilon_2}({-}2a{+}\alpha{-}\epsilon_1{-}\epsilon_2;1) \\ - \gamma_{\epsilon_1,\epsilon_2}(2a{-}\alpha;1) - \gamma_{\epsilon_1,\epsilon_2}({-}2a{-}\alpha;1)\Big).
	\end{multline}
	In the main text we mainly need two special cases of $\mathcal{Z}(a,\alpha;\epsilon_1,\epsilon_2|\mathfrak{q})$, namely the self-dual case ($\epsilon_1+\epsilon_2=0$) and  the Nekrasov-Shatashvili case ($\epsilon_2\rightarrow 0$).
	
	\paragraph{Self-dual case.}
	In this case we impose the condition $\epsilon_1+\epsilon_2=0$. It is also convenient to fix some rescaling freedom and impose $\epsilon_1=-1$, $\epsilon_2=1$. As a consequence we change the notation and use $a=\sigma$, $\alpha=m$. We have
	\begin{equation}
		\mathcal{Z}(a,\alpha;1,-1|\mathfrak{q})= \mathfrak{q}^{\sigma^2} (2\pi)^{m} \frac{G(1+2\sigma-m)G(1-2\sigma-m)}{G(1+2\sigma)G(1-2\sigma)} 
	\varphi(\mathfrak{q})^{1-2m^2} \mathcal{Z}^{U(2)}_{\mathrm{inst}}(\sigma,m, -1,1|\mathfrak{q}).
	\end{equation}
	Here for the $\gamma_{\epsilon_1,\epsilon_2}$ we used the transformation (see e.g. \cite[App. A,B]{Bershtein:2018zcz})
	\begin{equation}
		\exp(\gamma_{1,-1}(x;1))=\exp(-\gamma_{-1,-1}(x+1))=\Gamma_2(x+1;1;1)^{-1}=G(x)\re^{-\zeta'(-1)}(2\pi)^{1-x/2}.
	\end{equation}
	In the main text we use the self-dual limit of the function $\mathcal{Z}^{SU(2)}_{\mathrm{inst}}$ and we denote it by $Z$:
	\be \label{zsd2s}{Z}(\sigma,m,\mathfrak{q})=	\varphi(\mathfrak{q})^{1-2m^2} \mathcal{Z}^{U(2)}_{\mathrm{inst}}(\sigma,m|\mathfrak{q})=\mathcal{Z}^{SU(2)}_{\mathrm{inst}}(\sigma,m|\mathfrak{q}).\ee
	The first few orders read
	\be\label{c1} 
		Z(\sigma,m,\mathfrak{q})=1+\left(1+\frac{\left(m^2-1\right) m^2 }{2 \sigma ^2}\right)\mathfrak{q}+\mathcal{O}(\mathfrak{q}^2).
	\ee

	\paragraph{Nekrasov-Shatashvili limit.}
	In the limit $\epsilon_2 \rightarrow 0$, we have 
	\begin{multline}\label{ns2s}
	\lim_{\epsilon_2\rightarrow 0}\epsilon_2\log \mathcal{Z}(\sigma,\mu+\frac{1}{2};1,\epsilon_2|\mathfrak{q})=F^{\mathrm{NS}}(\sigma,\mu,\mathfrak{q})\\=	- \sigma^2\log\mathfrak{q}+F^{\mathrm{NS}}_{\mathrm{1-loop}}(\sigma,\mu,\mathfrak{q})+F^{\mathrm{NS}}_{\mathrm{inst}}(\sigma,\mu,\mathfrak{q})+	\left(2\mu^2-\frac12\right)\log(\varphi(\mathfrak{q})),
	\end{multline}
	where
	\begin{multline}\label{eq:FNS 1-loop}
		F^{\mathrm{NS}}_{\mathrm{1-loop}}(\sigma,\mu,\mathfrak{q})= - 	\psi^{(-2)}(\frac12+2\sigma-\mu)- \psi^{(-2)}(\frac12-2\sigma-\mu)\\ +\psi^{(-2)}(1+2\sigma)+\psi^{(-2)}(1-2\sigma)+(\mu+\frac{1}{2})\log(2\pi),
	\end{multline}
	and 
	\begin{multline}\label{nsi2s}
	F^{\mathrm{NS}}_{\mathrm{inst}}(\sigma,\mu,\mathfrak{q})=\lim_{\epsilon_2 \rightarrow 0} \mathcal{Z}^{U(2)}_{\mathrm{inst}}(\sigma,\mu+\frac12;1,\epsilon_2|\mathfrak{q})=\frac{( 4 \mu^2-1) (3 + 4 \mu^2 - 16 \sigma^2)}{
		8(1-4\sigma^2)}\mathfrak{q}
	\\ 
	+\left(\frac{3}{4} (4 \mu^2{-}1)+
	\frac{(4 \mu^2{-}1)^2}{64}  
	\left(\frac{3 \left(1{-}4 \mu^2\right)^2}{4 \left(1{-}4 \sigma^2\right)^2}-
	\frac{(1{-}4 \mu^2)^2}{(1{-}4 \sigma^2)^3}+
	\frac{16 \mu^4{-}72 \mu ^2{-}15}{4 (4 \sigma^2{-}1)}+
	\frac{(9{-}4 \mu^2)^2}{16 (1{-}\sigma^2)}\right)
	\right)\mathfrak{q}^2+\mathcal{O}(\mathfrak{q}^3).
	\end{multline}
	The formula for $F^{\mathrm{NS}}_{\mathrm{1-loop}}$ is somehow ambiguous, it depends on the branch of the function $\psi^{(-2)}$. Fortunately $F^{\mathrm{NS}}$ appears in the main text only through $ \mathfrak{q}$ and $\sigma$ derivatives. The term $F^{\mathrm{NS}}_{\mathrm{1-loop}}$ do not depend on $\mathfrak q$. For the $\sigma$ derivative we have
	\be \ba \label{eq:FNS 1-loop derivative}
		\partial_{\sigma}F^{\mathrm{NS}}_{\mathrm{1-loop}}(\sigma,\mu,\mathfrak{q})=&-2 \log\Gamma\left(-\mu +2 \sigma +\frac{1}{2}\right)+2 \log\Gamma\left(-\mu -2 \sigma +\frac{1}{2}\right)
		\\
		&-2 \log\Gamma(1-2 \sigma )+2 \log\Gamma(1+2 \sigma ).
	\ea\ee
	Formally speaking, this function also has monodromy, but now its exponent $\exp( \partial_{\sigma}F^{\mathrm{NS}}_{\mathrm{1-loop}})$ is well defined, so the eqs. \eqref{eq:etaMinus} and \eqref{eq:6} make sense. {The formula \eqref{eq:FNS 1-loop derivative} is actually used in the main text.}

	In order to obtain the formula \eqref{eq:FNS 1-loop} it is useful to consider $\gamma^{\mathrm{NS}}(x)=\lim_{\epsilon_2 \rightarrow 0}\epsilon_2 \gamma_{1,\epsilon_2}(x;1)$. By using the expansion of the exponent in terms of Bernoulli numbers, we get the asymptotic series (for $x>0$)
	\begin{multline}
	\gamma^{\mathrm{NS}}(x)= \left(\frac{3}{4}-\frac{1}{2} \log x\right)x^2+\frac{1}{2}(1-\log x)x-\frac1{12}\log x+\sum_{m=3} \frac{B_m}{m(m-1)(m-2)}x^{2-m}\\
	= -\psi^{(-2)}(x+1)+\frac{1}{2}(x+1)\log (2\pi)+\log(A),
	\end{multline}
	where $A$ is Glaisher--Kinkelin constant.

	We used the polygamma function $\psi^{(-2)}(x)$ in the formula \eqref{eq:FNS 1-loop}, but it can be also written by using Barnes $G$ functions (as in self-dual case) thanks to the formula (for $x>0$)
	\begin{equation}
		G(x+1)=\exp\left(-\psi^{(-2)}(x)+x \log\Gamma(x)-\frac{1}{2}x^2+\frac{1+\log(2\pi)}{2}x\right).
	\end{equation}

\section{Some proofs}\label{proofs}

\subsection{Proof of the relation \eqref{eq:-2 1Hirota}} \label{app:proof -2}
The proof is similar to the one in \cite{Shchechkin:2020ryb}. We will use the notation $\beta_0^0$ and $\beta_1^0$ for the functions which appear in the algebraic blowup relations \eqref{bua}
\begin{equation}
	\beta_0^0(\mathfrak{q})=\frac{\theta_3(0| 2\tau)}{\varphi(\mathfrak{q})},\quad 
	\beta_1^0(\mathfrak{q})=\frac{\theta_2(0| 2\tau)}{\varphi(\mathfrak{q})}.
\end{equation}
It follows from the blowup relations \eqref{bua} and \eqref{bu3},\eqref{bu4} that for $j=0,1$
\begin{multline}\label{eq:be cor}
	\sum _{n \in \mathbb{Z}+\frac{j}{2}} \big( \partial_{\log\mathfrak{q}}\mathcal{Z}(a+n\epsilon_1,\alpha;\epsilon_1,\epsilon_2-\epsilon_1|\mathfrak{q})\big) \mathcal{Z}(a+n\epsilon_2,\alpha;\epsilon_1-\epsilon_2,\epsilon_2|\mathfrak{q})
	\\
	=
	\frac{1}{\epsilon_1-\epsilon_2}\left(\epsilon_1 \beta_j^{1,1}(\mathfrak{q})+\alpha\beta_j^{1,2}(\mathfrak{q})-\epsilon_2 \beta_j^0(\mathfrak{q}) \partial_{\log \mathfrak{q}}\right) \mathcal{Z}(a,\alpha;\epsilon_1,\epsilon_2|\mathfrak{q}).
\end{multline}
We will use Nekrasov functions depending on different $\epsilon$ parameters. Let us denote 
\begin{equation}
	\begin{aligned}
		\mathcal{Z}^{(1)}(a)&=\mathcal{Z}(a,\alpha;\epsilon_1,\epsilon_2-\epsilon_1|\mathfrak{q}),&\quad		\mathcal{Z}^{(2)}(a)&=\mathcal{Z}(a,\alpha;\epsilon_1{-}\epsilon_2,2\epsilon_2{-}\epsilon_1|\mathfrak{q}),
		\\		\mathcal{Z}^{(3)}(a)&=\mathcal{Z}(a,\alpha;\epsilon_1{-}2\epsilon_2,\epsilon_2|\mathfrak{q}),&\quad 	\mathcal{Z}^{(4)}(a)&=\mathcal{Z}(a,\alpha;\epsilon_1{-}\epsilon_2,\epsilon_2|\mathfrak{q}).
	\end{aligned}
\end{equation}
By $\partial^{(l)}$ we denote the operator $\partial_{\log \mathfrak{q}}$ acting on the argument of  $\mathcal{Z}^{(l)}$, for $l=1,2,3,4$.
We also denote 
\begin{equation}
	\widehat{\mathcal{Z}}(a)=\sum_{2r \in \mathbb{Z}} \hspace{-0.2cm}\Big(\epsilon_1 \partial^{(1)}+(2\epsilon_2{-}\epsilon_1)\partial^{(2)}-\epsilon_2\gamma_0(\mathfrak{q})\Big)\, \mathcal{Z}^{(1)}(a+r\epsilon_1)\mathcal{Z}^{(2)}(a+r(2\epsilon_2{-}\epsilon_1)).
\end{equation}
Then equation \eqref{eq:-2 1Hirota} is equivalent to $\widehat{\mathcal{Z}}=0$. We have
%
\begin{multline}\label{eq:ZZZ=0}
	\hspace{-0.3cm} \sum_{2s \in \mathbb{Z}} \widehat{\mathcal{Z}}(a+s\epsilon_1) \mathcal{Z}^{(3)}(a+2s \epsilon_2)
	\\
	=\hspace{-0.2cm}\sum_{2r,2s \in \mathbb{Z}} \hspace{-0.2cm}\Big(\epsilon_1 \partial^{(1)}+(2\epsilon_2{-}\epsilon_1)\partial^{(2)}-\epsilon_2\gamma_0\Big)\, \mathcal{Z}^{(1)}(a+(r{+}s)\epsilon_1)\mathcal{Z}^{(2)}(a+r(2\epsilon_2{-}\epsilon_1)+s \epsilon_1)\mathcal{Z}^{(3)}(a+2s \epsilon_2)
	\\
	=\hspace{-0.4cm}\sum_{2n \in 2\mathbb{Z}+j, j=0,1} \hspace{-0.4cm}\Big(\epsilon_1 \beta_j^0\partial^{(1)}-\left((\epsilon_1{-}\epsilon_2) \beta_j^{1,1}+\alpha\beta_j^{1,2}-\epsilon_2 \beta_j^0 \partial^{(4)}\right)-\epsilon_2\gamma_0 \beta_j^0\Big) \mathcal{Z}^{(1)}(a+n\epsilon_1)\mathcal{Z}^{(4)}(a+n\epsilon_2)
	\\
	=\sum_{j=0,1}\Big( \beta_j^0 \big((\epsilon_1{+}\epsilon_2) \beta_j^{1,1}+\alpha\beta_j^{1,2}\big)-\big((\epsilon_1{-}\epsilon_2) \beta_j^{1,1}+\alpha\beta_j^{1,2}\big)\beta_j^0-  \epsilon_2 \gamma_0 (\beta_j^0)^2    \Big) \mathcal{Z}(a,\alpha;\epsilon_1,\epsilon_2|\mathfrak{q})
	\\
	=\epsilon_2 \Big(\partial_{\mathfrak{q}} \big((\beta_0^0)^2 +(\beta_1^0)^2 \big) - \gamma_0 \big((\beta_0^0)^2 +(\beta_1^0)^2 \big)  \Big) \mathcal{Z}(a,\alpha;\epsilon_1,\epsilon_2|\mathfrak{q})=0.
\end{multline}
Here for shortness we omit the  $\mathfrak{q}$ dependence in the $\beta$ and $\gamma$ functions. In the first transformation we change variables to $n=r+s$ and use \eqref{eq:be cor} for $\partial^{(2)}$. In the second transformation we used the blowup relations \eqref{bua},\eqref{bu3},\eqref{bu4}. Then we used the relation $\beta_j^{1,1}(\mathfrak{q})=\partial_{\log \mathfrak{q}}\beta_j^0(\mathfrak{q})$ and the definition of $\gamma_0(\mathfrak{q})$, $\beta_j^0(\mathfrak{q})$.

If we decompose 
\begin{equation}
	\widehat{\mathcal{Z}}(a)=\mathfrak{q}^{2a^2/(\epsilon_1-2\epsilon_2)\epsilon_1}\sum_{2N\in \mathbb{Z}_{\geq 0}} \widehat{\mathcal{Z}}_N(a) \mathfrak{q}^N,
	\quad
	\mathcal{Z}^{(3)}(a)=\mathfrak{q}^{a^2/(2\epsilon_2-\epsilon_1)\epsilon_2}\sum_{N\in \mathbb{Z}_{\geq 0}} \mathcal{Z}^{(3)}_N(a) \mathfrak{q}^N,
\end{equation}
then using $\mathcal{Z}^{(3)}_0(a)\neq 0$ and \eqref{eq:ZZZ=0} we get by induction that $\widehat{\mathcal{Z}}_N(a)=0$ for any $N$.

\subsection{Proofs of the relations \eqref{eq:2tau=tau 2} and \eqref{eq:2tau=tau 4}} \label{app:2tau=tau}
\begin{proof}[Proof of the relation \eqref{eq:2tau=tau 2}]
The proof is just a computation based on the power series expansion of theta function. Let $y=\re^{2\pi \ri z}$ 
\begin{multline}
	\theta_3(z|2\tau)^2\, \big(\partial_{\log \mathfrak{q}}^2 \log \theta_3(z|2\tau)\big) +\theta_2(z|2\tau)^2\, \big(\partial_{\log \mathfrak{q}}^2 \log \theta_2(z|2\tau)\big)
	\\
	= \Big(\theta_3(z|2\tau)\,\partial_{\log {\mathfrak{q}}}^2\theta_3(z|2\tau)- \big(\partial_{\log {\mathfrak{q}}}\theta_3(z|2\tau)\big)^2+\theta_2(z|2\tau)\,\partial_{\log {\mathfrak{q}}}^2\theta_2(z|2\tau)- \big(\partial_{\log {\mathfrak{q}}}\theta_2(z|2\tau)\big)^2\Big)
	\\
	=\frac12\sum_{2m,2n\in \mathbb{Z}+j, j=0,1}\mathfrak{q}^{m^2+n^2} y^{n+m}\Big(n^4-2n^2m^2+m^4\Big)
	=
	\frac12\sum_{a,b\in \mathbb{Z}}\mathfrak{q}^{(a^2+b^2)/2} y^{a} a^2b^2
	\\
	=\frac12\Big(\sum_{a\in \mathbb{Z}}a^2\mathfrak{q}^{a^2/2} y^{a}\Big)\, \Big(\sum_{b\in \mathbb{Z}}b^2\mathfrak{q}^{b^2/2} \Big)=2\big(\partial_{\log \mathfrak{q}}\theta_3(z|\tau)\big)\,\big(\partial_{\log \mathfrak{q}} \theta_3(0|\tau)\big).
\end{multline}
Here we changed variables in the sum by $a=n+m$ and $b=n-m$.
\end{proof}

\begin{proof}[Proof of the relation \eqref{eq:2tau=tau 4}]
	The notations are as before,
	\begin{multline}
		\theta_3(z|2\tau)^2\, \big(\partial_{\log \mathfrak{q}}\partial_{2 \pi \ri z} \log \theta_3(z|2\tau)\big) +\theta_2(z|2\tau)^2\, \big(\partial_{\log \mathfrak{q}}\partial_{2 \pi \ri z} \log \theta_2(z|2\tau)\big)=
		\\
		= \Big(\theta_3(z|2\tau)\,\partial_{2 \pi \ri z}\partial_{\log {\mathfrak{q}}}\theta_3(z|2\tau)- \big(\partial_{2 \pi \ri z}\theta_3(z|2\tau)\big)\big(\partial_{\log {\mathfrak{q}}}\theta_3(z|2\tau)\big)+[\theta_3 \leftrightarrow \theta_2]\Big)
		\\
		=\frac12\sum_{2m,2n\in \mathbb{Z}+j, j=0,1}\mathfrak{q}^{m^2+n^2} y^{n+m}\Big(n^3-n^2m-nm^2+m^3\Big)
		=
		\frac12\sum_{a,b\in \mathbb{Z}}\mathfrak{q}^{(a^2+b^2)/2} y^{a} ab^2
		\\
		=\frac12\Big(\sum_{a\in \mathbb{Z}}a\mathfrak{q}^{a^2/2} y^{a}\Big)\, \Big(\sum_{b\in \mathbb{Z}}b^2\mathfrak{q}^{b^2/2} \Big)=\big(\partial_{2 \pi \ri z}\theta_3(z|\tau)\big)\,\big(\partial_{\log \mathfrak{q}} \theta_3(0|\tau)\big).
	\end{multline}
\end{proof}

\bibliographystyle{JHEP}
\bibliography{biblio}

\footnotesize

\bigskip
\noindent \textsc{Landau Institute for Theoretical Physics, Chernogolovka, Russia,\\
	Center for Advanced Studies, Skoltech, Moscow, Russia,\\
	HSE -- Skoltech International Laboratory of Representation Theory and Mathematical Physics, HSE University, Moscow, Russia}

\emph{E-mail}:\,\,\textbf{mbersht@gmail.com}\\

\noindent \textsc{Center for Advanced Studies, Skoltech, Moscow, Russia,\\
	HSE -- Skoltech International Laboratory of Representation Theory and Mathematical Physics, HSE University, Moscow, Russia}

\emph{E-mail}:\,\,\textbf{pasha145@gmail.com}\\

\noindent \textsc{Section de Math\'{e}matiques, Universit\'{e} de Gen\`{e}ve, 1211 Gen\`{e}ve 4, Switzerland,\\
	Theoretical Physics Department, CERN, 1211 Geneva 23, Switzerland}

\emph{E-mail}:\,\,\textbf{alba.grassi@cern.ch}

\end{document}